\documentclass{article}[11pt]
\usepackage[utf8]{inputenc}
\usepackage{amsmath,amsthm,amsfonts,amssymb,amscd}
\usepackage{multirow,booktabs}
\usepackage[table]{xcolor}
\usepackage{fullpage}
\usepackage{dsfont}
\usepackage{multicol}
\usepackage{lastpage}
\usepackage{enumitem}
\usepackage{fancyhdr}
\usepackage{mathrsfs}
\usepackage{wrapfig}
\usepackage{setspace}
\usepackage{calc}
\usepackage{multicol}
\usepackage{cancel}

\usepackage{empheq}
\usepackage{framed}
\usepackage[most]{tcolorbox}
\usepackage{xcolor}

\usepackage{graphicx}
\usepackage[colorlinks=true, allcolors=blue]{hyperref}
\usepackage[explicit]{titlesec}
\usepackage{authblk}

\colorlet{shadecolor}{orange!15}
\theoremstyle{definition}

\newcommand\aref[1]{Appendix~\ref{#1}}

\newcommand\eref[1]{Eq.~\ref{#1}}
\newcommand\sref[1]{Section~\ref{#1}}
\newcommand\fref[1]{Fig.~\ref{#1}}

\newcommand\tr{\mathrm{Tr}\,}
\newcommand\dd{\mathrm{d}}
\newcommand\ee{\mathrm{e}}

\newcommand\sgn{\mathrm{sgn}}

\usepackage{tikz}
\usepackage{pgfplots}
\usetikzlibrary{decorations.markings,calc,fadings,decorations.pathreplacing, patterns, decorations.pathmorphing,positioning,arrows,arrows.meta,matrix}
\usepgfplotslibrary{fillbetween}

\tikzset{ 
    A/.style={
        matrix of nodes,
        row sep= 0cm,
        column sep= 0cm,
        nodes={
            rectangle,
            draw=black,
            fill=white,
            align=center
        },
        minimum height = 0.3cm,
        text depth = 0.2cm,
        text height = 0.3cm,
        text width = 0.5cm,
        nodes in empty cells,
        every even row/.style={
            nodes={fill=green!30}
        },
        row 2 column 3/.style={
        nodes={fill=blue!30}
        }
    }
}

\title{Grokking phase transitions in learning local rules with gradient descent}
\author{Bojan Žunkovič \footnote{\href{mailto:bojan.zunkovic@fri.uni-lj.si}{bojan.zunkovic@fri.uni-lj.si}}}
\affil{Faculty of computer and information science, University of Ljubljana, Ljubljana, Slovenia}

\author{Enej Ilievski}
\affil{Faculty of mathematics and physics, University of Ljubljana, Ljubljana, Slovenia}

\date{}

\begin{document}
\maketitle

\begin{abstract}
We discuss two solvable grokking (generalisation beyond overfitting) models in a rule learning scenario. We show that grokking is a phase transition and find exact analytic expressions for the critical exponents, grokking probability, and grokking time distribution. Further, we introduce a tensor-network map that connects the proposed grokking setup with the standard (perceptron) statistical learning theory and show that grokking is a consequence of the locality of the teacher model. As an example, we analyse the cellular automata learning task, numerically determine the critical exponent and the grokking time distributions and compare them with the prediction of the proposed grokking model. Finally, we numerically analyse the connection between structure formation and grokking.
\end{abstract}

\tableofcontents

\section{Introduction}
Despite recent progress in understanding the double descend phenomena \cite{belkin2019reconciling, nakkiran2021deep, krogh1991simple, pezeshki2021multi} we still do not have a complete theory of generalisation in over-parameterised models. Two recent empirical observations, neural collapse~\cite{papyan2020prevalence} and grokking (generalisation beyond over-fitting)~\cite{power2022grokking}, can help us understand the training and generalisation properties of over-parameterised models. 

Neural collapse occurs in the terminal phase of training, i.e. the phase with zero train error. It refers to the collapse of the $N-$dimensional, last-layer features (input to the last/classification layer)~\cite{papyan2020prevalence} to a $(C-1)$-dimensional equiangular tight frame (ETF) structure, where $C$ is the number of classes. The feature vectors converge towards the vertices of the ETF structure such that features for each class are close to one vertex. Also, the distance between the vertices is much larger than all intra-class feature variances. We can partially understand neural collapse within the unconstrained features and local elasticity models \cite{kothapalli2022neural}. However, its role in generalisation, relation to grokking, and appearance of different latent space structures are still not completely understood.

Grokking also occurs during the terminal phase of training. When training on algorithmic datasets past the zero train error, a sudden decrease of the test error from approximately one to zero is observed~\cite{power2022grokking}. The grokking phenomenon has been discussed within an effective theory approach~\cite{liu2022towards}, where an empirical connection between representation/structure formation and generalisation has been made. An empirical study \cite{thilak2022slingshot} established a relation between grokking and training loss spikes and weight norm increase. However, no exactly solvable model exhibiting the grokking phenomenon has been discussed so far. Further, it is not clear how to reconcile grokking with the standard generalisation theory based on statistical learning methods~\cite{engel2001statistical}. The statistical learning theory predicts (in a teacher-student setting) an algebraic (as $t^{-\nu}$, where $\nu=1$ for most learning rules) decrease of the generalisation error with training time $t$ (or a number of training samples)~\cite{engel2001statistical}.

Grokking and neural collapse (or latent-space structure formation in general) have been observed in over-parametrised models. However, we do not know what is the minimal framework within which we can understand these phenomena or if they are genuinely deep-network phenomena. We aim to formulate a simple solvable model of grokking and relate it to latent-space structure formation and other common deep-network training features, e.g. spikes in the training loss.

\textbf{Main contributions--} We have four main contributions:
\begin{itemize}
    \item We propose a simple learning scenario that exhibits grokking (\sref{sec:toy grokking model}). We study two solvable models where grokking is a \textit{phase transition} to zero test error and calculate exact critical exponents and grokking-time distributions.
    
    \item We discuss the teacher-student model within the tensor network approach and map the standard supervised statistical-learning scenario in the thermodynamic limit to the proposed grokking setup (\sref{sec:tensor network model}). 
    
    \item We numerically study grokking and structure formation on the example of learning a 1D cellular automaton rule 30 (\sref{sec:tensor network model}). We show that sudden spikes in the training loss correspond to structural changes in the latent space representation of the data.
    
    \item Our analytical results and numerical experiments show a significant difference between $L_1$ and $L_2$ regularisations. The $L_1$ regularised models have a larger grokking probability, shorter grokking time, shorter generalisation time, and smaller effective dimension compared to $L_2$ regularised models. 
\end{itemize}

\textbf{Broader impact--} The proposed exactly-solvable grokking models are a step towards theoretical understanding of the late learning phase and generalisation benefits of the terminal phase of training. The introduced tensor-network map connects the standard teacher-student setup in the thermodynamic limit with the proposed grokking setup. It offers a new tool for studying generalisation properties of local rules (local teacher-student models), which could lead to more complex learning dynamics (compared to the standard infinite-range rules). 

Although based on simple models, our results can be relevant also for deep learning training practice. We conjecture that good generalisation is more likely in models with latent space data distributions with small effective dimension. Our results hint that $L_1$ regularisation can improve the generalisation properties of deep models compared to $L_2$ regularisation. Further, we show that spikes in the loss (which often occur during training of deep neural networks) correspond to latent space structural changes that can be beneficial or detrimental for generalisation. Assuming this is the case also in deep networks, we can use the information about the latent space effective dimension to revert the model to a state before the spike or continue training with the current model. 

\section{Related work}
A sudden transition from zero to 100\% accuracy on algorithmic datasets in over-fitted transformer models has been first described in \cite{power2022grokking} and named grokking. In the grokking phase, a formation of simple structures reflecting the properties of the problem have been observed. This finding contradicts the common practice of early stopping and supports recent observations on the benefits of the terminal phase of training~\cite{soudry2018implicit,belkin2019reconciling,merrill2022extracting} and the double descend phenomena~\cite{belkin2019reconciling, d2020triple,nakkiran2021deep,pezeshki2022multi}. In \cite{liu2022towards}, an effective theory of grokking has been proposed. Within the effective theory we can calculate the critical training size to observe grokking. The authors relate grokking with a good representation (or structure formation) and introduce it as a phase between generalisation and memorisation. We go beyond these findings since we obtain exact solutions for the proposed setup and calculate even the grokking-time probability density function (PDF). A systematic experimental study of the grokking phenomena has been presented in \cite{thilak2022slingshot}. A sling-shot mechanism (related to edge of stability \cite{cohen2021gradient}) has been proposed as a necessary condition for grokking. The sling-shot mechanism refers to the occurrence of cyclic spikes in the training loss and steps in the weight norms during training. The sling-shot behavior is not restricted to algorithmic datasets but is present also in various common classification tasks \cite{thilak2022slingshot}. We find a similar behaviour, i.e. that the grokking coincides with train loss spikes. Moreover, we connect training loss spikes with discontinuous step-like evolution of the effective dimension of the latent space representation of the data, which indicate structural changes of the latent space representation. 

A particular structure formation common in deep classification neural networks is the neural collapse (NC). It refers to four empirical/numerical observations in training deep neural network classifiers~\cite{papyan2020prevalence}:
\begin{itemize}
    \item\textbf{(NC1) Variability collapse --} variations of within class features become negligible
    \item\textbf{(NC2) Convergence to equiangular tight frame (ETF)--} class mean vectors form an equal-sized angles between any given pair
    \item\textbf{(NC3) Convergence to self-duality--} the class means and linear classifiers converge to each other, up to rescaling 
    \item\textbf{(NC4) Simplification to nearest class center--}  the network classifier converges to a classifier that selects the class with the nearest train class mean.
\end{itemize}
The role of the loss function, the regularisation,  the batch normalisation, and the optimizer have been discussed within the unconstrained features model~\cite{mixon2020neural,fang2021exploring,kothapalli2022neural} and the local elasticity assumption~\cite{kothapalli2022neural}. The relation of NC to generalisation properties has been discussed in \cite{zhu2021geometric,hui2022limitations,kothapalli2022neural}. However, no relation to grokking has been discussed so far. Although we do not observe NC as defined above, our findings regarding the spikes in the training loss and latent-space data structure might also be relevant for the NC dynamics.

Our main technical tools are tensor networks which are models obtained by contracting many low-dimensional tensors. Tensor networks have been very successful in modelling many-body quantum systems. Recently, they have also been applied to machine learning tasks. In particular to classification problems \cite{stoudenmire2016supervised,stoudenmire2018learning, efthymiou2019tensornetwork, liu2019machine,martyn2020entanglement,meng2020residual,chen2021residual,kong2021quantum}, generative modelling \cite{cheng2019tree,stokes2019probabilistic,sun2020generative,liu2021tensor}, sequence and language modelling \cite{pestun2017tensor,guo2018matrix, bradley2020modeling, bradley2020language, zunkovic2022Deep}, anomaly detection \cite{wang2020anomaly,streit2020network}. Besides, tensor networks have been used as tools to advance machine learning theory by a derivation of interesting generalisation bounds \cite{bradley2020modeling}, information theoretical insights \cite{cohen2016expressive,deng2017quantum,levine2017deep,glasser2019expressive}, and new connections between machine learning and physics \cite{chen2018equivalence,dymarsky2021tensor,adhikary2021quantum}. Particularly relevant for latent space structure formation is the connection between recurrent neural networks (RNN) and matrix product states~\cite{wu2022tensor}. In \cite{merrill2022extracting} benefits of the terminal phase of training for state automata extraction from RNNs (and hence matrix product state tensor networks) have been discussed. The authors find internal state space compression and increased extraction in the terminal phase of training. This is similar to our findings of reduced effective dimension in the latent (internal) space. In contrast to \cite{merrill2022extracting}, we introduce a new tensor network, similar to the tensor-network attention model~\cite{zunkovic2022Deep}, and study grokking and structure formation in a teacher-student learning setup. 

Finally, our work is related to the statistical-mechanics theory of supervised learning~\cite{engel2001statistical}. In the supervised perceptron teacher-student case, an algebraic decrease of the generalisation error with the training set size (training time) has been predicted~\cite{engel2001statistical}. A first-order phase transition has been derived only in a restricted setting of discrete weights~\cite{engel2001statistical}. These results are typically based on the replica method~\cite{gardner1989three} which requires the thermodynamic limit, where both the number of samples and the dimension are large and their ratio is fixed. Outstanding recent results in this direction concern the analysis of optimal generalisation errors of generalised linear models~\cite{barbier2019optimal, carleo2019machine}. We study the same teacher-student scenario but with a restriction to a local teacher (still within the thermodynamic limit). The locality of the teacher/rule enables us to map the problem to a finite-dimensional latent space where we discuss grokking (a second-order phase transition) and latent-space structure formation.

\section{Perceptron grokking}
\label{sec:toy grokking model}
We consider a simple binary classification problem that exhibits the grokking phenomena. Let us assume we have a dataset $\mathcal{D}$ consisting of $(\tilde{x}^i,y^i)\in\mathcal{D}$, with two linearly separable classes ($y^i\in \{-1,1\}$) and $D-$dimensional features $\tilde{x}^i\in\mathds{R}^D$. More precisely, the probability densities for the positive ($P^+$) and the negative ($P^-$) class are linearly separable in $\mathds{R}^D$. Our model class is a simple perceptron in $D$ dimensions, namely
\begin{align}
    \label{eq:model}
    f(\tilde{x}) = \sgn(\hat{y}),\quad \hat{y}=w\cdot \tilde{x}+b,
\end{align}
where $w,x \in \mathds{R}^D$ and $b\in\mathds{R}$. We sample $N$ positive and $N$ negative samples and then train the model with gradient descend
\begin{align}
    \frac{\partial \theta}{\partial t} &=-\frac{\partial \mathcal{R}}{\partial\theta}, \\ \nonumber
    \mathcal{R}&= \frac{1}{2N}\sum_{i=1}^{2N} \frac{1}{2}|\hat{y}^i-y^i|^2 + \lambda_1||\theta||_1+\frac{\lambda_2}{2}||\theta||_2,
\end{align}
where $\lambda_1$, $\lambda_2$ denote regularisation parameters, $\theta$ denotes the collection of all model parameters $w$, $b$, and $||\bullet||_{1,2}$ denote the one and two norm. By construction, the setup displays the grokking phenomena due to the separability assumption. The grokking probability, grokking time, and the critical exponent depend on the setup details.

The suggested setup is relevant in the transfer learning scenario, where we initially train a model on one task and then retrain only the final classification layer on another task. Additionally, in \sref{sec:tensor network model} we construct a tensor network map that connects the standard teacher-student statistical learning scenario in the thermodynamic limit to the setup described above.

In the following, we will explicitly calculate the model parameter dynamics, the test error dynamics, the critical exponent, the grokking probability, and the grokking-time probability density function (PDF) for particular choices of the dimension $D$ and data probability densities $P^\pm$.

\subsection{1D exponential model}
\label{sec:model 1D}
We start by considering a simple, one-dimensional model where we obtain all results in closed form. Although the model is not applicable to the real-world scenario it captures several qualitative features and provides a starting point to study more realistic models. 

\begin{figure}[!htb]
\centering
\begin{tikzpicture}[domain=-6:6]

\begin{axis}[
   		xmin=-5, xmax=5,
   		ymin= -1/32, ymax= 1/5,
   		axis lines = middle,
   		xlabel = {$\tilde{x}$},
   		xtick = \empty,
   		ytick = \empty]

\addplot [domain=1:4, samples=100, thick, name path = R, color=blue!50] {(1/8)*exp(-(x-1))};
\addplot [domain=-4:-1, samples=100, thick, name path = L, color=orange!50] {(1/8)*exp(x+1)};

\path [name path=xaxis] (axis cs: -5,0) -- (axis cs: 5,0);
\addplot[blue!10,opacity=0.5] fill between [of= R and xaxis, soft clip={domain=1:4}];
\addplot[orange!10,opacity=0.5] fill between [of= L and xaxis, soft clip={domain=-4:-1}];

\node[color=blue, font=\footnotesize] at (axis cs: 2.4,0.1) {\large $P^{+}(\tilde{x})$};
\node[color=orange, font=\footnotesize] at (axis cs: -2.4,0.1) {\large $P^{-}(\tilde{x})$};

\draw[black!80,dashed] (axis cs: 1,1/6) -- (axis cs: 1,0) node[below] {\large $\varepsilon$};
\draw[black!80,dashed] (axis cs: -1,1/6) -- (axis cs: -1,0) node[below] {\large $-\varepsilon$};
\draw[teal,dashed] (axis cs: 2,1/16) -- (axis cs: 2,0) node[below] {\large $b$};

\draw [fill=blue,blue!20] (axis cs: 1,-0.002) rectangle (axis cs: 2,0.002); 

\end{axis}

\end{tikzpicture}
\caption{A schematic representation of the linearly separable random 1D model. The model is represented by $b$. The samples between $\epsilon$ and $b$ (marked by thick blue line) are incorrectly classified.}
\label{fig:1d model}
\end{figure}
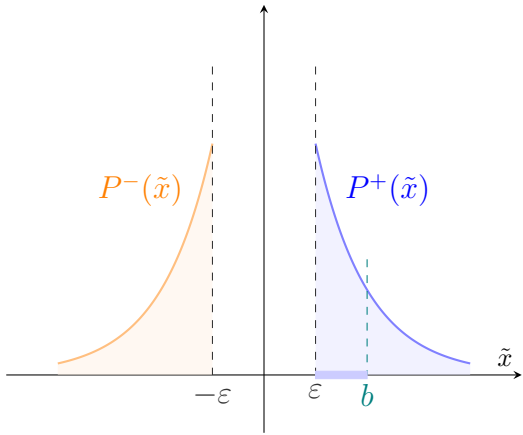

The dataset distribution is shown in \fref{fig:1d model}. Positive and negative samples follow the same probability, i.e. $P^+(\tilde{x})=P^-(-\tilde{x})$. The minimal distance between the positive and negative samples is $2\epsilon$, therefore $P^\pm(|\tilde{x}|\leq\epsilon)=0$.

Since the input $x$ is one dimensional \eref{eq:model} reduces to 
\begin{align}
    f(x)=\sgn(x-b),
\end{align}
where $b$ is the sole model parameter (we fix the weight $w=1$). As described above, we train the model with gradient descend and loss
\begin{align}
    \label{eq:loss general}
    \mathcal{R}=\frac{1}{2N}\sum_{i=1}^{2N}\frac{1}{2}((\tilde{x}^i-b)-y^i)^2+\frac{\lambda_2b^2}{2}+\lambda_1 |w|.
\end{align}
We also assume that the training dataset is balanced, i.e. $\sum_{i=1}^{2N}y^i=0$.

\subsubsection{Test error dynamics}
First, we calculate the model parameter dynamics governed by the negative gradient of the loss function
\begin{align}
    \frac{\partial b}{\partial t}=-\frac{\partial \mathcal{R}}{\partial b}=\frac{1}{2N}\sum_{i=1}^{2N}(\tilde{x}^i-b-y^i)-\sgn(b)\lambda_1-\lambda_2 b=\bar{x}-\sgn(b)\lambda_1-(1+\lambda_2)b,
    \label{eq:1d update}
\end{align}
where $\bar{x}=\frac{1}{2N}\sum_{i=1}^{2N}\tilde{x}^i$ denotes the average over training inputs. The solution to \eref{eq:1d update} with the initial condition $b(0)$ is
\begin{align}
    b(t)= \bar{x}_\lambda-\left(\bar{x}_\lambda-b(0)\right)\ee^{-(1+\lambda_2)t},\quad \bar{x}_\lambda=\begin{cases}
    \frac{\bar{x}-\lambda_1}{1+\lambda_2}, & b(t)\geq 0\\
    \frac{\bar{x}+\lambda_1}{1+\lambda_2}, & b(t)<0
    \end{cases}.
    \label{eq:b solution}
\end{align}
In the following, we assume that $b(0)>0$.
To have a nontrivial train- and test-error dynamics, we also assume that $x_{\rm min}<b(0)$, where $x_{\rm min}$ denotes the minimum of the positive samples in the training dataset.

To obtain explicit expressions for the test error we choose the exponential distribution of the samples $P^+(x)=\ee^{-(x-\epsilon)}\Theta(x-\epsilon)$, where $\Theta(x)$ denotes the Heaviside step function. In this case, the cumulative probability to get a sample with $b<x$ is given by $P(x>b)=\ee^{\epsilon-b}$ and the test error is
\begin{align}
    \label{eq:1d test error}
    E(t)=\begin{cases} \frac{1}{2}(1-\ee^{\epsilon-b(t)})& b(t)>\epsilon \\ 0 &{\rm else}\end{cases},
\end{align}
where $b(t)$ is determined by \eref{eq:b solution}. We are interested in the behavior of the test error just before the test error drops to zero. The time at which the test error becomes zero ($t_\epsilon$) is given by \begin{align}
    \label{eq:simple teps}
    t_\epsilon = \log\left(\frac{b(0)-\bar{x}_\lambda}{\epsilon-\bar{x}_\lambda}\right).
\end{align}
By expanding the test error (\eref{eq:1d test error}) around $t_\epsilon$ we get
\begin{align}
    E(t<t_\epsilon)\approx \frac{(\epsilon-\bar{x}_\lambda)}{2}(1+\lambda_2) (t_\epsilon- t).
\end{align}
Interestingly, the first-order coefficient, as well as the critical exponent, do not depend on the initial condition (assuming $b(0)>\epsilon$). We now calculate the average test error over different initial conditions by aligning the phase-transition points $t_\epsilon$. Since $P(-\bar{x})=P(\bar{x})$, we find the test error
\begin{align}
    \label{eq:1d critical exponent}
    \langle\langle E(t)\rangle\rangle\approx\frac{\epsilon_\lambda}{2}(t_\epsilon-t),\quad \epsilon_\lambda=\epsilon(1+\lambda_2)+\lambda_1,
\end{align}
where $\langle\langle \bullet\rangle\rangle$ denotes the average over all valid initial conditions $b(0)$ and training input averages $\bar{x}$. 

Grokking in the considered 1D exponential model is a second-order phase transition with the test-error critical exponent equal to one. The regularisation parameters and the distance between positive and negative class distributions change only the prefactor. In general, we expect that the critical exponent depends on the distribution as well as the training parameters, e.g. regularisation strength.

\subsubsection{Grokking probability}
\begin{figure}[!htb]
    \centering
    \includegraphics[width=0.55\textwidth]{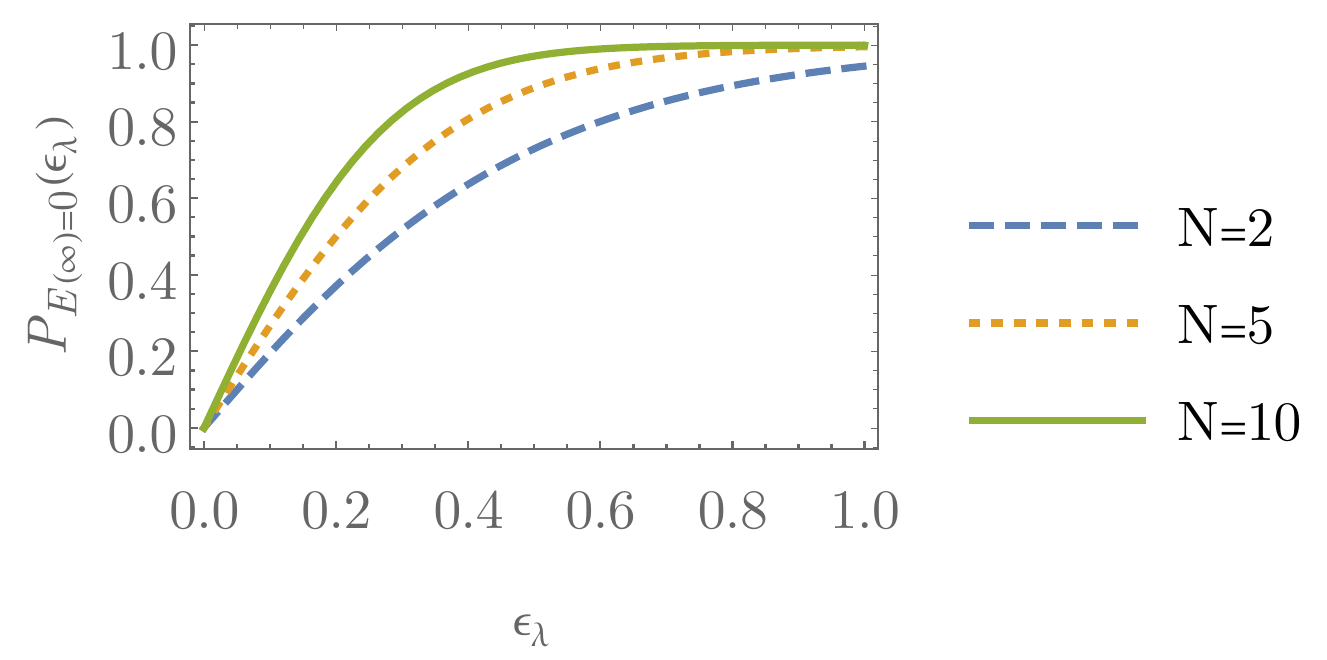}
    \caption{Exact numerical calculation for grokking probabilities for a different number of training samples $N=2$ (dashed blue line), 5 (dotted orange line), and 10 (full green line). Grokking probability increases with $N$ and $\epsilon_\lambda$.}
    \label{fig:1d grokking probability}
\end{figure}

We are also interested in the probability to sample a training dataset with which we can train the model to zero test error. We name this probability the \textit{grokking probability}. In the considered 1D case the final test error vanishes only if $|\bar{x}_\lambda|<\epsilon$. We express this condition for zero test/generalisation error of the trained model as
\begin{align}
    |\bar{x}|<\epsilon_\lambda,
\end{align}
where $\epsilon_\lambda$ is given by \eref{eq:1d critical exponent}. Since our training dataset has an equal number of positive and negative samples, we need to consider the distribution of the mean of $N$ independent exponentially distributed variables, which is given by the gamma distribution
\begin{align}
    P_N^{exp}(\bar{x})=\frac{N^N}{\Gamma(N)}\bar{x}^{N-1}\ee^{-N \bar{x}}\Theta(\bar{x}),
\end{align}
where $\Gamma(N)$ denotes the gamma function. First, we calculate the probability $P_N(\bar{x})$ to get the average $\bar{x}$
\begin{align}
    P_N(\bar{x})=&\int_{\bar{x}_+=0}^{\infty}\dd\bar{x}_+P_N^{\rm exp}(\bar{x}_+)\int_{\bar{x}_-=0}^{\infty}\dd\bar{x}_-P_N^{\rm exp}(\bar{x}_-)\delta\left(\bar{x}-(\bar{x}_+-\bar{x}_-)/2\right)\\ \nonumber
    =& \frac{2 N^{N+\frac{1}{2}}
   \bar{x}^{N-\frac{1}{2}} K_{N-\frac{1}{2}}(2 N
   \bar{x})}{\sqrt{\pi } \Gamma (N)},
\end{align}
where $K_n(z)$ denotes a modified Bessel function of the second kind. The probability to get the dataset with zero test error is then given by 
\begin{align}
    \label{eq:grokking probability}
    P_{E(\infty)=0}(\epsilon_\lambda,N)=&2\int_{\bar{x}=0}^{\epsilon_\lambda}P_N(\bar{x})\dd\bar{x}\\ \nonumber
    =&\sqrt{\pi } (-1)^N (B \epsilon_\lambda )^{2 N} \,
   _1\tilde{F}_2\left(N;N+\frac{1}{2},N+1;N^2
   \epsilon_\lambda ^2\right) \\ \nonumber
   &+\frac{\pi  (-1)^{N+1} N
   \epsilon_\lambda  \,
   _1\tilde{F}_2\left(\frac{1}{2};\frac{3}{2},\frac{3}{2}-N;N^2 \epsilon_\lambda ^2\right)}{\Gamma (Nd)},
\end{align}
where $\,
   _p\tilde{F}_q\left(a;b;z\right)$ is the regularized generalized hypergeometric function. The above expression (\eref{eq:grokking probability}) simplifies for a particular choice of $N$, e.g. for $N=2$ we get
\begin{align}
    P_{E(\infty)=0}(\epsilon_\lambda,N=2)=&1-(1+2\epsilon_\lambda)\ee^{-4\epsilon_\lambda}.
\end{align}
In \fref{fig:1d grokking probability} we show several numerically exact grokking probabilities. As expected, the grokking probability increases with the number of training samples and the effective separation between the two classes determined by $\epsilon_\lambda$.

The effect of the $L_1$ and $L_2$ regularisations on the trained model is different. The $L_2$ regularisation is multiplicative, and the $L_1$ regularisation is additive concerning the gap between the positive and negative samples $\epsilon$. Hence, in the case of a small gap, the $L_1$ regularisation becomes much more effective. In other words, for an infinitesimal gap ($\epsilon\ll1$) and finite $N$, the $L_1$ regularisation ensures that the probability of zero test error is finite. This is not the case when using the $L_2$ regularisation.

It would be interesting to see if $L_1$ regularisation is preferred to $L_2$ regularisation also in more realistic scenarios. In fact, we find a similar distinction between the $L_1$ and $L_2$ normalised models also in the more general grokking scenario discussed in \sref{sec:grokkin probability ball model}.

\subsubsection{Grokking time}
We define the \textit{grokking time} as the difference between $t_\epsilon$ (zero-test-error time) and the time at which the training error becomes zero. In our simple case, we have 
\begin{align}
    t_G=\frac{1}{1+\lambda_2}\log\left(\frac{\epsilon+x_{\rm min}-\bar{x}_\lambda}{\epsilon-\bar{x}_\lambda}\right).
    \label{eq:grokking time}
\end{align}
The grokking time does not depend on the initial condition as long as $b(0)>x_{\rm min}$. To find the grokking-time PDF, we need to calculate the distribution $P_N(\bar{x},x_{\rm min})$ and then consider only the part $|\bar{x}|\leq\epsilon_\lambda$. We provide the details of the calculation in the \aref{app:1D grokking time}. For a finite $N$ it is possible to obtain a closed form expression which, however, is not very instructive. Here we provide the unnormalised grokking-time PDF for $N=2$
\begin{align}
    P^{\rm unnorm}_{N=2,\epsilon,\lambda_1}(t)=&\frac{1}{8} e^{-\frac{4 e^{t} \left(2 e^{t}+5\right) \epsilon _{\lambda }}{e^{t}+1}-2
   t} \Bigg(\exp \left(\frac{4 e^{t} \epsilon _{\lambda } (3 \sinh (t)+\cosh
   (t)+4)}{e^{t}+1}+3 t\right)\\ \nonumber
   &-64 \epsilon _{\lambda }^2 e^{4
   \left(4-\frac{3}{e^{t}+1}\right) \epsilon _{\lambda }+3 t}-24 \epsilon _{\lambda } e^{2
   \left(\left(8-\frac{6}{e^{t}+1}\right) \epsilon _{\lambda }+t\right)}-8 \epsilon _{\lambda } e^{4
   \left(4-\frac{3}{e^{t}+1}\right) \epsilon _{\lambda }+t}\\ \nonumber
   &-2 e^{\frac{4 \left(4
   e^{t}+1\right) \epsilon _{\lambda }}{e^{t}+1}}-3 e^{4 \left(4-\frac{3}{e^{t}+1}\right)
   \epsilon _{\lambda }+t}-\left(e^{t}+1\right) e^{4 \left(2
   e^{t}-\frac{1}{e^{t}+1}+2\right) \epsilon _{\lambda }} \left(e^{t} \left(e^{t}-8
   \epsilon _{\lambda }-1\right)-2\right)\Bigg),
\end{align}
which we normalise by dividing with the appropriate grokking probability, see \eref{eq:grokking probability}.

\begin{figure}[!htb]
    \centering
    \includegraphics[width=0.305\textwidth]{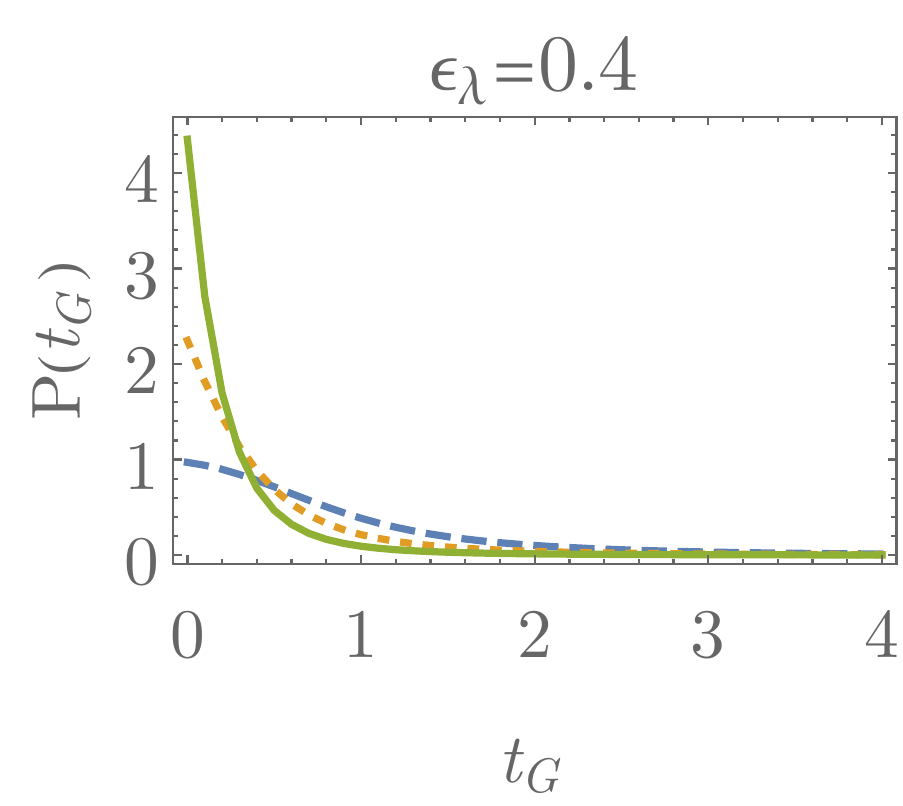}
    \includegraphics[width=0.32\textwidth]{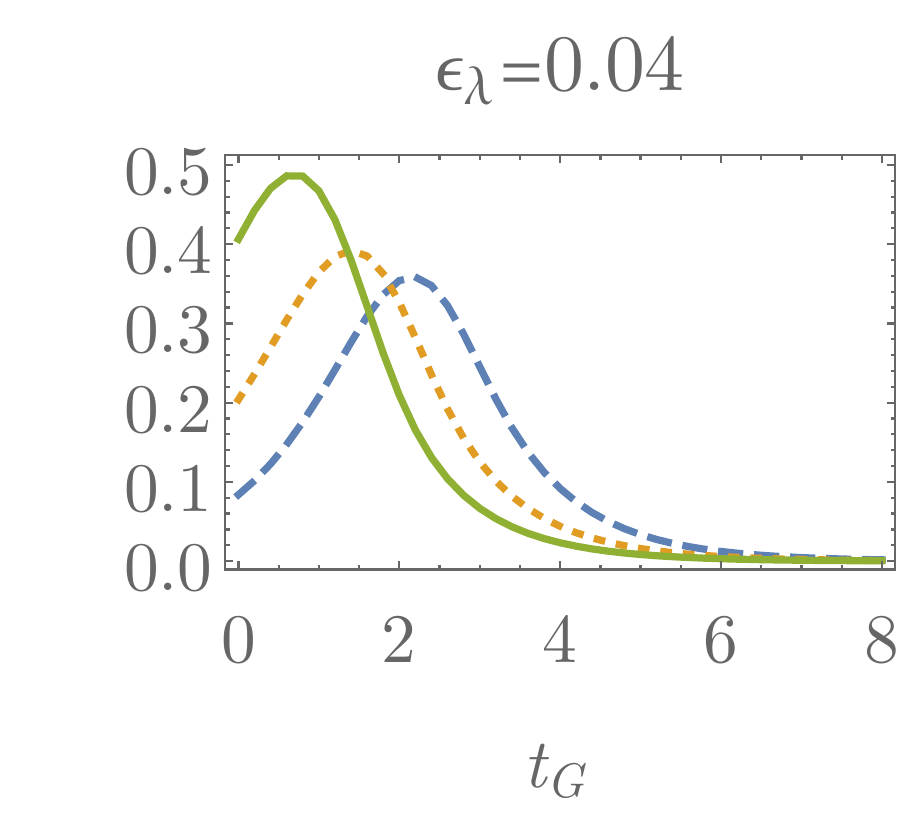}
    \includegraphics[width=0.34\textwidth]{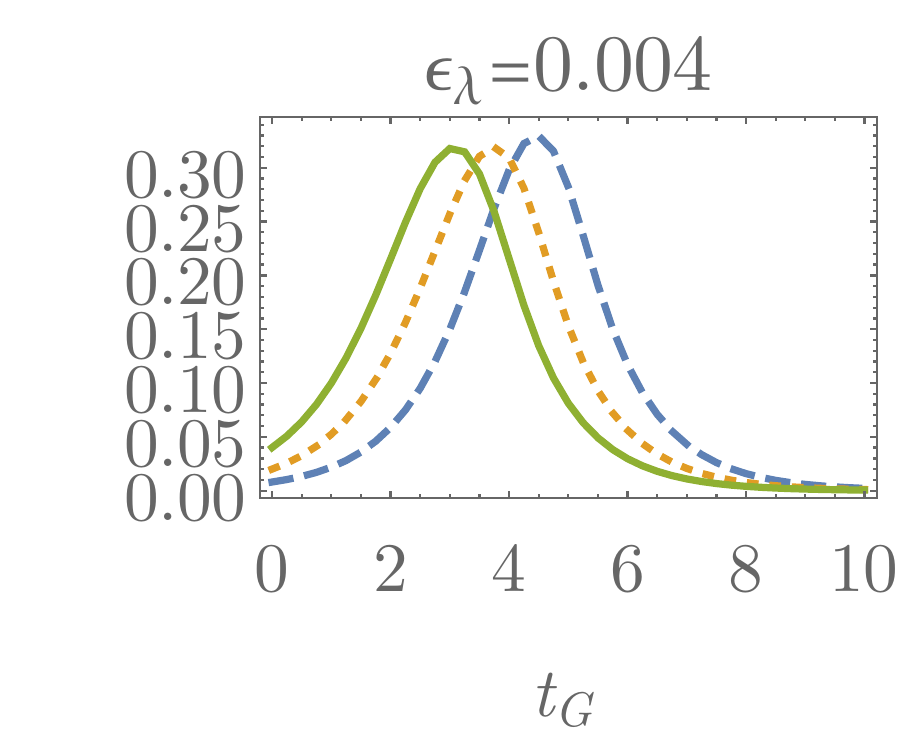}
    
    \includegraphics[width=0.4\textwidth]{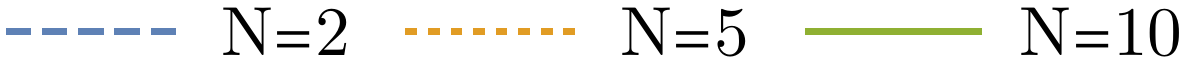}
    \caption{Numerically exact the grokking-time $t_{\rm G}$ PDF at different number of training samples $N=$2 (dashed blue line), 5 (dotted orange line), 10 (full green line). Left, middle, and right panel correspond to $\epsilon_\lambda=$0.4 (left), 0.04 (middle), 0.004 (right). Grokking time is shorter with increasing $N$ and $\epsilon_\lambda$.}
    \label{fig:1d grokking-time PDF}
\end{figure}

In \fref{fig:1d grokking-time PDF} we show several numerically exact grokking-time PDFs. The expected grokking time is smaller with increasing training size $N$ and effective class separation $\epsilon_\lambda$. This is consistent with the observations of \cite{power2022grokking,liu2022towards} where a shorter grokking time has been reported for increased number of training samples and a larger weight decay.

\subsection{D-dimensional uniform ball model}
\label{sec:uniform ball}
\begin{figure}[!htb]
\centering
\begin{tikzpicture}[scale=0.8,domain=-8:8]

\draw[fill=blue!10] (4,0) circle (2cm);
\draw[draw=white,fill=white, rotate around={14:(4,1)}] (1,1) rectangle (7,-3);
\draw[fill=orange!10] (-4,0) circle (2cm);
\draw[draw=white,fill=white, rotate around={14:(-4,-1)}] (-1,-1) rectangle (-7,3);

\draw[-latex,color=black] (-6.5,0) -- (6.5,0);
\draw[-latex,color=black] (0,-3.5) -- (0,3.5);
\draw (6.5,0.2) node[anchor=north west] {$\tilde{x}_{1}$};
\draw (-0.3,4.3) node[anchor=north west] {$\tilde{x}^\perp$};
\draw[color=black] (0pt,-10pt) node[left] {\footnotesize $0$};

\draw[teal!50] plot (\x,\x/4) node[right] {};
\draw[thick,blue!50] (4,0) circle (2cm);
\draw[thick,orange!50] (-4,0) circle (2cm);

\draw[black,-latex] (0,0) -- (4,0) node[midway,below,xshift=0.5cm] {$\varepsilon$};
\draw[-,black,rotate around={30:(4,0)}] (4,0) -- (6,0) node[midway,below,xshift=3pt,yshift=1pt] {\small $r$};
\draw[teal,-latex] (0,0) -- (1/2,-2) node[midway,right] {$w$};
\draw[blue] (4,-3.25) node[anchor=south] {$P^{+}(\tilde{x})$};
\draw[orange] (-4,3.25) node[anchor=north] {$P^{+}(\tilde{x})$};

\end{tikzpicture}
\caption{A two-dimensional projection of the $D-$dimensional uniform-ball model on the plane defined by the shift vector $\epsilon$ and the model vector $w$. The positive and negative samples are uniformly distributed in unit balls shifted away from the origin along the $\tilde{x}_1$ axis by $\pm \epsilon$, respectively. The model used to separate the classes is a linear model (determined by the vector $w$) going through the origin (green line). The volume of the shaded red and blue regions determines the test error.}
\label{fig:D-dim ball}
\end{figure}
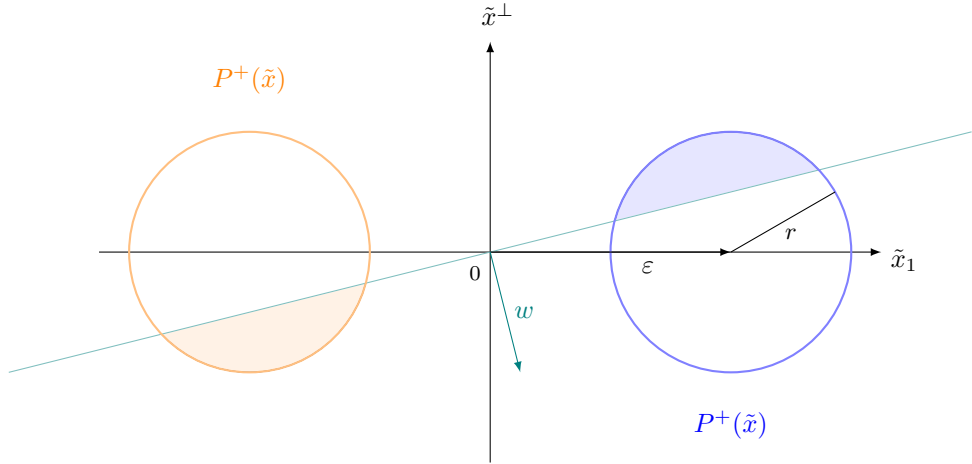

The second model we consider is shown in \fref{fig:D-dim ball}. The inputs are $D$ dimensional real vectors $\tilde{x}\in\mathds{R}^D$. Positive and negative samples are distributed uniformly in unit balls shifted from the origin by vectors $\pm\epsilon\in\mathds{R}^D$. We will assume that the shift is along the first dimension, i.e. $\epsilon_1=\varepsilon$, and $\epsilon_{j>1}=0$. The student model is a simple perceptron determined by the vector $w\in\mathds{R}^D$ (we set the bias to zero $b=0$)
\begin{align}
    \label{eq:perceptron}
    f(\tilde{x})=\sgn(\tilde{x}\cdot w).
\end{align}

We write the training loss with $L_1$ and $L_2$ regularisation as
\begin{align}
    \label{eq:loss ball}
    \mathcal{R}=&\frac{1}{2N}\sum_{i=1}^{2N}\frac{1}{2}(\tilde{x}^i\cdot w-y^i)^2+\frac{\lambda_2||w||_2^2}{2}+\lambda_1 |w|_1. \\ \nonumber
    =&\frac{1}{2}  w\cdot\left(\frac{1}{2N}\sum_{i=1}^{2N} \tilde{x}^i\otimes \tilde{x}^i+\lambda_2\mathds{1}_D\right) w- w\cdot\left(\frac{1}{2N}\sum_{i=1}^{2N} y^i \tilde{x}^i -\lambda_1\sgn( w)\right) + \frac{1}{2}\\ \nonumber
    =&\frac{1}{2} w\cdot G  w -  w\cdot  a +\frac{1}{2}, \\ \nonumber
    G=&\frac{1}{2N}\sum_{i=1}^{2N} \tilde{x}^i\otimes \tilde{x}^i+\lambda_2\mathds{1}_D ,\\ \nonumber
     a=&\frac{1}{2N}\sum_{i=1}^{2N} y^i \tilde{x}^i -\lambda_1\sgn( w).
\end{align}
Again we assume that the training dataset is balanced, i.e. $y^{i\leq N}=1$, $y^{i> N}=-1$, $\tilde{x}^{i\leq N}=x^i+\epsilon$, and $\tilde{x}^{i> N}=x^i-\epsilon$. In our model $x^i$ are distributed uniformly in a $D-$dimensional ball centered at the origin.

The presented model is relevant in the transfer learning setting \cite{weiss2016survey}, if only the last layer of a network is retrained, and with sigmoid activation functions in the penultimate layer. If the model transfers well to a new classification task, the latent-space distributions of the new classes are linearly separable and can be bounded by a $D-$dimensional ball. Since we do not know the details of the distributions, we assume the uniform distribution in the ball. Further, the positive and negative feature distributions might be embedded in a higher dimensional latent space. In this case, $D$ corresponds to the effective dimension of the data, which can be calculated from the covariance matrix. As we will see in the next section, the introduced $D-$dimensional ball model qualitatively reproduces the critical exponent and the grokking-time PDF in a local-rule learning problem.

\subsubsection{Test error dynamics}
To determine the test error dynamics we first specify the dynamics of the model parameters which is determined by the negative gradient of the loss function
\begin{align}
    \frac{\partial  w}{\partial t}=-\frac{\partial \mathcal{R}}{\partial  w}=-G w +  a.
    \label{eq:D-dim update}
\end{align}
The solution to \eref{eq:D-dim update} with the initial condition $w(0)$ is 
\begin{align}
     w(t)=  w^\lambda-\left( w^\lambda- w(0)\right)\ee^{-Gt},\quad  w^\lambda=G^{-1} a.
    \label{eq:D-dim solution}
\end{align}
If $N<D/2$ and $\lambda_2=0$, the matrix $G$ is not invertible. In this case, we use the pseudo-inverse.

The test error is given by the volume of an $\epsilon$-shifted, unit ball that is cut out by the plane defined by the vector $w$. The relevant parameter determining this volume is the distance $h$ between the plane and the origin of the ball. We find that $h=\varepsilon \frac{w_1}{||w||_2}$, where $\varepsilon=||\epsilon||_2$ and assuming $w_1>0$. The critical angle is given by $w_1/|| w||_2=\frac{1}{\varepsilon}$. For larger $w_1/|| w||_2$ the error is zero. For smaller values of $w_1/|| w||_2$ the error is given by
\begin{align}
    E_D(h)=\begin{cases} \frac{1}{2}-\frac{D \Gamma \left(\frac{D}{2}\right) }{2\sqrt{\pi }\Gamma \left(\frac{D+1}{2}\right)}\, _2F_1\left(\frac{1}{2},\frac{1-D}{2};\frac{3}{2};h^2\right)h&,~~ h\leq1\\ 0 &,~~ h>1\end{cases},
\end{align}
where ${}_2F_1(a,b;c,z)$ represents the Gaussian hypergeometric function.
For $h<1$ and close to the critical point $h\approx 1$ we find the following expression for the test error
\begin{align}
    \label{eq:ball error}
    E_D(t)\approx & \frac{D 2^{\frac{D-3}{2}} \Gamma \left(\frac{D}{2}\right)}{\sqrt{\pi } \Gamma \left(\frac{D+3}{2}\right)} (1-h(t))^{\frac{D+1}{2}} \\ \nonumber 
    =&\frac{D 2^{\frac{D-3}{2}} \Gamma \left(\frac{D}{2}\right)}{\sqrt{\pi } \Gamma \left(\frac{D+3}{2}\right)} \left(1-\epsilon \frac{w_1(t)}{|| w(t)||_2}\right)^{\frac{D+1}{2}}\\ \nonumber
    =&\frac{D 2^{\frac{D-3}{2}} \Gamma \left(\frac{D}{2}\right)}{\sqrt{\pi } \Gamma \left(\frac{D+3}{2}\right)} (k_G(t-t_\epsilon)) )^{\frac{D+1}{2}},
\end{align}
where $t_\epsilon$ is defined as time at which the test error vanishes, and the coefficient $k_G$ is given by the linear expansion of $w(t)$ around $t_\epsilon$. The critical exponent is hence determined only by the dimensionality of the feature distribution.

\subsubsection{Grokking probability}
\label{sec:grokkin probability ball model}
Next, we will calculate the probability of training a model with zero test error (\textit{grokking probability}) for a given number of positive/negative samples $N$. The condition for the final test error to be zero is given by
\begin{align}
    \label{eq:critical train ball}
    \frac{w^\lambda_1}{|| w^\lambda||_2}\geq\frac{1}{\epsilon}.
\end{align}
It will be useful to write the zero test error condition in terms of components of $w^\lambda$
\begin{align}
    \label{eq:critical train ball components}
    (\epsilon^2-1)\left(w_1^\lambda\right)^2\geq \sum_{j=2}^{D} \left(w_j^\lambda\right)^2=r,
\end{align}
where $r$ denotes the 2-norm squared of the final weights vector $w^\lambda$ with the first component equal to zero.

A general calculation of the grokking probability and the grokking-time PDF is not feasible since we would have to invert a random matrix $G$. Therefore, we consider the limit of many training samples $N\gg 1$, where the matrix $G$ decomposes into a diagonal part proportional to $\lambda_{2,D}=\lambda_2+\frac{1}{D+2}$ and an off-diagonal part proportional to $N^{-1/2}$. We provide the full derivation of the grokking probability in this limit in \aref{app:grokking ball N>>1}. Here we consider a simpler case, where we additionally assume that $N\gg\lambda_{2,D}\gg1$. In this case, $G$ is approximately proportional to the identity
\begin{align}
    G\approx\lambda_{2,D}\mathds{1}_D+\epsilon\otimes\epsilon.
\end{align}
The inverse is
\begin{align}
    [G^{-1}]_{i,j}\approx\begin{cases}
    \left(\lambda_{2,D}+\varepsilon^2\right)^{-1}&,~i=j=1 \\
    \lambda_{2,D}^{-1}&,~i=j\neq1 \\
    0 &,~i\neq j
    \end{cases}.
\end{align}
Finally we get
\begin{align}
    \label{eq:stationary condition ball}
     w^\lambda_{i=1} \approx \frac{a_1}{\varepsilon^2+\lambda_{2,D}},\quad w^\lambda_{i>1} \approx \frac{a_i}{\lambda_{2,D}}.
\end{align}

In the limit $N\gg1$ the probability of the mean of $2N$ random vectors distributed uniformly in a $D$-dimensional ball is well approximated by the normal distribution with zero mean and variance $\mathds{1}_{D}/2N(D+2)$,
\begin{align}
    P_{D,2N}(\bar{x})\approx\mathcal{N}_{0,\mathds{1}_D/2N(D+2)}(\bar{x}).
\end{align}

In the following, we separately describe the grokking probability in the case $\lambda_1=0$ and the case $\lambda_1>0$. 

\paragraph{Case $\mathbf{\lambda_1=0}$ --} Let us first consider the case without the $L_1$ regularisation, i.e. $\lambda_1=0$. The grokking probability is given by (see \aref{app:grokking ball N>>1})
\begin{align}
    \label{eq:grokking probability D ball}
    P_{E(\infty)=0}=&\int\dd \bar{x}\,P_{D,2N}(\bar{x})\Theta\left(\frac{w^\lambda_1(\bar{x})}{|| w^\lambda(\bar{x})||_2}- \frac{1}{\varepsilon}\right)\\ \nonumber
    \approx&\int_{-\varepsilon}^\infty\dd \bar{x}_1\,\mathcal{N}_{0,1/2N(D+2)}\int_0^{2N(D+2)(\varepsilon^2-1)\left(\frac{\bar{x}_1+\varepsilon}{1+\varepsilon^2/\lambda_{2,D}}\right)^2}\dd r\, \chi^2_{D-1}(r)\\ \nonumber
    =& \int_{-\varepsilon}^\infty\dd \bar{x}_1\,\mathcal{N}_{0,1/2N(D+2)} \,P\left(\frac{D-1}{2},\frac{N(D+2)\left(\varepsilon ^2-1\right) (\bar{x}_1+ \varepsilon )^2}{\left(\frac{\varepsilon ^2}{\lambda_{2,D} }+1\right)^2}\right),
\end{align}
where $\chi^2_{D-1}(r)$ is the standard Chi-square distribution and $P(s,t)$ is the regularized gamma function. We also introduced the sample average $\bar{x}=\frac{1}{2N}\sum_{i=1}^{2N}y^ix^i$. The full grokking probability without the additional assumption has essentially the same structure with more complicated expressions for the parameters of the distributions (see \aref{app:grokking ball N>>1}).

\begin{figure}[!htb]
    \centering
    \includegraphics[width=0.245\textwidth]{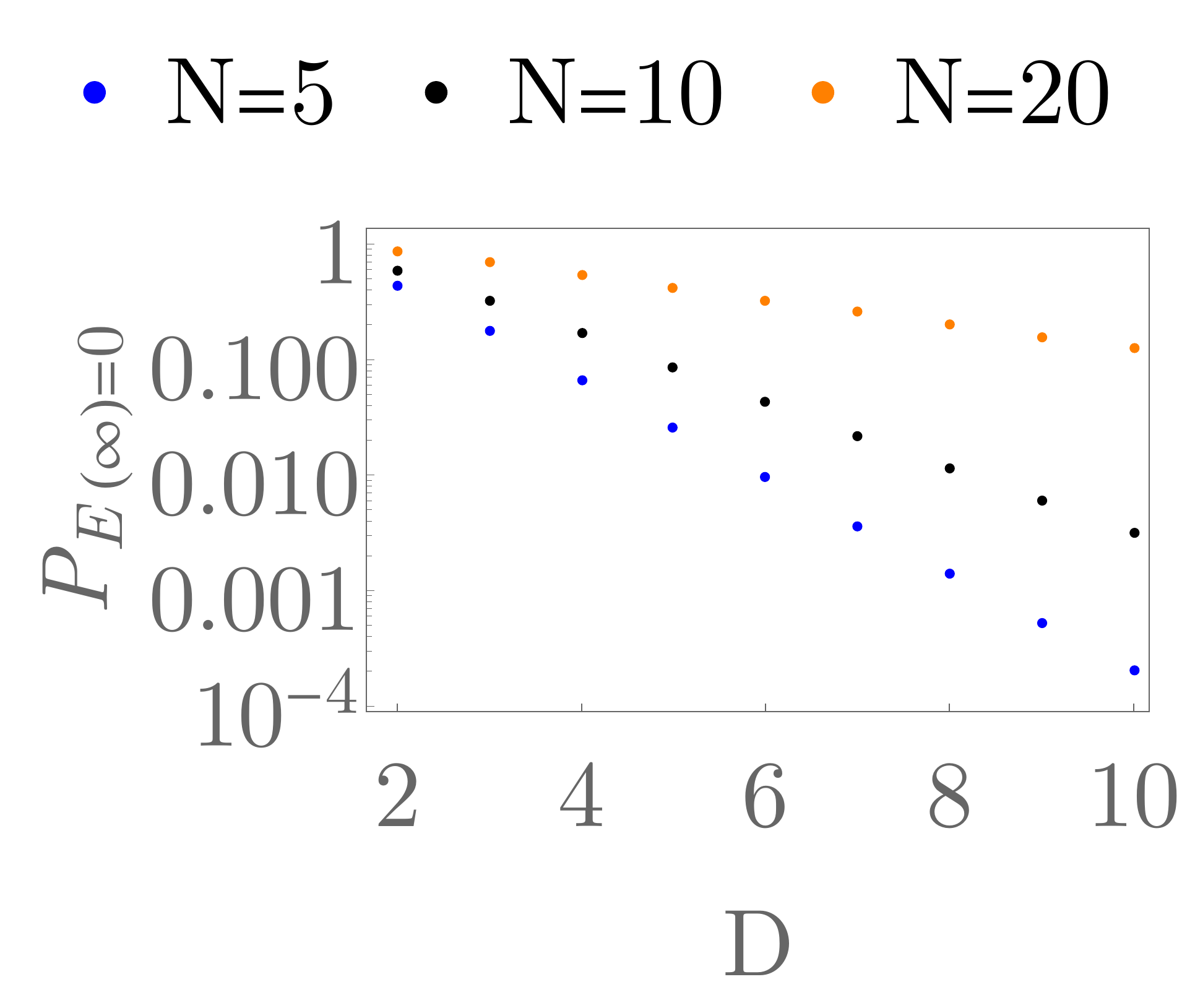}    \includegraphics[width=0.23\textwidth]{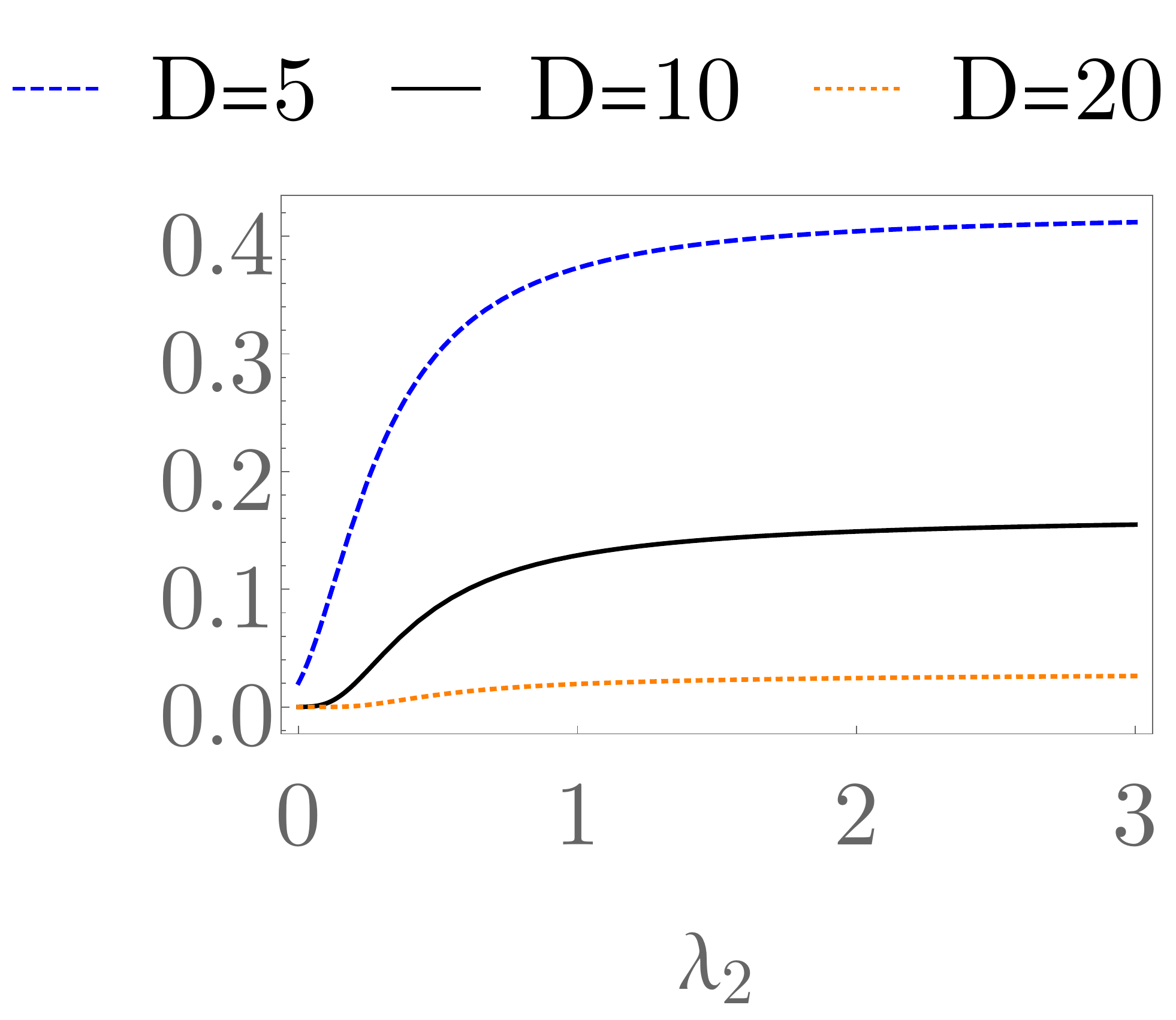}    \includegraphics[width=0.243\textwidth]{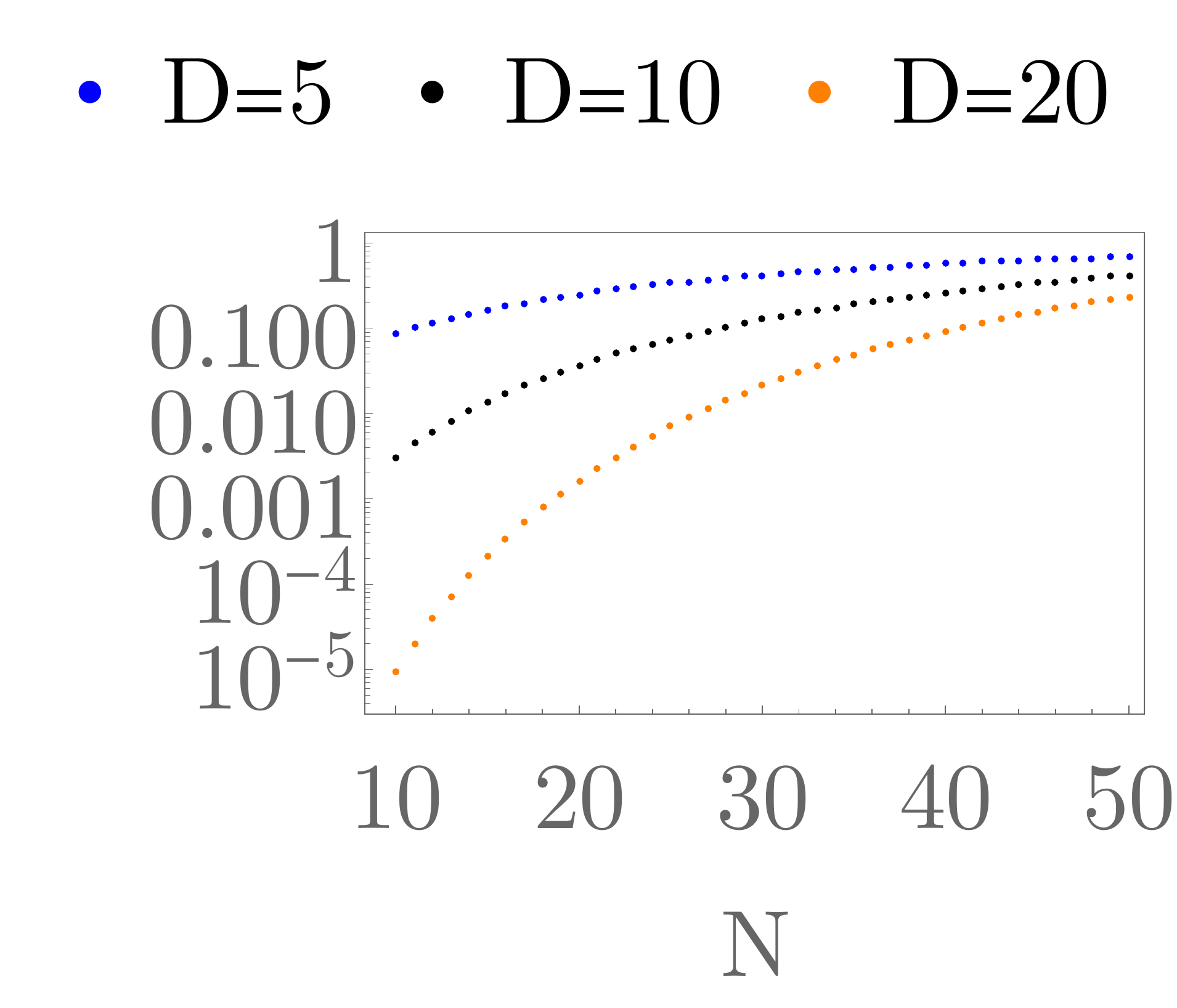}    \includegraphics[width=0.233\textwidth]{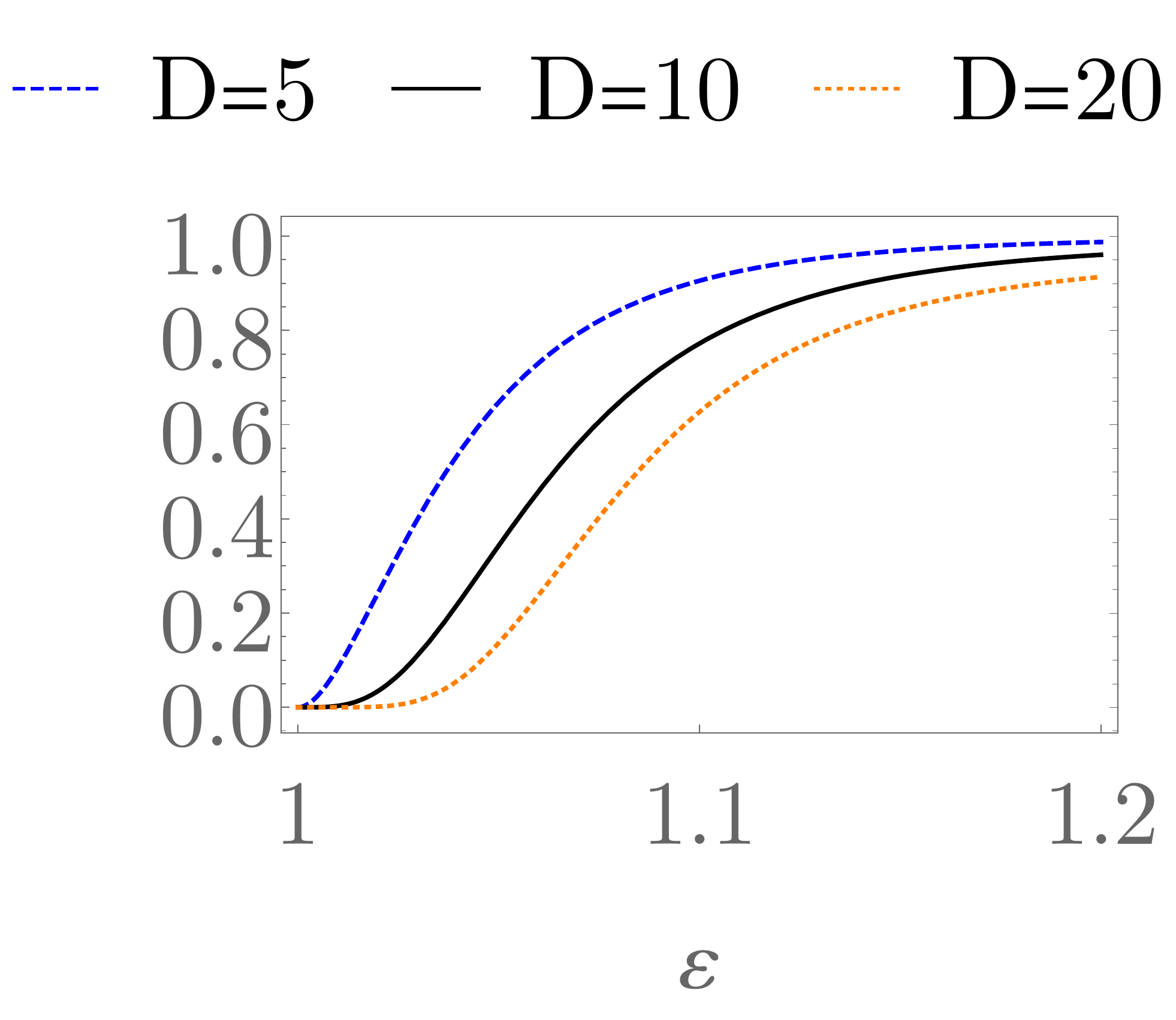}
    \caption{Grokking probability as a function of $D$, $\lambda_2$, $N$, and $\varepsilon$. We find that larger dimension $D$ decreases the grokking probability. In contrast, larger regularisation strength $\lambda_2$ increases the grokking probability. As expected, increased class separation $\varepsilon$ and number of training samples $N$ also increases the grokking probability. If not specified in the panels, additional parameters are set to: $D=10$, $\lambda_2=0.1$, $N=10$, and $\varepsilon=1.01$.}
    \label{fig:grokking probability ball model}
\end{figure}

For a given set of parameters $\lambda_2$, $D$, and $\varepsilon$ we can efficiently numerically evaluate the integral in \eref{eq:grokking probability D ball} (and the full formula reported in \aref{app:grokking ball N>>1}). In \fref{fig:grokking probability ball model} we show the full grokking probability as a function of $D$, $\lambda_2$, $N$, and $\varepsilon$. As expected, the grokking probability is larger with increasing distance $\varepsilon$ and number of samples $N$. We also observe that the grokking probability exponentially decreases with the dimensionality of the latent-space data distribution $D$. Therefore, latent-space distribution with a low effective dimension is preferred for better generalisation. This result partially explains the observation in \cite{liu2022towards} which relates grokking to structure formation and effective dimension decrease at the transition. We expect that low effective dimension in the latent space increases generalisation in a more general setting, beyond the simple grokking scenario described in this section. In other words, models with latent space distributions with small effective dimension will more likely lead to good generalisation. Finally, by increasing the $L_2$ regularisation strength $\lambda_2$ the grokking probability increases up to a maximum that depends on the remaining parameters. These results provide, some justification of the numerical observation in \cite{power2022grokking, liu2022towards, thilak2022slingshot} that weight decay increases the parameter region where grokking is observed.

\paragraph{Case $\mathbf{\lambda_1>0}$ --}
Let us again consider the limit $N\gg\lambda_2\gg1$. If $\lambda_1>0$ the stationary solution $w_j^\lambda$ depends on the sign of $|\bar{x}_j|-\lambda_1$, where $\bar{x}=\frac{1}{2N}\sum_{i=1}^{2N}x^i$. The $j-$th component of the stationary vector is
\begin{align}
    w^\lambda_1=\begin{cases} 0&,~\lambda_1\geq|\bar{x}_1+\varepsilon| \\ \frac{\bar{x}_1+\varepsilon-\lambda_1\sgn(\bar{x}_1)}{\lambda_{2,D}+\varepsilon^2}&,~\mbox{else} \end{cases}, \quad w^\lambda_{j\neq1}=\begin{cases} 0&,~\lambda_1\geq|\bar{x}_j| \\ \frac{\bar{x}_j-\lambda_1\sgn(\bar{x}_j)}{\lambda_{2,D}}&,~\mbox{else} \end{cases}.
\end{align}
The number of non-vanishing components of the stationary solution $w^\lambda$ depends on the value of $\lambda_1$. Therefore, we get (in the $N\gg1$ limit) an additional sum over the number of non-zero elements in the $w^\lambda$,
\begin{align}
    \label{eq:full P_e D-ball}
    P_{E(\infty)=0}=\int^\infty_{\lambda_1-\varepsilon}\dd \bar{x}_1&\mathcal{N}_{0,1/2N(D+2)}\Bigg[ (1-p_\lambda)^{D-1} \\ \nonumber
    &+\sum_{k=1}^{D-1}\binom{D-1}{k}p_\lambda^k(1-p_\lambda)^{D-1-k}\int_0^{2N(D+2)(\varepsilon^2-1)\left(\frac{\bar{x}_1+\varepsilon-\lambda_1}{1+(\varepsilon/\lambda_{2,D})^2}\right)^2}\dd r R_k(r) \Bigg],
\end{align}
where $p_\lambda=1-\text{erf}\left(\sqrt{N(D+2)} \lambda_1 \right)$ is the probability of the variable $|\bar{x_i}|$ (for $i>1$) to be larger than $\lambda_1$ and $R_k(r)$ is the PDF of the sum of squares of $k$ random variables sampled from the truncated normal distribution. 
Since half-Gaussian distribution has a longer tail as the truncated Gaussian at $\lambda_1$, we can lower bound (or estimate) the grokking probability by using the Chi-squared distribution instead of $R(r)$. In the limit $\lambda_1=0$ we recover \eref{eq:grokking probability D ball}. In the case $\lambda_1>0$, we can approximate the inner integrals in \eref{eq:grokking probability D ball} by the regularised gamma function and efficiently numerically evaluate \eref{eq:grokking probability D ball}. 

Further, by discarding the sum over $k\geq1$ in \eref{eq:full P_e D-ball} we obtain a lower bound on the grokking probability
\begin{align}
    \label{eq:grokking probability D ball bound}
    P_{E(\infty)=0}\geq&\int^\infty_{\lambda_1-\varepsilon}\dd\bar{x}_1\mathcal{N}_{0,1/2N(D+2)}(1-p_\lambda)^{D-1}\\ \nonumber
    =&\frac{1}{2}\left(1+\text{erf}\left(\sqrt{N(D+2)} (\varepsilon-\lambda_1 )\right)\right) \left(\text{erf}\left(\sqrt{N(D+2)} \lambda_1\right)\right)^{D-1}.
\end{align}
We find a similar distinction between the $L_1$ and $L_2$ regularisations as in the simple 1D case. At $\varepsilon=1$ and $\lambda_1=0$ the grokking probability vanishes for any value of $\lambda_2$. In contrast, for $\lambda_1>0$ the grokking probability can increase even above $90\%$ for any $D\geq2$. Interestingly, the grokking probability increases with the dimensionality of the data distribution $D$. In fact, if we send $D\rightarrow\infty$ the grokking probability becomes 100\% if $0<\lambda_1<\epsilon$. This result is a consequence of the concentration of measure of the uniform distribution "around the equator". Similarly, by using the lower bound \eref{eq:grokking probability D ball bound} we estimate the best value of $\lambda_1$ for any $\epsilon$, $D$ and $N$ and find that the grokking probability maximum is always larger than $0.915$. In contrast, in the $\lambda_1=0$ case, the grokking probability becomes exponentially small with $D$, independent of the remaining parameter values. We make similar observations also if we relax the condition $\lambda_2\gg1$ (see \aref{app:grokking ball N>>1}). 

The discussed results could be applicable more generally. It would be interesting to check if $L_1$ weight regularisation in the last (classification) layer significantly improves the generalisation of deep models compared to the $L_2$ regularisation. The works \cite{power2022grokking,liu2022towards} do not study the differences between $L_1$ and $L_2$ regularisations. In \cite{liu2022towards} a consistent observation has been made, namely larger weight decay leads in most cases to a larger parameter region where grokking is observed. We confirm this expectation on a simple model discussed in \sref{sec:constant attention tensors} and \sref{sec:full model training results}.

\subsubsection{Grokking time}
\label{sec:grokking time n-ball}

To calculate the grokking time, we first determine the condition for the zero train error. In contrast to the simple 1D case, this condition depends non-trivially on the training dataset and on the initial condition $w(0)$. To simplify the calculation, we calculate the distribution of the upper bound on the grokking time in the limit $N\gg1$. We obtain the most conservative estimate for zero train error by selecting the training sample $\tilde{x}$ that forms the smallest angle with the plane defined shift vector $\epsilon$. We write this condition in terms of the cosine of the angle as
\begin{align}
    \label{eq:critical test ball}
    \frac{w_1}{|| w||_2}\geq\xi_{\rm train}=\max_i\frac{\sqrt{||x^i+\varepsilon||_2^2-(x^i_1+\varepsilon)^2}}{||x^i+\varepsilon||_2},
\end{align}
where $\xi_{\rm train}$ denotes the cosine of the smallest angle between the plane defined by $\epsilon$ and any training sample $\tilde{x}^i$.
We will consider only the zeroth-order solution in $1/\sqrt{N}$, where the grokking probability becomes 100\%. Namely, we also discard terms proportional to $1/\sqrt{N}$. In this limit the stationary solution is proportional to $\epsilon$, i.e. $w^\lambda\approx \frac{\epsilon}{\lambda_D+\varepsilon^2}$. The time dependent model parameters simplify to
\begin{align}
    \label{eq:w(t) 1}
    w_1(t)\approx& \frac{\varepsilon}{\lambda_D+\varepsilon^2} + \left(w_1(0)-\frac{\varepsilon}{\lambda_D+\varepsilon^2}\right)\ee^{-(\lambda_{2,D}+\varepsilon^2)t},\\ 
    \label{eq:w(t) j}
    w_j(t)\approx& w_j(0)\ee^{-\lambda_{2,D}t},\quad j>1.
\end{align}
Since we consider only the leading (zeroth) order in $\frac{1}{\sqrt{N}}$, the value of $\lambda_1$ does not have such a dramatic effect as in \sref{sec:grokkin probability ball model}. Therefore, we will study only the case $\lambda_1=0$. To further simplify the calculation we will also assume  $\lambda_{2,D}=\lambda_2+\frac{1}{D+2}\ll1$. In this limit we find
\begin{align}
    \label{eq:xi train}
    \xi_{\rm train}&\approx\max_i \sqrt{1-\left(\frac{x^i_1+\varepsilon}{||x^i+\epsilon||_2}\right)^2}\approx \frac{x_{\rm max}}{\sqrt{x_{\rm max}^2 + \varepsilon^2}}\approx \frac{1}{\sqrt{1+\varepsilon^2}},
\end{align}
where $x_{\rm max} = \max_i||x^i||_2$. Similarly, \eref{eq:w(t) 1} and \eref{eq:w(t) j} simplify to
\begin{align}
    \label{eq:w(t) approx 1}
    w_1(t)\approx& \frac{1}{\varepsilon} + \left(w_1(0)-\frac{1}{\varepsilon}\right)\ee^{-\varepsilon^2t},\\ 
    \label{eq:w(t) approx j}
    w_j(t)\approx& w_j(0)\ee^{-\lambda_{2,D}t},\quad j>1.
\end{align}
We find that the first component of $w$ relaxes much faster as the remaining components. Therefore, the parameter path can be approximated by two straight lines/paths. Along the first path, $w_1(t)$ quickly relaxes towards the stationary value $w_1^\lambda\approx1/\varepsilon$. Then, along the second path, the remaining parameters slowly relax towards the stationary value $w_{j>1}^\lambda\approx0$. This leads to two different zero train/test error conditions.

First, we consider the case when grokking occurs during the fast relaxation (first path). In this case, the condition for grokking to occur reads
\begin{align}
    1\geq\frac{1}{\varepsilon^2}+||w^{\perp}(0)||_2^2,
\end{align}
where $w^{\perp}(t)$ is obtained from $w(t)$ by setting $w_1$ to zero.  The zero train/test error is achieved after time 
\begin{align}
    t = \frac{1}{\varepsilon^2}\ln\left(\frac{\frac{1}{\varepsilon}-w_1(0)}{\frac{1}{\varepsilon}-\frac{\xi}{\sqrt{1-\xi^2}}||w^{\perp}(0)||_2}\right).
\end{align}
We assume that $w_1(0)<w^\lambda_1\approx \frac{1}{\varepsilon}$ and obtain the final expression for the grokking time 
\begin{align}
    t_{\rm G}=\frac{1}{\varepsilon^2}\ln\left(\frac{\frac{1}{\varepsilon}-\frac{\xi_{\rm train}}{\sqrt{1-\xi_{\rm train}^2}}||w^{\perp}(0)||_2}{\frac{1}{\varepsilon}-\frac{\xi_{\rm test}}{\sqrt{1-\xi_{\rm test}^2}}||w^{\perp}(0)||_2}\right) \approx
    \frac{1}{\varepsilon^2}\ln\left(\frac{1-||w^{\perp}(0)||_2}{1-\frac{\varepsilon}{\sqrt{\varepsilon^2-1}}||w^{\perp}(0)||_2}\right).
\end{align}
The grokking time in the considered limit depends only on the initial condition $w^{\perp}(0)$, i.e. on the initial distribution of the classifier weights. We assume that the initial model weights are sampled independently from a normal distribution with zero mean and unit variance. Setting $r=||w^{\perp}(0)||_2^2$, the variable $r$ follows the $\chi^2_{D-1}$ distribution. Therefore, we express the grokking-time PDF as
\begin{align}
    P_{\rm fast}(t_{\rm G})\approx \chi_{D-1}^2(r(t_{\rm G}))\frac{\partial r(t_{\rm G})}{\partial t_{\rm G}},
\end{align}
where 
\begin{align}
    r(t_{\rm G}) = \frac{\left(\varepsilon ^2-1\right) \left(e^{t_{\rm G} \varepsilon ^2}-1\right)^2 \left(\varepsilon  \left(\varepsilon  e^{2 t_{\rm G} \varepsilon ^2}+2 \sqrt{\varepsilon ^2-1} e^{t_{\rm G} \varepsilon
   ^2}+\varepsilon \right)-1\right)}{\left(\varepsilon ^2 \left(e^{2 t_{\rm G} \varepsilon ^2}-1\right)+1\right)^2}.
\end{align}
Above result represents only one part of the grokking probability and hence the distribution
$P_{\rm fast}(t_{\rm G})$ is not normalised. In fact, integrating $P_{\rm fast}(t)$ over the whole domain we obtain the probability to start with the initial condition where grokking occurs during the fast relaxation
\begin{align}
    \label{eq:fast probability}
    p_{\rm fast}=\int_{w}P(w) \Theta\left(1-\frac{1}{\varepsilon^2}-|w^\perp|^2\right)\dd w=\int_0^{1-\frac{1}{\varepsilon^2}}\chi^2_{D-1}(r)\dd r,
\end{align}
where $\chi^2_{D-1}$ is the standard Chi-squared distribution.

The second part of the grokking-time PDF comes from the initial conditions where the zero test/train error is obtained during the slow relaxation process. In this case, we assume that the value $w_1(t)$ is stationary, i.e. $w_1(t)\approx w_1^\lambda\approx \frac{1}{\varepsilon}$. The remaining model parameters evolve according to \eref{eq:w(t) approx j}. The time at which the train/test error vanishes reads
\begin{align}
    t_{\rm train/test} = \frac{1}{\lambda_{2,D}}\ln\left(
    \frac{||w^\perp(0)||_2}{\sqrt{1-\xi_{\rm train/test}^2}w_1^\lambda}\right).
\end{align}
After simplification we find the grokking time in the slow relaxation regime  
\begin{align}
    t_{\rm G} = \frac{1}{2\lambda_{2,D}}\ln\left(\frac{\varepsilon^4}{\varepsilon^4-1}
    \right).
\end{align}
Interestingly, the grokking time is independent of the initial condition. Therefore, the distribution of the slow-relaxation grokking time is trivial, i.e. proportional to a Dirac delta distribution with the weight $1-p_{\rm fast}$, where $p_{\rm fast}$ is the probability of initialising the parameters with grokking during the fast relaxation given in \eref{eq:fast probability}.

\begin{figure}[!htb]
    \centering
    \includegraphics[width=0.34\textwidth]{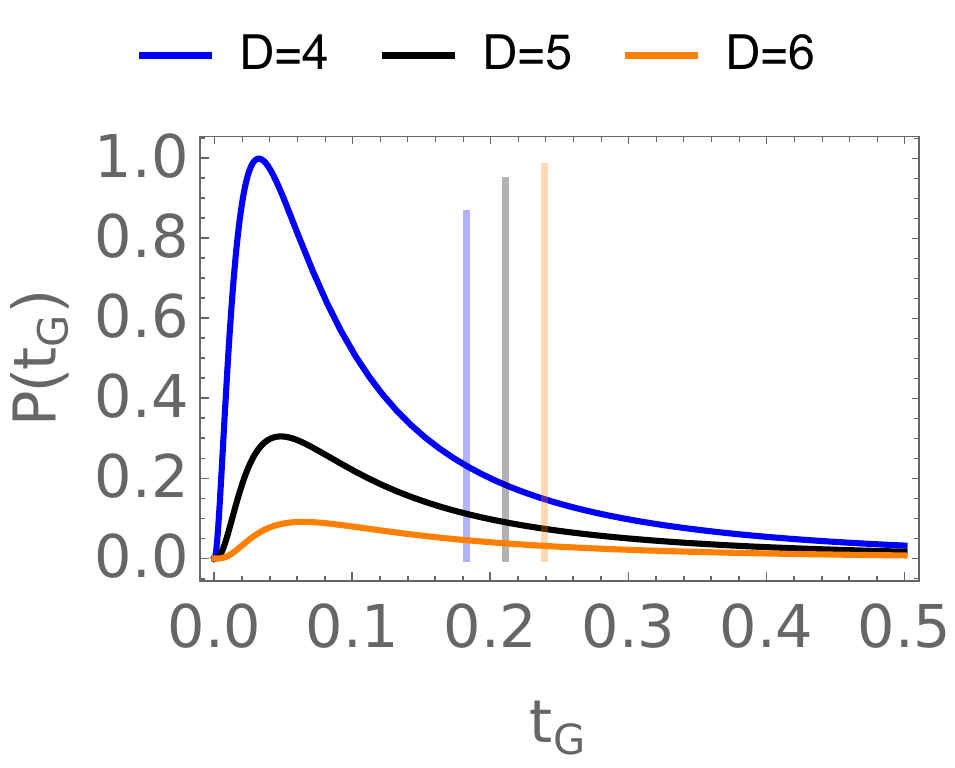}
    \includegraphics[width=0.323\textwidth]{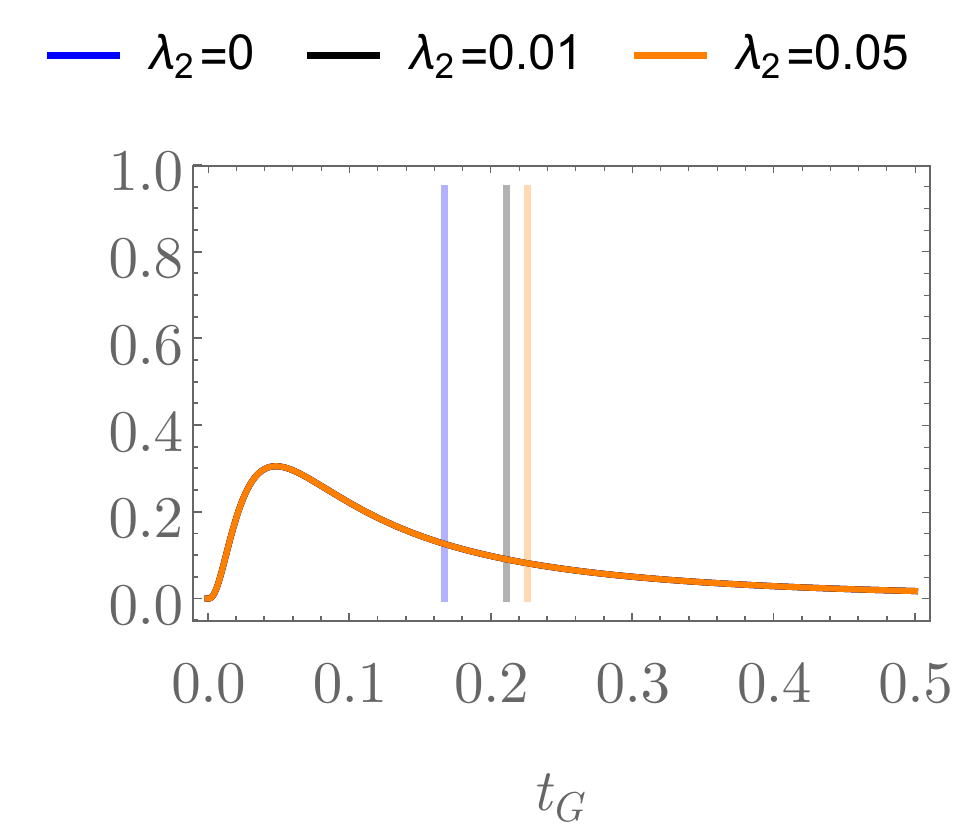}
    \includegraphics[width=0.323\textwidth]{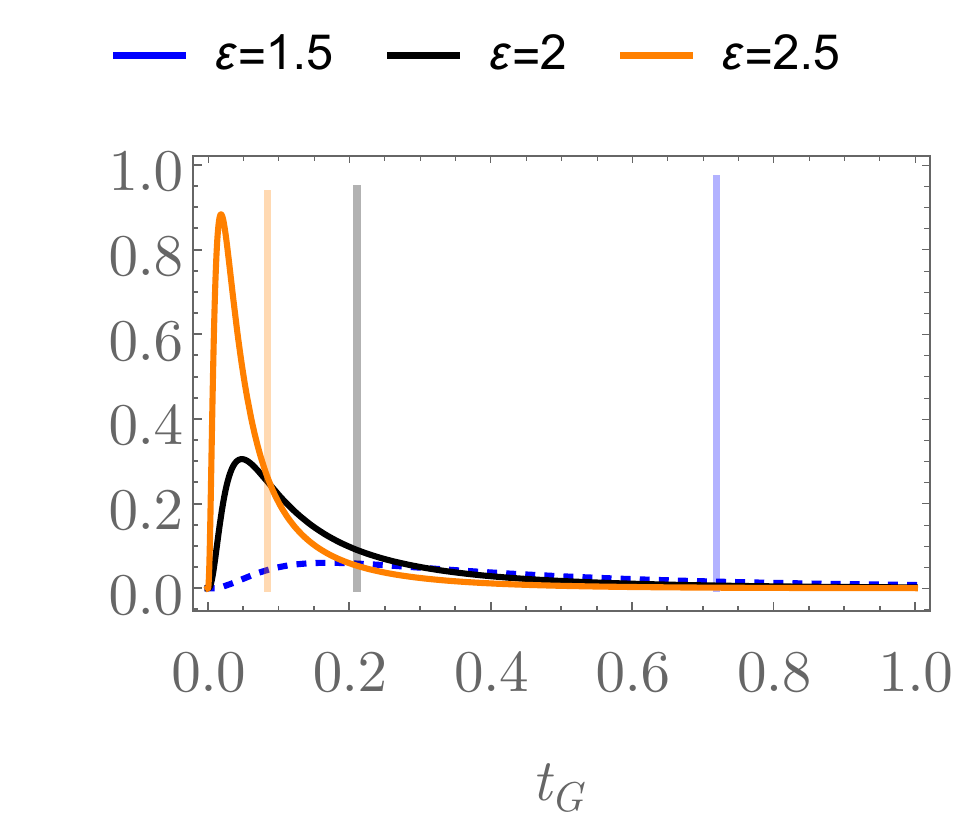}
    \caption{Grokking-time PDF for several values of $D$, $\varepsilon$, and $\lambda_2$. The short relaxation grokking-time PDFs are represented by full lines. The Dirac-delta long-relaxation grokking time is represented by vertical bars. The position of the bar is the position of the Dirac-delta function and the height of the bar represents the weight of the Dirac-delta part of the distribution. If not specified in the panels, additional parameters are set to: $D = 5$, $\lambda_2=0.01$, and $\varepsilon=2$.}
    \label{fig:grokking time ball model}
\end{figure}

By combining the grokking-time PDFs for the fast and the slow relaxation we obtain the grokking-time PDF in the limit $\lambda_2\ll\varepsilon^2$. In \fref{fig:grokking time ball model} we show the grokking-time PDFs for several parameters sets in the considered limit. Increasing the input size $D$ reduces the probability of fast-relaxation grokking times and increases the slow-relaxation grokking time. While the fast-relaxation grokking time does not depend on the regularisation strength $\lambda_2$, smaller regularisation leads to increased slow-relaxation grokking time. On the contrary, larger class separation decreases both fast- and slow-relaxation grokking times. 
We do not expect the analytically obtained grokking-time PDF to quantitatively describe real experiments, particularly because it is a zeroth-order large $N$ solution. However, the bimodal structure and the qualitative parameter dependence should also be present in more realistic scenarios. We will discuss one such example in \sref{sec:ca grokking results}.

\subsubsection{Critical exponents for a general isotropic data PDF}
\label{sec:general critical exponents}
By assuming isotropic probability densities on a compact domain $P^\pm$ in $\mathds{R}^D$ we can relate the data PDF close to the domain boundary ($D-$dimensional sphere) with the critical exponent. For an isotropic data probability density we can write the test error close to the grokking transition as
\begin{align}
    E_{\rm test}(\delta h)\approx \frac{1}{2}I_{2\delta h-\delta h^{-2}}\left(\frac{D-1}{2},\frac{1}{2}\right) \int_{0}^{\delta h}\dd r \rho(r), 
\end{align}
where, $I_z(a,b)$ is the regularized incomplete beta function, $\rho(r)$ is a probability distribution to find a sample with $||x||_2=1-r$, and $\delta h \approx k (t_\epsilon-t)$, and coefficient $k$ is determined by expanding \eref{eq:D-dim solution} around $t_\epsilon$. If the density $\rho(r)$ admits a Taylor expansion around zero, i.e. $\rho(r)=\rho_0 + \rho_1 r +\mathcal{O}( r ^2)$, we find 
\begin{align}
    E_{\rm test}(\delta h)\approx& \frac{2^{\frac{D-1}{2}} \delta h^{\frac{D-1}{2}}}{(D-1) B\left(\frac{D-1}{2},\frac{1}{2}\right)} \left(\rho_0 \delta h + \frac{1}{2}\rho_1 \delta h^2\right) \\ \nonumber
    \propto & \rho_0 \delta h^{\frac{D+1}{2} } +\frac{1}{2}\rho_1 \delta h^{\frac{D+2}{2}}\\ \nonumber
    \propto & (t-t_\epsilon)^{\frac{D+1}{2}},
\end{align}
where $B(a,b)$ is the Euler beta function. Obtained critical exponent $\nu=\frac{D+1}{2}$ is universal for isotropic probability densities that do not vanish at the ball boundary and is consistent with the result in \eref{eq:ball error}. If in addition the density $\rho(\delta h)$ has an algebraic divergence, e.g. $\rho(\delta h)\approx \rho_\xi \delta h^{-\xi}$ where $0<\xi<1$ we get
\begin{align}
    \label{eq:critical divergent sphere}
    E_{\rm test}(t)\propto (t-t_\epsilon)^{\frac{D+1-2\xi}{2}}.
\end{align}

The critical exponent of the test error reveals the behaviour of the sample density at the boundary of the sample domain. While both the grokking probability and the grokking-time PDF depend on the details of the model's initial parameters and the evolution, the critical exponent $\nu$ depends only on the data distribution at the boundary of the domain. Therefore, we expect that \eref{eq:critical divergent sphere} describes the critical exponent quantitatively also in a more general setting. In this case we might have to relax the condition $0<\xi<1$ to accommodate a more general divergence of the data distribution at the sample domain boundary.

\section{Learning local rules with shallow tensor networks}
\label{sec:tensor network model}
In standard rule-learning theory the teacher-student model describes a setting where the student model has to learn a rule given by the teacher model, see ~\cite{engel2001statistical}. In the simplest scenario where the teacher and the student models are perceptrons of the form \eref{eq:perceptron} we use statistical mechanics methods to calculate the expected generalisation error for a given number of training samples (or training time). This is achieved in the thermodynamic limit where the input size $M$, and the number of training samples $N$ go to infinity such that $N=\alpha M$. The teacher and student weights are sampled uniformly on an $M$-sphere. In this setup, one can use the replica trick~\cite{gardner1989three} to calculate the test-error behaviour as a function of $\alpha$. One finds $E_{\rm test}\propto \frac{1}{\alpha}$ when $\alpha\rightarrow\infty$. Although sudden transitions to zero generalisation error are possible, they are a consequence of a restriction on the phase space of parameters, e.g. in the Ising perceptron the parameters can take only values $\pm1$.

In summary, the standard rule-learning theory does not describe the grokking phenomenon and it is not clear how to reconcile the standard algebraic decay to zero test error with the grokking phase transition observed in deep models and presented in \sref{sec:toy grokking model}. 

In this section, we fill this gap by introducing a local-rule learning scenario and a tensor-network map, allowing to interpolate between the standard mean-field like theory and the local, grokking setup. In particular, we introduce a local teacher and a tensor-network student setup which displays the grokking behaviour described in the previous section without any restriction on the values of the student model parameters. The tensor-network techniques will provide a correspondence between the standard teacher-student setup in the thermodynamic limit and the setup described in the \sref{sec:toy grokking model}. The grokking phase transition is then a consequence of the locality of the learned rule.

\subsection{Local teacher model}
In the standard statistical-learning scenario, we determine the output of the teacher model (the rule) by \eref{eq:perceptron} (see \cite{engel2001statistical}). In this case all values of the input contribute to the final result. In the thermodynamic limit this leads to a mean-field like behaviour, i.e. the value of the input at any particular position has only infinitesimal influence on the result/rule.

\begin{figure}[!htb]
    \centering
    \includegraphics[width=0.6\textwidth]{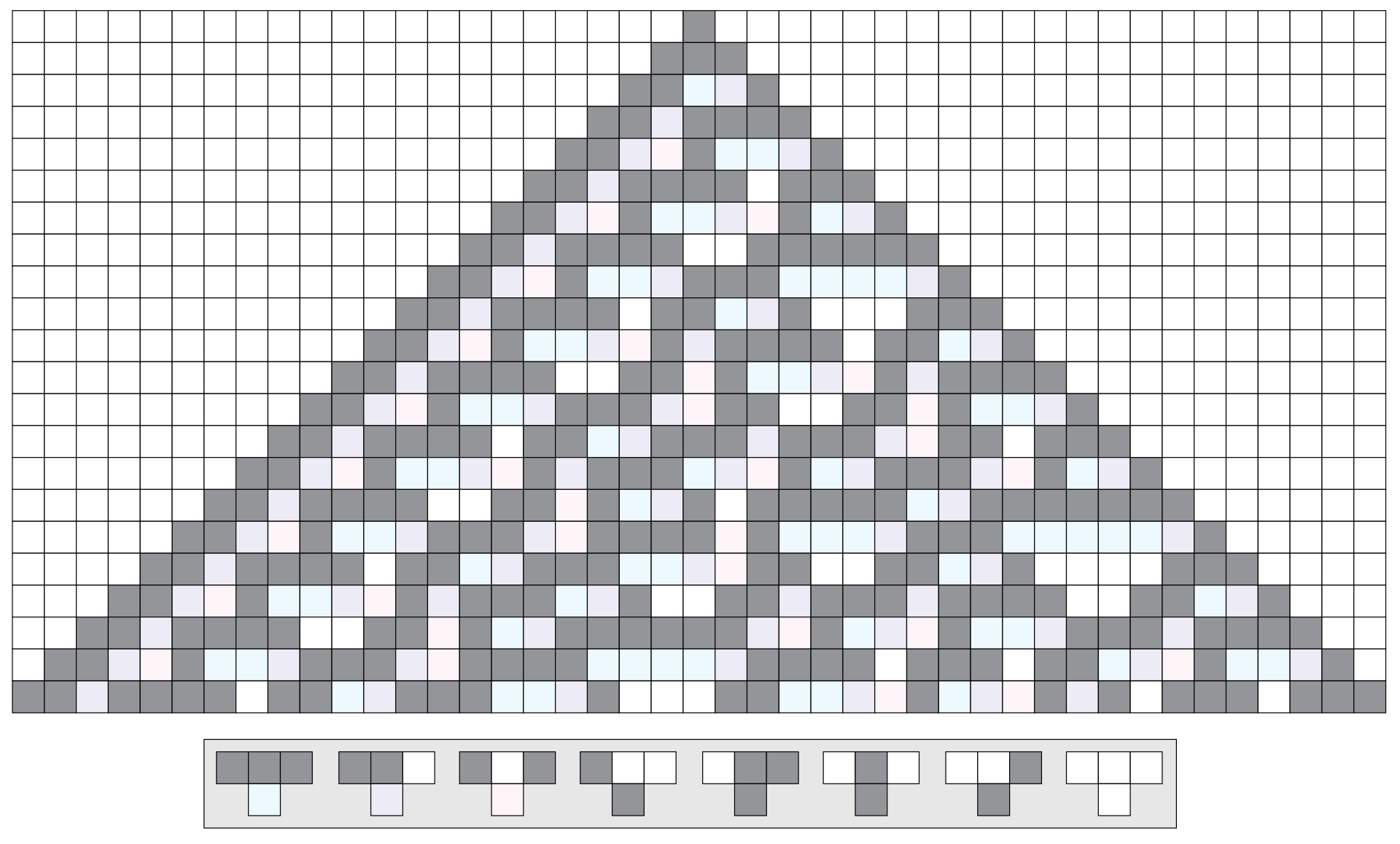}
    \caption{Diagramatic representation of the rule-30 cellular automaton. In our convention, the black cells represent 1 and the white cells represent -1. The diagram is taken from Wikipedia \cite{rule30}.}
    \label{fig:rule 30}
\end{figure}

We will study the opposite, local scenario $x\rightarrow y$, where $x,y\in\{-1,1\}^M$. The $i$-th component of the output vector $y_i$ will depend only on a $K$-neighborhood of the input at position $i$
\begin{align}
    \label{eq:local rule}
    y_i = \mathrm{rule}(x_{i-K},\ldots,x_{i},\ldots x_{i+K}).
\end{align}
We call such model a $K-$local model. The \eref{eq:local rule} describes a well-known cellular automata computational paradigm. Cellular automata are a universal discrete space-time dynamical systems with a finite set of possible states at each position \cite{wolfram1983statistical, wolfram2002new}. We define a cellular automaton by a set of rules which transform one configuration of states into another configuration. We will consider the rule 30 one-dimensional automaton ($K=1$) \cite{wolfram1983statistical,wolfram2002new}, which exhibits chaotic behaviour and is defined by the rule $y_i=\mbox{rule30}(x_{i-1},x_i,x_{i+1})$. The next state of the cell $i$, i.e. $y_i$, is determined by the current configuration at cells $i-1$, $i$, and $i+1$, i.e. $x_{i-1},x_i,x_{i+1}$, as follows 
\small{
\begin{align}
\begin{tabular}{c|c|c|c|c|c|c|c|c}
     $\mathbf{x_{i-1},x_i,x_{i+1}}$ & -1,-1,-1 & -1,-1,1 & -1,1,-1 & -1,1,1 & 1,-1,-1 & 1,-1,1 & 1,1,-1 & 1,1,1 \\
     \hline
     ${\bf y_i=\mbox{rule30}(x_{i-1},x_i,x_{i+1})}$ & -1& 1& 1& 1& 1& -1& -1& -1
\end{tabular}.
\end{align}
}
We show an example time evolution of the rule 30 cellular automaton in \fref{fig:rule 30}. The initial condition is represented by the first line, black cells represent the value 1, and white cells represent the value -1. Our aim will be to learn one step of this evolution.

The rule-30 automaton has already been discussed in the context of sequence-to-sequence prediction with tensor networks \cite{guo2018matrix,efthymiou2019tensornetwork,zunkovic2022Deep}, however, no grokking phenomena have been reported. To study the effect of the neighbourhood size $K$, we shall consider a rule defined by $K$ consecutive applications of rule 30. We will refer to such rule as a $K$--local rule. 

In summary, we modify the standard perceptron teacher-student setup by restricting the teacher model to local instead of global rules. The teacher will be modelled by a local map transforming a sequence $x$ into the sequence $y$. The task will be to approximate the chosen map by training on a finite set of input samples of length $M$. For a finite $M$ we can choose open and closed boundary conditions. Open boundary conditions refer to the case when $y_1$ and $y_M$ are calculated as if $x_{0}=x_{M+1}=-1$. In the case of closed boundary conditions we have $x_0=x_M$ and $x_{M+1}=x_1$. The test set will include all possible input sizes from $M_{\rm test}=3,4,\ldots,\infty$. We will determine the error as the ratio of incorrectly predicted values $y_i$.

Besides the change from a global to a local rule, we will also modify the student model. Instead of the standard perceptron student model, we will use the uniform tensor-network attention model.

\subsection{Uniform tensor-network student model}
\label{sec:student model}
The simplest student model discussed in the literature is a perceptron model which is not applicable to our problem, since we will discuss inputs of different sizes. The standard architectures applicable to variable size inputs are the recurrent neural network (RNN) model and the convolutional neural network model. However, we found it convenient to use a tensor network approach, which enables us to construct a bridge between the teacher-student rule learning scenario and the grokking model discussed in the previous section.\footnote{Due to a connection between RNNs and tensor networks \cite{wu2022tensor} we expect that one can rephrase our tensor network model in the language of RNNs.} Before introducing the tensor-network attention layer and the student model we will summarise the basic properties of tensor networks applied to machine learning \cite{stoudenmire2016supervised, stoudenmire2018learning}.

\subsubsection{Short introduction to tensor network methods}

A tensor network is a tensor that is represented as a contraction of two or more tensors. The tensors that are contracted typically have much smaller number of dimensions (indices) and hence less parameters. A trivial example of a tensor network is a scalar product of two vectors, where the second vector is a result of a matrix vector multiplication,  $c= u\cdot A v$
\begin{eqnarray}
    &\vcenter{\hbox{\includegraphics[width=16px]{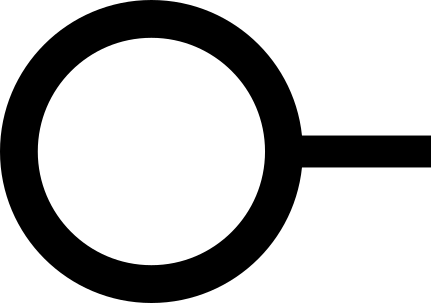}}}=v,\quad \vcenter{\hbox{\includegraphics[width=16px]{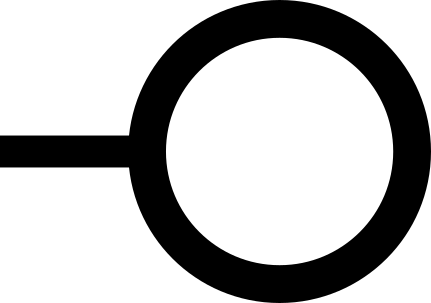}}}=u,\quad \vcenter{\hbox{\includegraphics[width=22px]{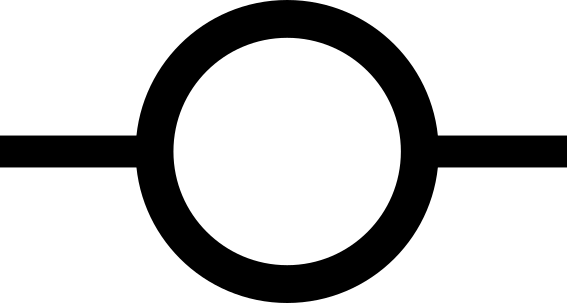}}}=A,&\\ \nonumber
    & \vcenter{\hbox{\includegraphics[width=40px]{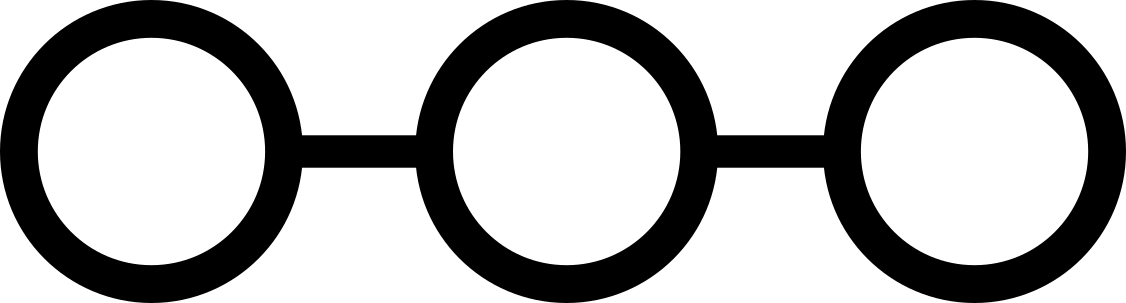}}} = c .&
\end{eqnarray}
We introduced a diagrammatic notation, which makes longer tensor contractions more transparent. A tensor in this notation is represented as a circle with legs. The direction of the legs is typically not important. The number of legs determines the dimensionality of the tensor, e.g. a number has zero legs, a vector has one leg, a matrix has two legs etc. The most prominent tensor network, related to RNNs, is the matrix product state (MPS) obtained by contracting 3-dimensional tensors
\begin{align}
    \psi=\vcenter{\hbox{\includegraphics[width=100px]{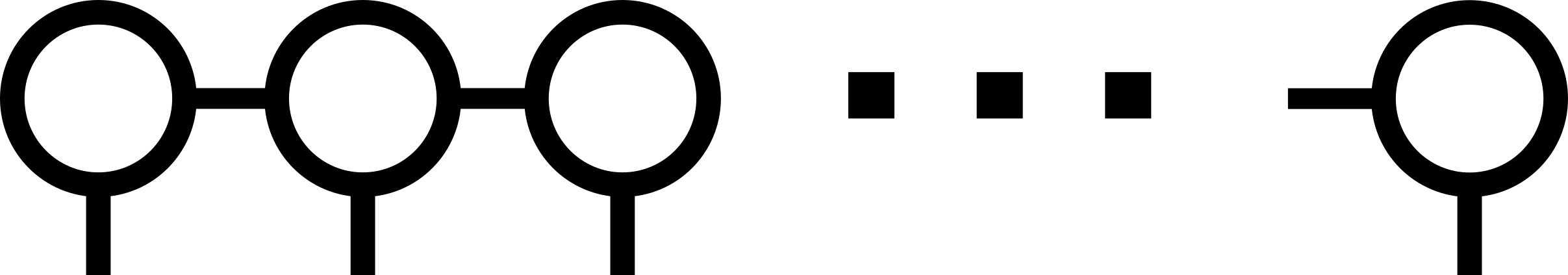}}}.
\end{align}
To use a tensor network as a machine learning model, we have to transform the inputs such that they can be contracted with the tensor-network model in order to produce a scalar output. We do that by using an embedding function and transform the elements of the input vector with a vector transformation
\begin{align}
    \phi(x_j)=\vcenter{\hbox{\includegraphics[width=12px]{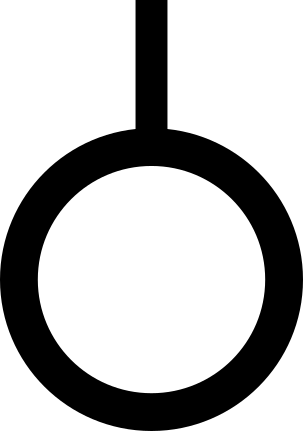}}}.
\end{align}
The entire input vector is then transformed as
\begin{align}
    \Phi(x)=\phi(x_1)\otimes\phi(x_2)\otimes\ldots \otimes\phi(x_M)=\vcenter{\hbox{\includegraphics[width=100px]{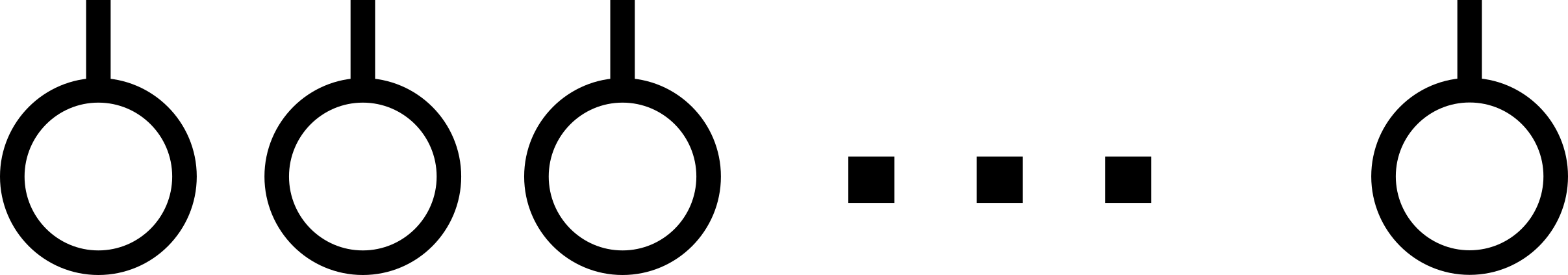}}}.
\end{align}
Formally, $\Phi(x)$ is an exponentially large vector with a compact MPS representation and will never be used directly. The output of the MPS model is then a contraction of the embedded input elements with the MPS tensor-network model
\begin{align}
    \psi \cdot \Phi(x)= \vcenter{\hbox{\includegraphics[width=100px]{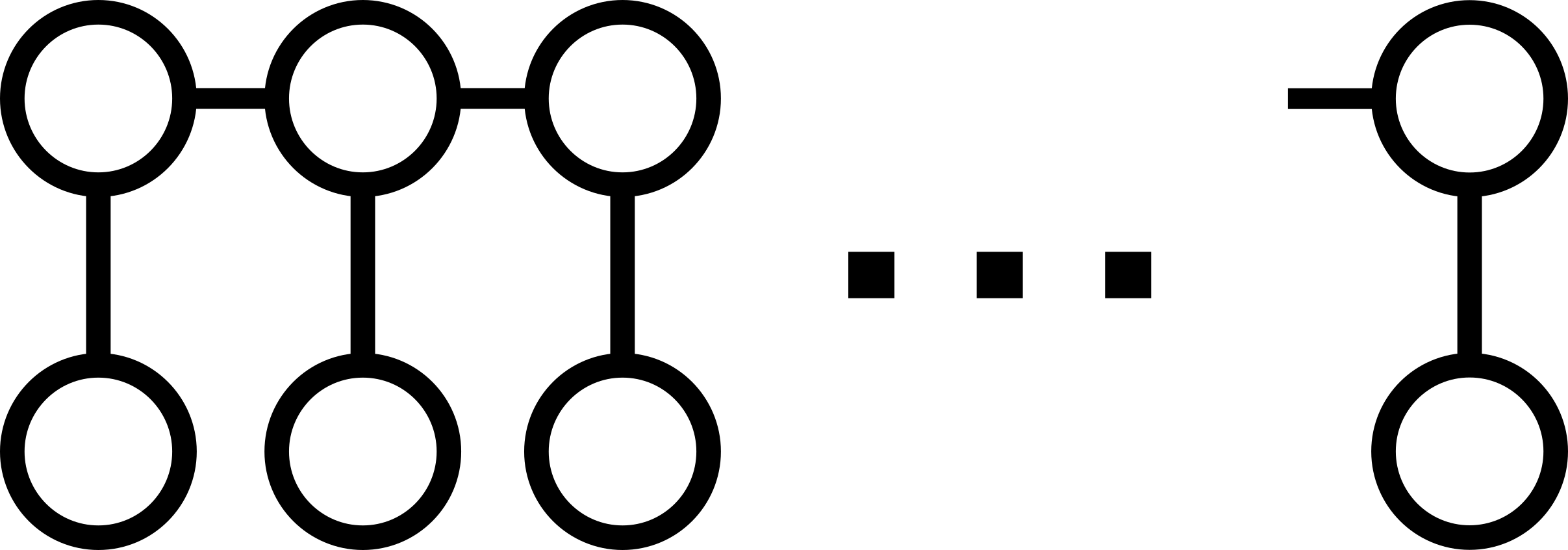}}}
\end{align}
We can produce a vector output by adding one dimension to one of the MPS tensors. The presented setup has all main parts of the typical tensor-network model. It is differentiable with respect to tensor-network parameters and applicable to the standard training methods based on gradient descent.

\subsubsection{Tensor-network attention model}
In this section, we will introduce a simplified version of the tensor network proposed in \cite{zunkovic2022Deep}. As in the introductory example above, the entire model has two parts: an embedding layer and a tensor-network attention layer. Since the input is binary, we define the embedding layer with a local embedding function $\phi(x_i):\{-1,1\}\rightarrow \mathds{R}^2$ as 
\begin{align}
  \phi(-1)=\begin{pmatrix}1\\0\end{pmatrix},\quad   \phi(1)=\begin{pmatrix}0\\1\end{pmatrix}.
\end{align}
After the embedding, we apply the tensor-network attention determined by two parameter tensors $A,B\in \mathds{R}^{d\times d\times 2}$. We call the tensor $A$ the attention tensor and the tensor $B$ the classification tensor. The names of the tensors $A$ and $B$ reflect their role in the tensor-network attention layer. As we will describe below (see also \sref{sec:tensor network map}), for a given position the tensor $A$ determines the context of the input which is then linearly classified by the tensor $B$.

First, we construct matrices $\mathcal{A}(i)$ by contracting the attention tensor $A$ with the local embedding vectors $\phi(x_i)$
\begin{align}
    \mathcal{A}_{\mu,\nu}(i)= \sum_{j=1}^2 A_{\mu,\nu,j}\phi(x_i)_j.
\end{align}
Then, we use the matrices $\mathcal{A}(i)$ to construct the left and right context matrices $H^{\rm L,R}(i)$
\begin{align}
    H^{\rm L}(1)&=\mathds{1}_d, & H^{\rm L}(i)&=H^{\rm L}(i-1)\mathcal{A}(i-1),\\
    H^{\rm R}(M)&=G, & H^{\rm R}(i)&=\mathcal{A}(i+1)H^{\rm R}(i+1).
\end{align}
The matrix $G$ determines the boundary conditions of the model. In the case of closed boundary conditions $G=\mathds{1}_d$. In the case of open boundary conditions $G=v^{\rm L}\otimes v^{\rm R}$, where the boundary vectors $v^{\rm L,R}\in \mathds{R}^d$ are additional model parameters. Alternatively, the boundary vectors $v^{\rm L,R}$ can be determined as left and right eigenvectors of the matrix $A_0$ corresponding to the largest eigenvalue. The final local weight vector $w(i)$ is then obtained by contracting the tensor $B$ with the normalised left and right context matrices $H_{\rm N}^{\rm L,R}=H  ^{\rm L,R}/||H^{\rm L,R}||_2$,
\begin{align}
    \label{eq:tn w}
    w(i)_j = \tr\left( H_{\rm N}^{\rm L}(i) B_j H_{\rm N}^{\rm R}(i)\right),\quad j=1,2,\quad i=1,\ldots, M,
\end{align}
where $B_j$ denotes the matrix with elements $[B_j]_{\mu,\nu}=B_{\mu,\nu,j}$. We calculate the attention layer output at position $i$ as
\begin{align}
    \label{eq:tn y}
    \hat{y}_i= w(i)\cdot\phi(x_i).
\end{align}
The final model output is then obtained by using the sign nonlinearity $f(x)=\sgn(\hat{y})$. The described tensor-network layer is a generalisation of the linear-dot attention mechanism (see \cite{zunkovic2022Deep}). Therefore, we refer to it as a tensor-network attention.

It is instructive to present the tensor-network attention layer in a diagramatic form by using the following definitions
\begin{align}
    \phi(x_i)=\,&\vcenter{\hbox{\includegraphics[width=18px]{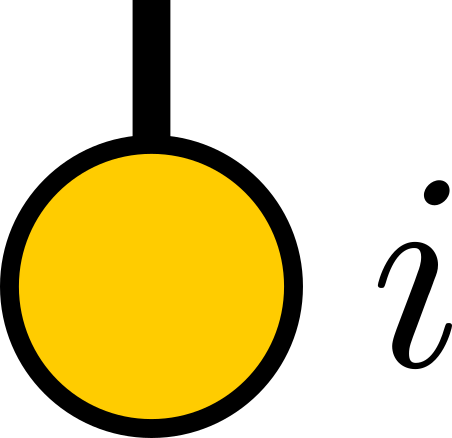}}}, \quad
    A = \vcenter{\hbox{\includegraphics[width=22px]{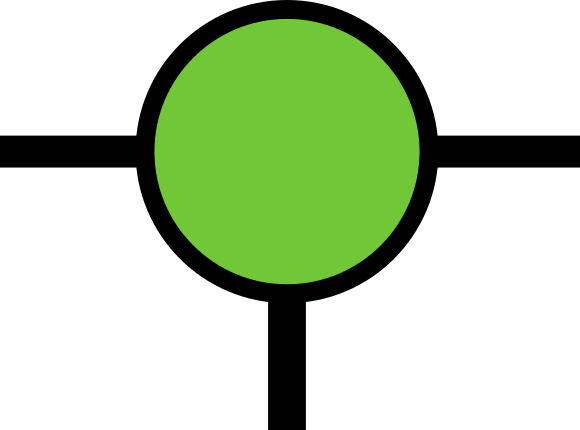}}},\quad
    B =\vcenter{\hbox{\includegraphics[width=22px]{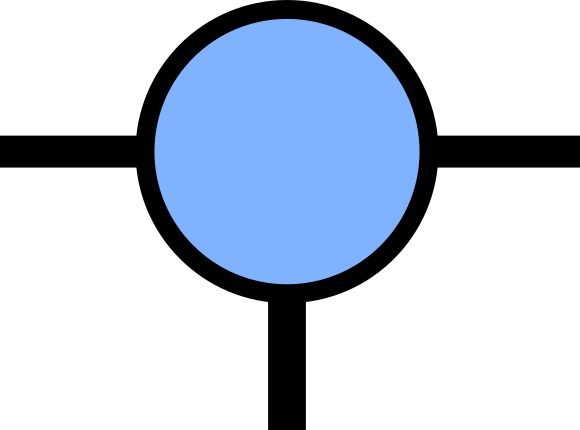}}} ,\quad
    \mathds{I}_D =\vcenter{\hbox{\includegraphics[width=22px]{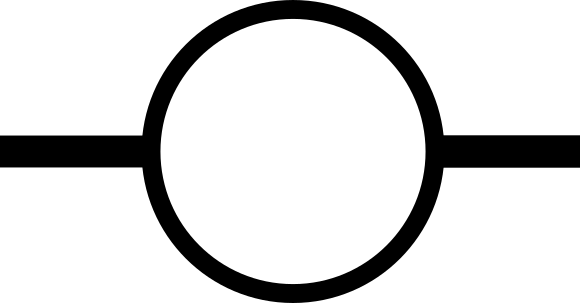}}} ,\quad
    G =\vcenter{\hbox{\includegraphics[width=22px]{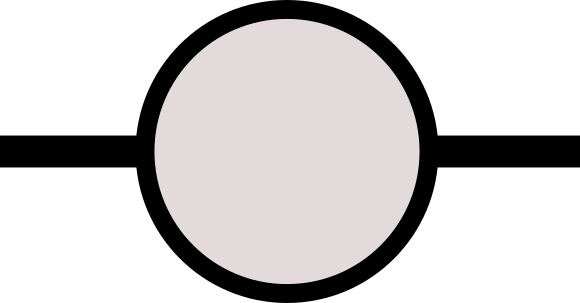}}}.
\end{align}
We compactly write the entire transformation of an input at the position $i$ as
\begin{align}
    \label{eq:tn model}
    \mathcal{A}(j) =\,& \vcenter{\hbox{\includegraphics[width=22px]{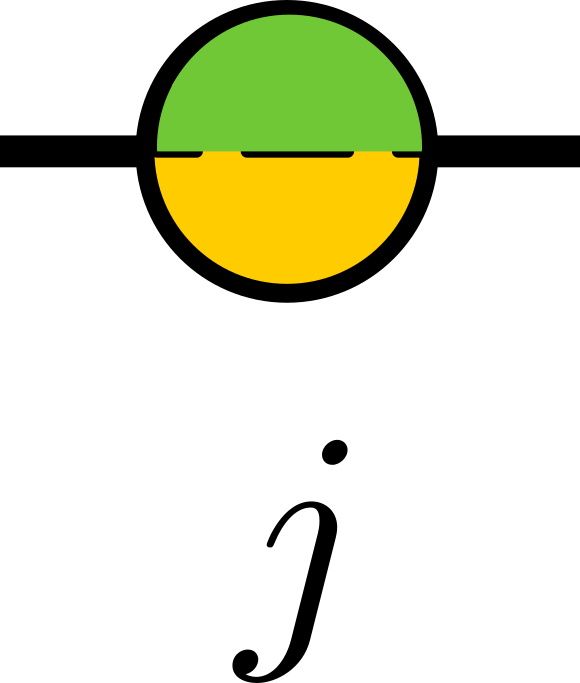}}} = \vcenter{\hbox{\includegraphics[width=22px]{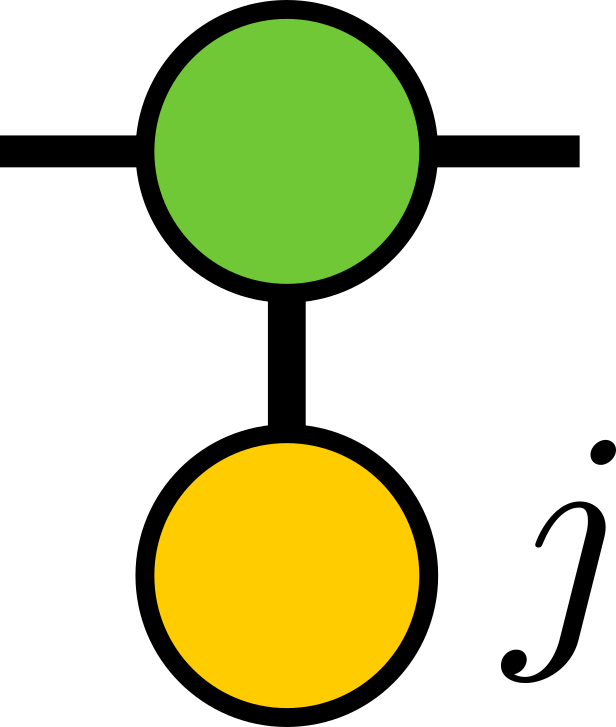}}}~,\\ \nonumber
    H^{\rm L}(i)  =\,& \vcenter{\hbox{\includegraphics[width=22px]{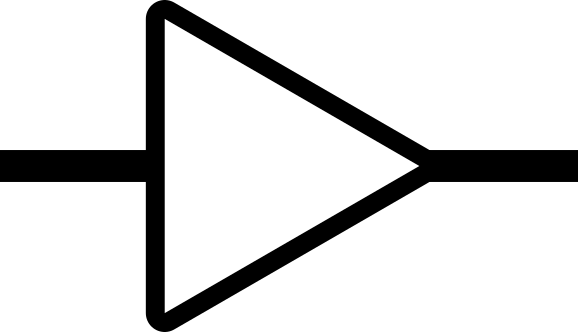}}} = \vcenter{\hbox{\includegraphics[width=110px]{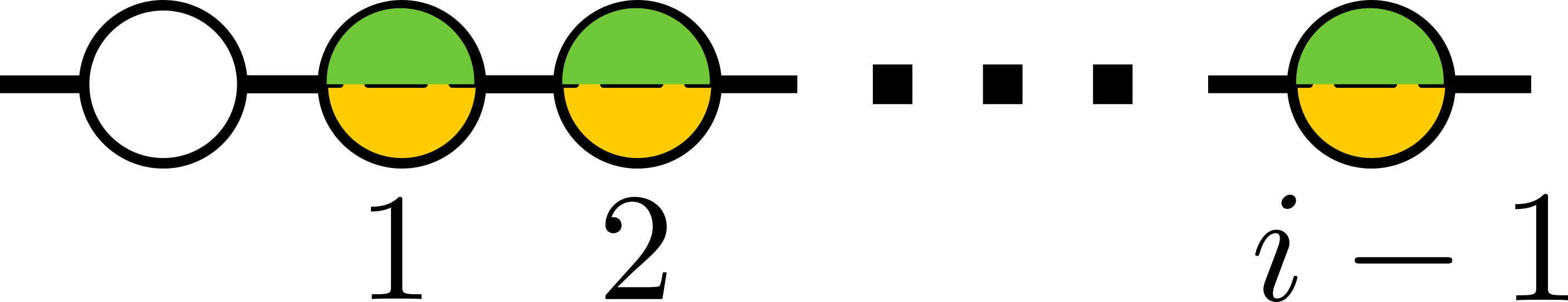}}}~,\\ \nonumber
    H^{\rm R}(i)  =\,& \vcenter{\hbox{\includegraphics[width=22px]{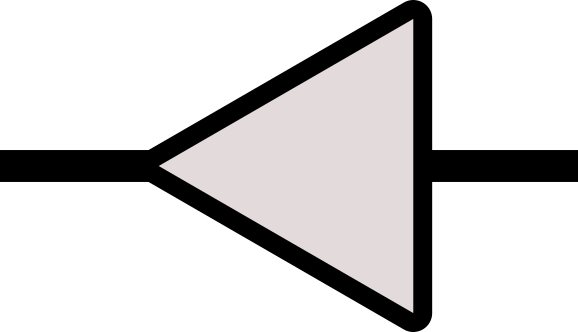}}} = \vcenter{\hbox{\includegraphics[width=110px]{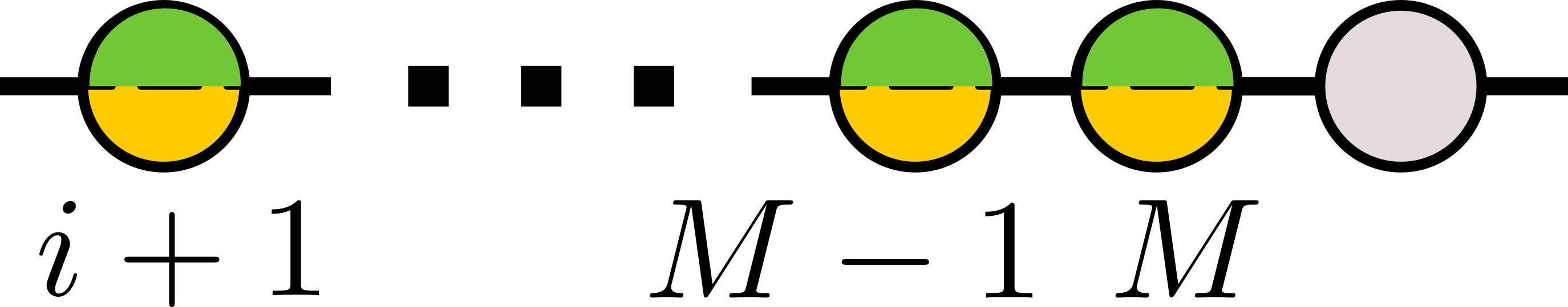}}}~,\\ \nonumber
    \hat{y}_i=\,&\vcenter{\hbox{\includegraphics[width=60px]{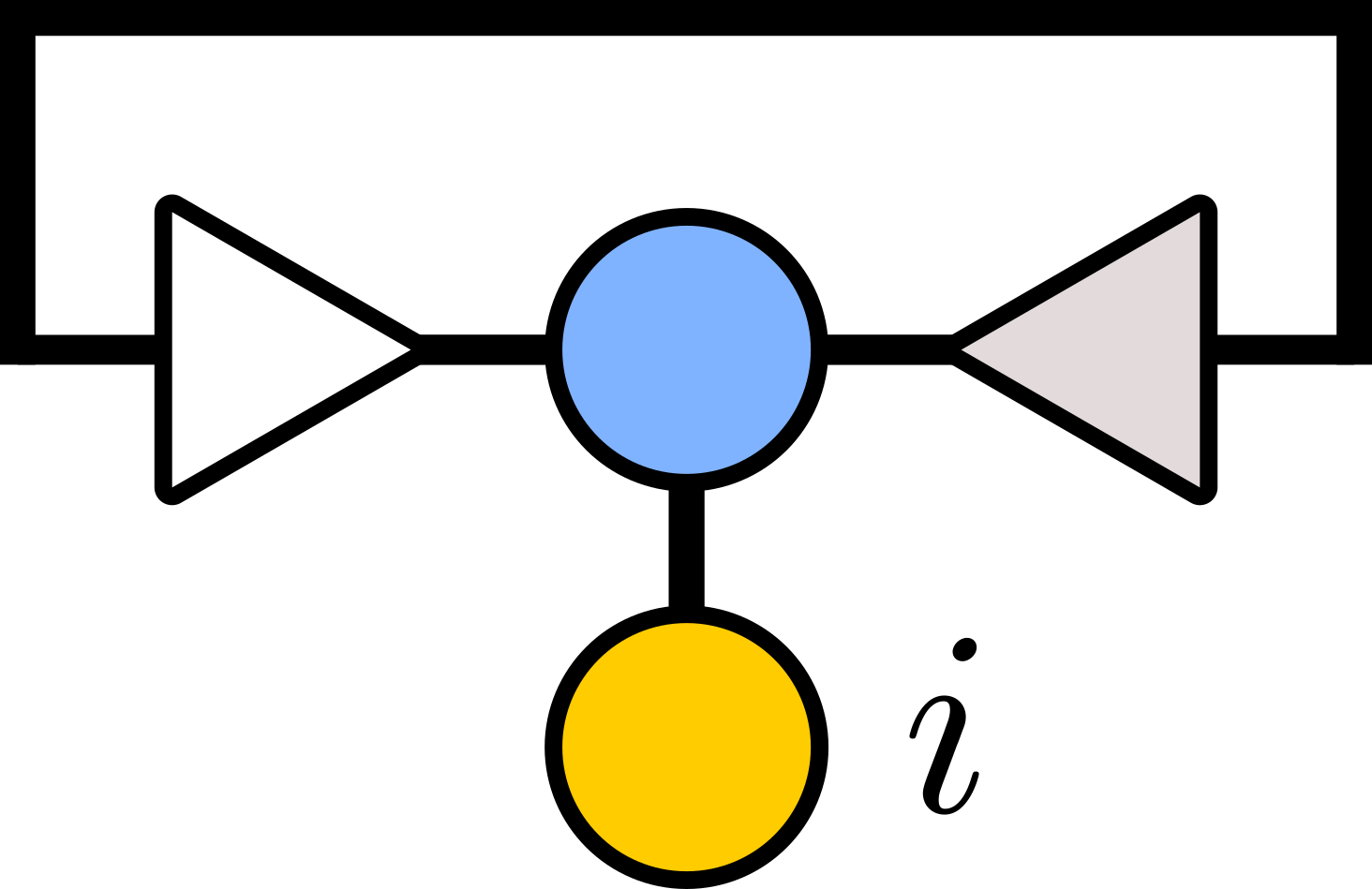}}}~.
\end{align}

\subsubsection{Tensor network map}
\label{sec:tensor network map}
\begin{figure}[!htb]
\centering

\begin{tikzpicture}

\begin{scope}[xshift=8cm]
\draw[thick] (0,0) circle (2cm);
\draw (-2,0) arc (180:360:2cm and 0.5cm);
\draw[dashed] (-2,0) arc (180:0:2cm and 0.5cm);

\draw[dashed,green!20,fill=green!10,rotate=10] (0.3,1.5) ellipse (0.6cm and 0.3cm);
\draw[dashed,blue!20,fill=blue!10,rotate=60] (-1,-0.75) ellipse (0.7cm and 0.5cm);
\draw[latex-latex] (0,1.2) -- (0.2,-0.6) node[midway,right] {$2\varepsilon$};

\coordinate (S1) at (-0.3,1.5);
\filldraw[green!50] (S1) circle (2pt);
\coordinate (S2) at (0.1,1.7);
\filldraw[green!50] (S2) circle (2pt);
\coordinate (S3) at (0.3,1.6);
\filldraw[green!50] (S3) circle (2pt);

\coordinate (S4) at (0.2,-0.8);
\filldraw[blue!50] (S4) circle (2pt);
\coordinate (S5) at (0.5,-1.2);
\filldraw[blue!50] (S5) circle (2pt);
\coordinate (S6) at (0.2,-1.4);
\filldraw[blue!50] (S6) circle (2pt);
\coordinate (S7) at (-0.1,-1.6);
\filldraw[blue!50] (S7) circle (2pt);

\coordinate (P1) at (-7.3,3);
\coordinate (P2) at (-6.6,3);
\coordinate (P3) at (-7.6,-1.5);

\coordinate (P4) at (-7,-1.5);
\coordinate (P5) at (-5.8,3);
\coordinate (P6) at (-6.2,-1.5);
\coordinate (P7) at (-5.4,-1.5);

\draw[gray!50] (S1) to[out=110,in=80] (P1);
\draw[gray!50] (S2) to[out=100,in=80] (P2);
\draw[gray!50] (S3) to[out=-140,in=90] (P3);

\draw[gray!50] (S4) to[out=160,in=70] (P4);
\draw[gray!50] (S5) to[out=170,in=80] (P5);
\draw[gray!50] (S6) to[out=170,in=85] (P6);
\draw[gray!50] (S7) to[out=-120,in=90] (P7);

\draw (-7.8,2.9) node[left] {\large $x$};
\draw (-7.8,2.1) node[left] {\large $y$};
\draw (-6.7,1.5) node[below] {\large $M=3$};

\draw (-8.1,-1.6) node[left] {\large $x$};
\draw (-8.1,-2.4) node[left] {\large $y$};
\draw (-6.7,-3) node[below] {\large $M=4$};

\draw (1,5) node[left] {\Large $x\in \{-1,1\}^{\otimes M} \xrightarrow{H^{\rm R}(i)H^{\rm L}(i)\otimes \Phi(x_{i})} \mathbb{R}^{2d^{2}}$};

\end{scope}

\begin{scope}[xshift=1.5cm,yshift=2.5cm]
\matrix (top) [A]
{
$1$ & $-1$ & $1$ \\
$-1$ & $-1$ & $1$ \\
};
\end{scope}

\begin{scope}[xshift=1.5cm,yshift = -2cm]
\matrix (bottom) [A,every even row/.style={
            nodes={fill=blue!30}
        },
    	row 2 column 1/.style={
        nodes={fill=green!30}
        }]
{
$1$ & $-1$ & $-1$ & $1$ \\
$1$ & $1$ & $1$ & $1$ \\
};
\end{scope}

\end{tikzpicture}
    
\caption{The tensor network map from $\{-1,1\}^M$ to $\mathds{R}^{2d^2}$ implemented with \eref{eq:tn map}, with $2\varepsilon$ denoting the distance between the closest positive and negative samples.}
\label{fig:tn map}
\end{figure}
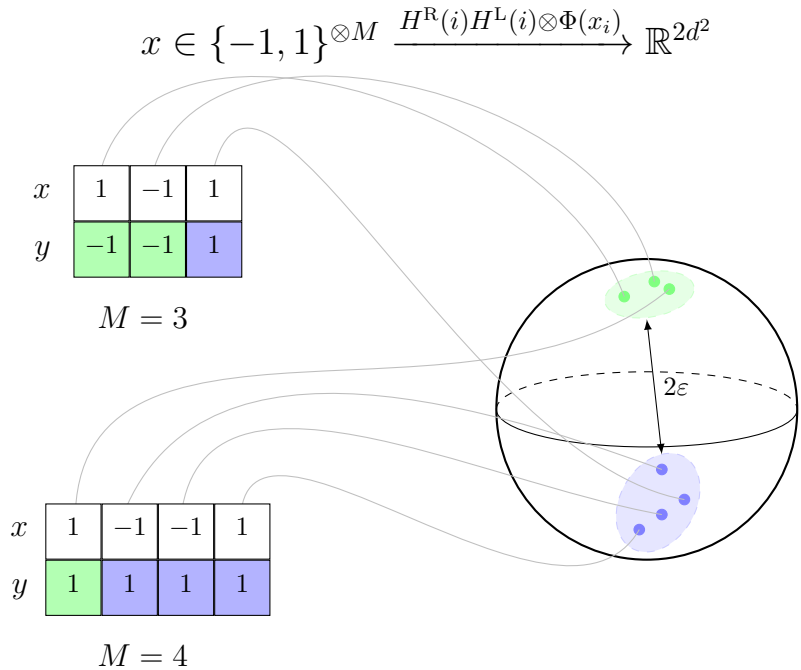

The described tensor-network attention model also implements a map from inputs of variable length $M$ to vectors of length $2d^2$. In the case of fixed attention tensors $A$ all possible infinitely many inputs define a PDF of vectors
$z_i(x)\in\mathds{R}^{2d^2}$, where 
\begin{align}
 \label{eq:tn map}
 z_i(x)=H_{\rm N}^{\rm R}(i)H_{\rm N}^{\rm L}(i)\otimes\phi(x_i).
\end{align}
We show a schematic representation of the map in \fref{fig:tn map}. By considering $z_i(x)$ as input features we can interpret the model defined by \eref{eq:tn y} as a perceptron defined by the weight tensor $B$, namely
\begin{align}
    \label{eq:perceptron map}
    \hat{y} = z_i(x)\cdot \vec{B}.
\end{align}
In the above formula $\vec{B}$ denotes the vectorised classification tensor $B$. 
By setting $D=2d^2$, we have mapped the local-rule learning problem in the thermodynamic limit to a (grokking) classification problem of the form discussed in \sref{sec:toy grokking model}. Interestingly, for a $K-$local rule, we can find $4^K$-dimensional matrices $A$ for which the transformed problem is solvable by a simple perceptron model and exhibits the grokking phenomena. Therefore, the $1/\alpha$ dependence on the training set size obtained from the standard rule-learning theory seems to be a consequence of the mean-field type infinite-range rule. For any local rule, we will observe grokking.

\subsection{Simulation details and results}
\label{sec:ca grokking results}
In this section, we present the results of training the uniform tensor-network student model on local algorithmic datasets. First, we consider the setting where the tensors $A$ are fixed and map the inputs to separable distributions as shown in \fref{fig:tn map}. We compare the numerical results with the predictions of the grokking model presented in \sref{sec:toy grokking model}. Second, we train the entire student model, i.e. the tensor $A$ and the tensor $B$. In this case, we also discuss structure formation.

\subsubsection{Constant attention tensors}
\label{sec:constant attention tensors}
We now discuss the simulation results  obtained by fixing the attention tensors $A$. Namely, we use the proposed tensor-network model as a map from $\{-1,1\}^M$ to $\mathds{R}^{2d^2}$ as discussed in \sref{sec:tensor network map} and shown in \fref{fig:tn map}. We choose the left and the right boundary vectors $v^{\rm L,R}$ to be the eigenvectors of $A_0$ corresponding to the largest eigenvalue. We also fix the bond dimension $d=2$ and study the 1--local rule, which facilitates the comparison with the results discussed in \sref{sec:toy grokking model}.

We determine the attention tensors $A$ by independently sampling each element according to the normal distribution with zero mean and unit variance. Since not all attention vectors lead to solvable problems, we perform rejection sampling by checking if the final model parameters given by \eref{eq:D-dim solution} have zero test error. Once we obtain a solvable instance of the attention tensor $A$, we do not change its parameters during training. 

\begin{figure}[!htb]
    \centering
    \includegraphics[width=0.4\textwidth]{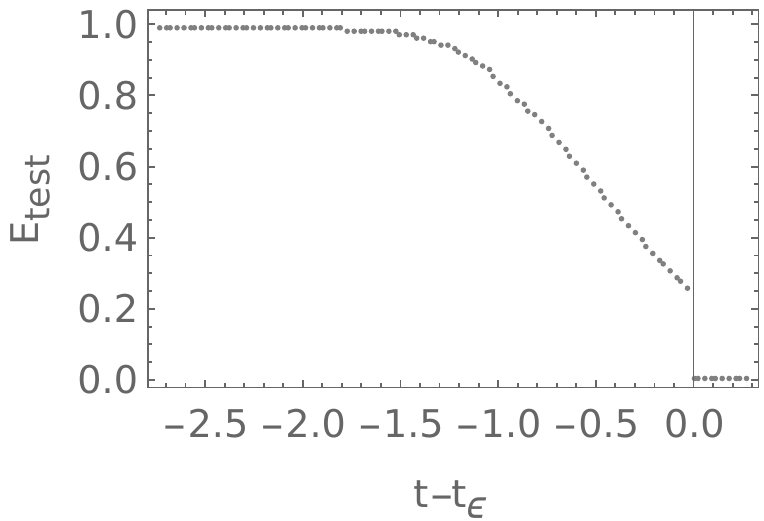}
    \caption{The first-order phase transition when learning with the exact attention tensors $A$ given in \eref{eq:exact A}. The jump in the transition is 1/4.}
    \label{fig:exact attention}
\end{figure}

\paragraph{Exact 1--local attention} We first consider the learning dynamics in the case of exact attention tensors
\begin{align}
    A_0=\left(
\begin{array}{cccc}
 1 & 1 & 0 & 0 \\
 0 & 0 & 0 & 0 \\
 1 & 1 & 0 & 0 \\
 0 & 0 & 0 & 0 \\
\end{array}
\right),\quad A_1=\left(
\begin{array}{cccc}
 0 & 0 & 0 & 0 \\
 0 & 0 & 1 & 1 \\
 0 & 0 & 0 & 0 \\
 0 & 0 & 1 & 1 \\
\end{array}
\right).
\label{eq:exact A}
\end{align}
The minimal bond dimension of the exact solution can be reduced to 2 if we generalise the model and train different left and right attention tensors $A$. In the exact, 1-local attention case, the vectors $z_i(x)$ contain only information about the state of the neighbouring positions. Since the smallest size $M=3$ contains all eight possible inputs, larger training set size $M$ does not change the results. After averaging over many initialisations of the classifier part of the network $B$ we obtain the average test error shown in \fref{fig:exact attention}. We observe a first-order transition with a jump of 1/4 in the test/train error. Here the factor 1/4 comes from the fact that the neighbourhood of any given position has four possible different values.

In the following, we discuss results obtained by randomly sampling the attention tensors $A$. We show results for three different attention tensors $A$, namely \texttt{Example  1}, \texttt{Example  2}, and \texttt{Example  3} reported in \aref{app:fixed attention}.

\begin{figure}[!htb]
    \centering
    \includegraphics[width=0.4\textwidth]{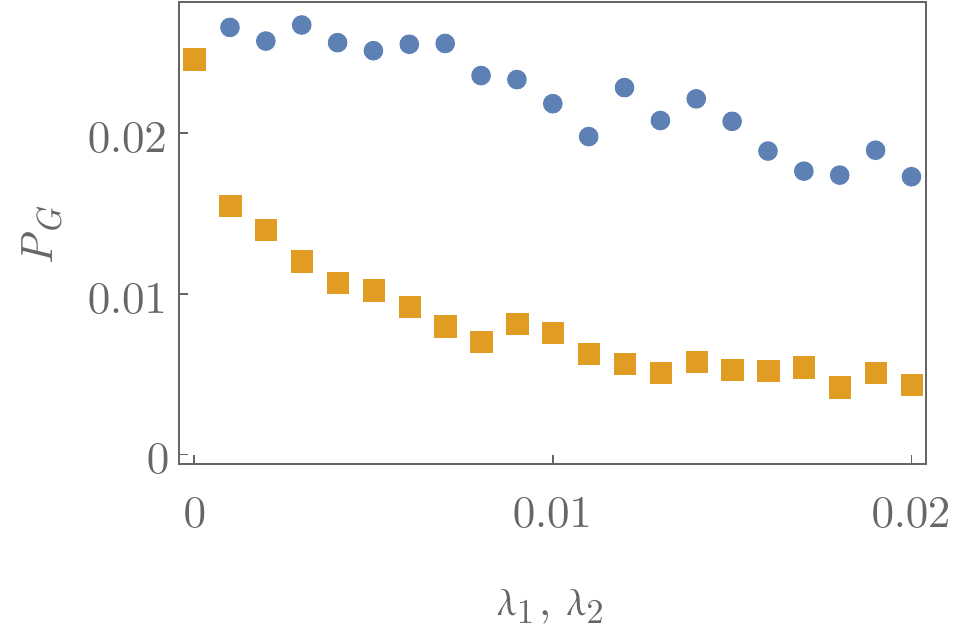}
    \caption{Grokking probability ($P_{\rm G}$), representing the fraction of the attention tensors $A$ that map the 1--local rule to linearly separable data. We show the dependence of $P_{\rm G}$ on the $L_{1}$ regularisation strength ($\lambda_1$ blue circles) and the $L_2$ regularisation strength ($\lambda_2$ orange squares). The $L_1$ regularised models have larger grokking probability compared to the $L_2$ regularised models. We used 20k random attention tensors to estimate the grokking probability.}
    \label{fig:grokking probability fixed attention}
\end{figure}

\paragraph{Grokking probability} We estimate the grokking probability as the fraction of the sampled attention tensors $A$ that leads to linearly separable feature space data for the studied rule. In contrast to the grokking probabilities discussed in \sref{sec:toy grokking model}, we fix the training set to contain all possible samples of length $M=3$. In \fref{fig:grokking probability fixed attention} we show the dependence of the grokking probability with respect to regularisation strengths $\lambda_{1,2}$. We observe that the $L_2$ regularisation decreases the grokking probability while the $L_1$ regularisation first slightly increases the grokking probability and then decreases compare to models without regularisation. In all cases the $L_1$ regularised model has larger grokking probability as the $L_2$ regularised model with the same regularisation strength. Larger grokking probability for $L_1$ regularised models is another indicator that $L_1$ regularisation could lead to better generalisation compared to the $L_2$ regularisation.

\begin{figure}[!htb]
    \centering
    \includegraphics[width=0.32\textwidth]{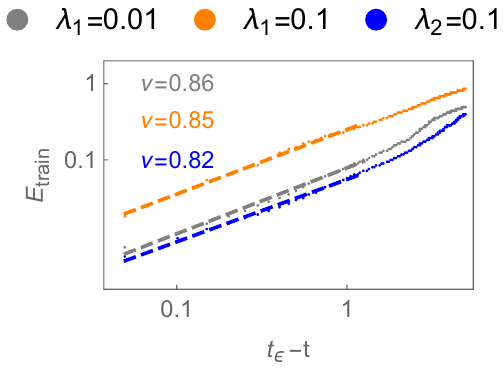}
    \includegraphics[width=0.32\textwidth]{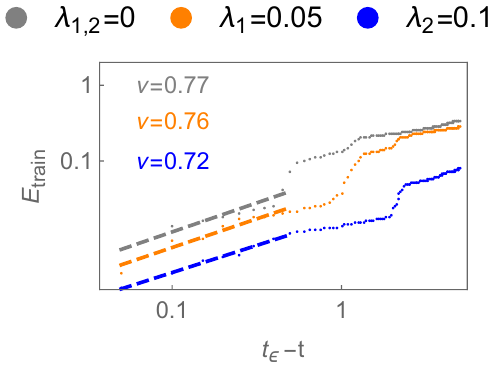}
    \includegraphics[width=0.31\textwidth]{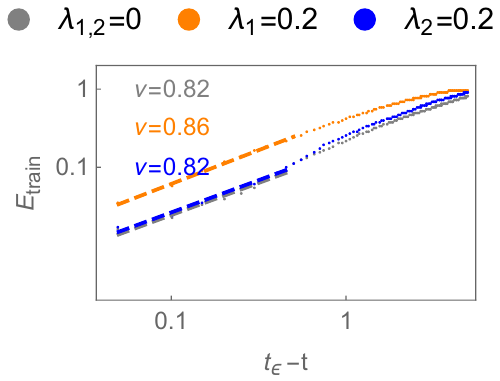}
    \caption{Average test error during training with fixed attention vectors (log-log plot). From left to right we report results for \texttt{Example  1}, \texttt{Example  2}, and \texttt{Example  3} attention tensors $A$ given in \aref{app:fixed attention}. The fitted critical exponents $\nu$ are shown in the plots and only mildly depend on $\lambda_{1,2}$.}
    \label{fig:fixed attention error}
\end{figure}

\paragraph{Critical exponent $\nu$} Sampled attention vectors $H^{\rm L,R}(i)$ also contain information about the input beyond only the neighbouring sites. Moreover, information about the neighbours is not complete. Therefore, we observe a second-order transition, as discussed in \sref{sec:toy grokking model}. In \fref{fig:fixed attention error} we show the average test error for three different but fixed attention tensors obtained by rejection sampling. The exact values of the attention vectors are reported in \aref{app:fixed attention}. We find that the critical exponent $\nu$ does not depend on the regularisation strengths $\lambda_{1,2}$ and is in all cases smaller than one, which is in agreement with the predictions of the simple grokking model discussed in \sref{sec:toy grokking model}.

Besides the test-error critical exponent we estimate several properties of the feature distributions. In particular, we calculate the effective dimension $D_{\rm eff}$, the divergence exponent $\xi$ of the sample PDF at the boundary of the domain, and the distance between positive and negative samples $\varepsilon$. These quantities are calculated from the training-dataset features $z_i(x)$. To calculate the effective dimension $D_{\rm eff}$ we first find $\sigma_k$ defined as the fraction of the variance explained by the $k$th principal component of the training dataset features $z_i(x)$. Then we calculate the entropy $S$ of the ratios $\sigma_k$ defined as $S=-\sum_k\sigma_k\log\sigma_k$. Finally, the effective dimension is obtained as the exponent of the entropy, i.e. $D_{\rm eff}=\ee^S$. We report the effective dimensions for the considered \texttt{Examples 1-3} in Table \ref{tab:distribution data}.

\begin{figure}[!htb]
    \centering
    \includegraphics[width=0.32\textwidth]{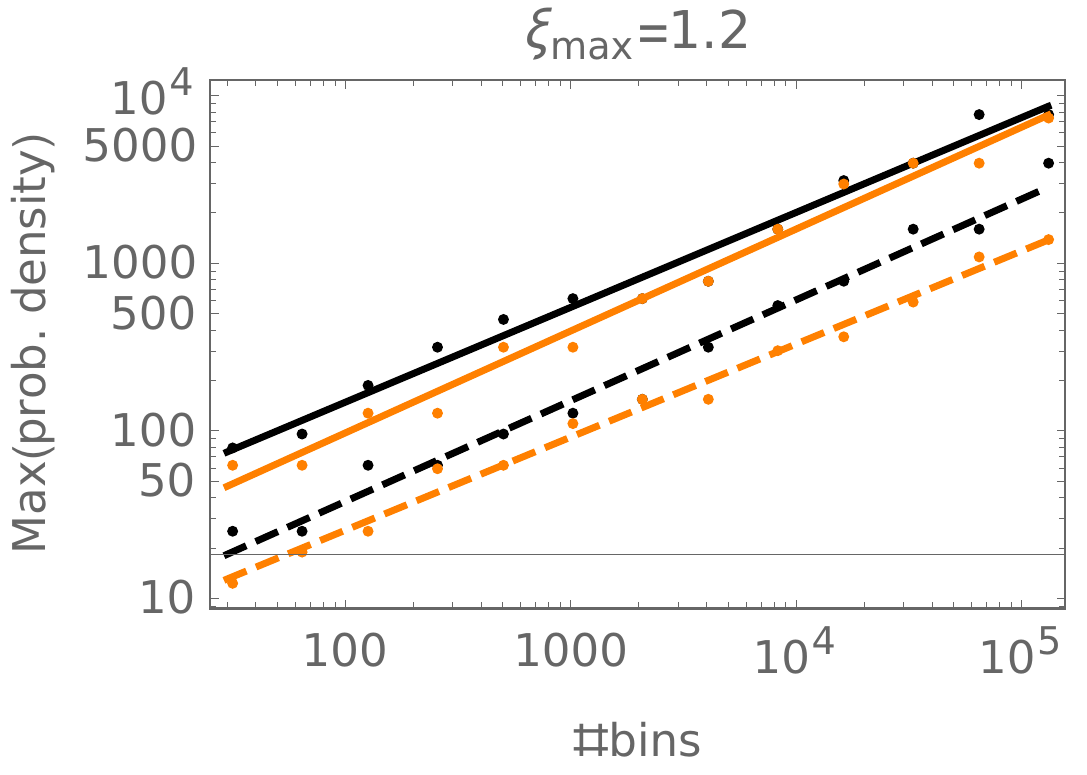}
    \includegraphics[width=0.32\textwidth]{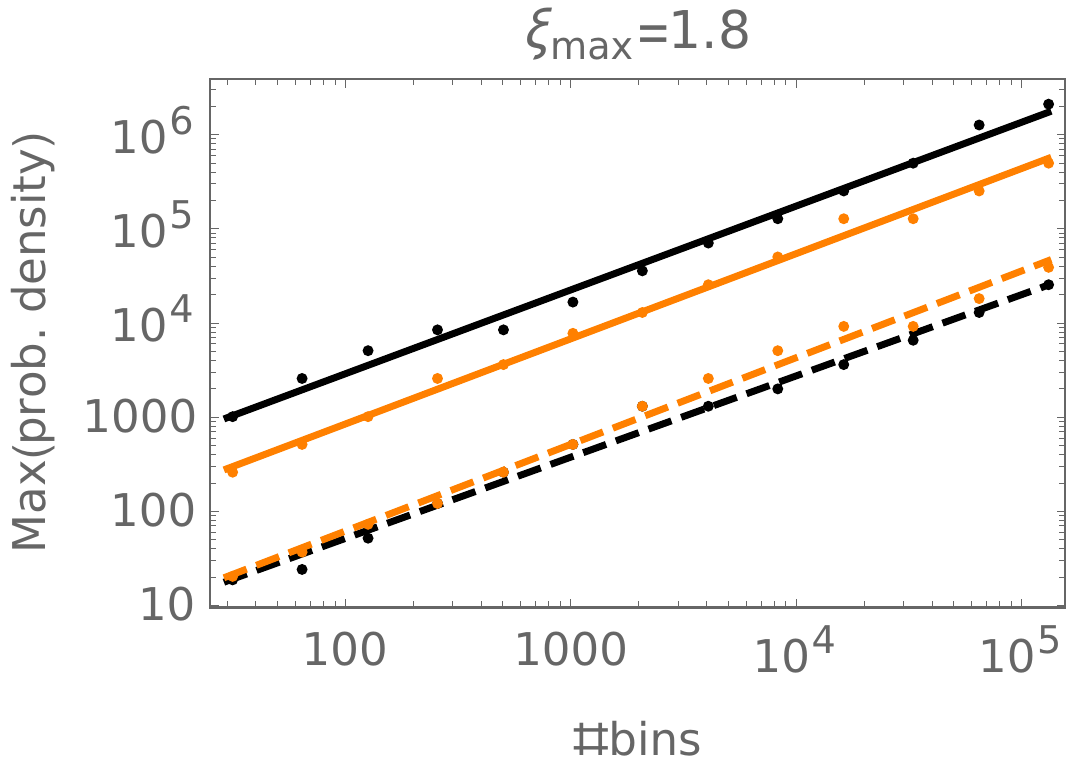}
    \includegraphics[width=0.32\textwidth]{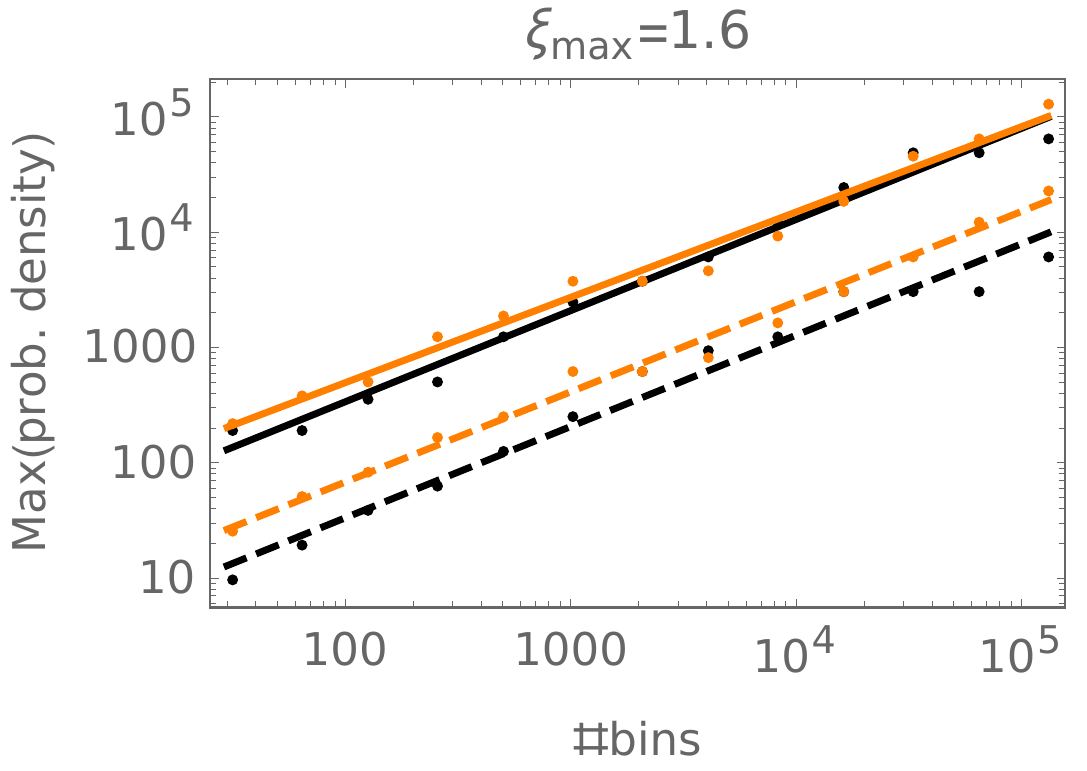}
    \caption{The estimated PDF maximum of the positive (black) and negative (orange) samples. The dashed line fits correspond to the left context vectors $v^{\rm L}H^{\rm L}_{\rm N}(i)$ and the full line fits to the right context vectors $H^{\rm R}_{\rm N}(i)v^{\rm R}$. The final value reported in the panel title is obtained as a maximum sum of the left and right divergence exponents.  From left to right we report results for \texttt{Example  1}, \texttt{Example  2}, and \texttt{Example  3} attention tensors $A$ given in \aref{app:fixed attention}.}
    \label{fig:density fixed attention}
\end{figure}

We also use the vectors $z_i(x)$ to estimate the divergence exponent of the sample PDF at the boundary of the domain. In the considered case, the vectors $z_i(x)$ are a tensor product of three vectors. Therefore, we estimate the divergence in the PDF by focusing separately on each of the components of the vector $z_i(x)$. One of the vectors is a constant vector determined by the embedding function and does not contribute to the divergence exponent. The remaining, important parts are the left and the right context vectors, namely $v^{\rm L}H^{\rm L}_N(i)$ and $H^{\rm R}_N(i)v^{\rm R}$. We consider the normalised context vectors which, in addition, have size two (since we fix $d=2$). Therefore, they are uniquely determined by the angle with the first component and we can accordingly estimate the divergence at the boundary of the domain by studying the PDF of the angle. We estimate the divergence exponent by looking at the behaviour of the estimated PDF at the boundary with an increasing number of bins. The final exponent is obtained as a sum of the exponents obtained from the left and the right-attention part of the feature vector $z_i(x)$. We find (see \fref{fig:density fixed attention}) that the PDF diverges algebraically with the powers reported in Table \ref{tab:distribution data}. 

In \sref{sec:general critical exponents} we derived a relation between the exponents $\nu$, $\xi$ and the effective dimension for a simple $D-$dimensional ball model. Interestingly, we find that the relation given by \eref{eq:critical divergent sphere} obtained from a simple spherically symmetric model is reasonably close in two out of the three considered cases (see Table \ref{tab:distribution data}). 

We also estimate the class separation from the actual feature space distribution and report it in the units of the intra-class variance (see Table \ref{tab:distribution data}). 
\begin{table}[!htb]
    \centering
    \begin{tabular}{c|c|c|c|c|c}
        & $D_{\rm eff}$ & $\nu$ & $\xi^*$ & $\xi=\tfrac{1}{2}(D_{\rm eff}-2\nu +1)$  (\eref{eq:critical divergent sphere})  & $\epsilon$\\
            \hline
        \texttt{Example  1} & 3.0 & 0.85 & 1.2 & 1.15  & 1.45 \\ 
        \texttt{Example  2} & 3.8 & 0.75 & 1.8 & 1.65 & 1.46  \\
        \texttt{Example  3} & 3.0 & 0.84 & 1.6 & 1.16  & 1.6 
    \end{tabular}
    \caption{Critical exponent $\nu$ and numerically calculated characteristic parameters of the feature vector $z_i(x)$ distribution. We also compare the numerically estimated divergence of the sample PDF at the boundary $\xi^*$ with the prediction of the spherical model (\eref{eq:critical divergent sphere}).}
    \label{tab:distribution data}
\end{table}

\begin{figure}[!htb]
    \centering
    \includegraphics[width=0.326\textwidth]{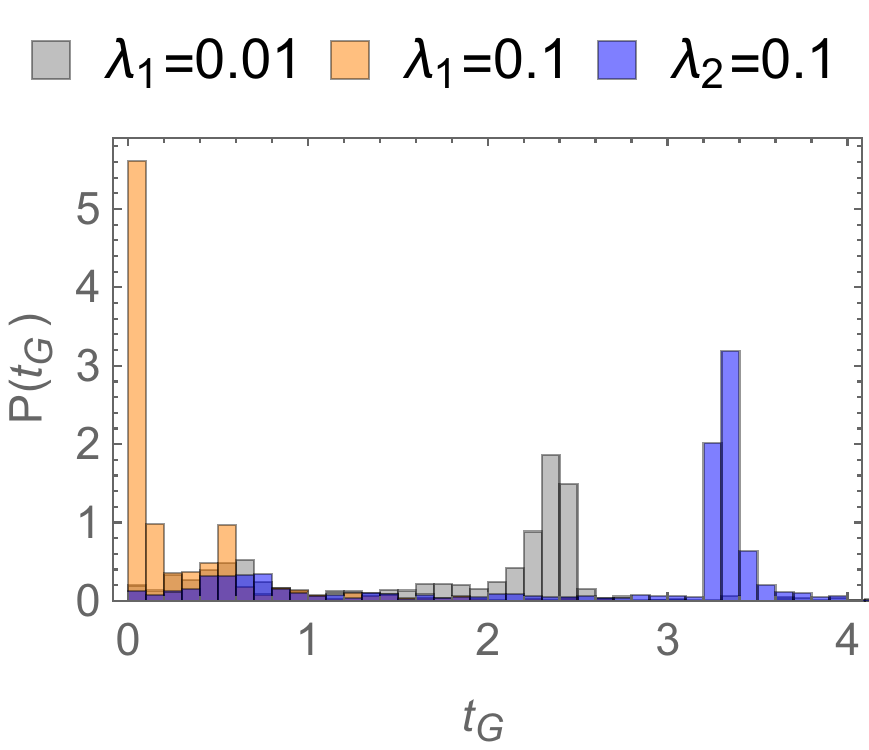}
    \includegraphics[width=0.326\textwidth]{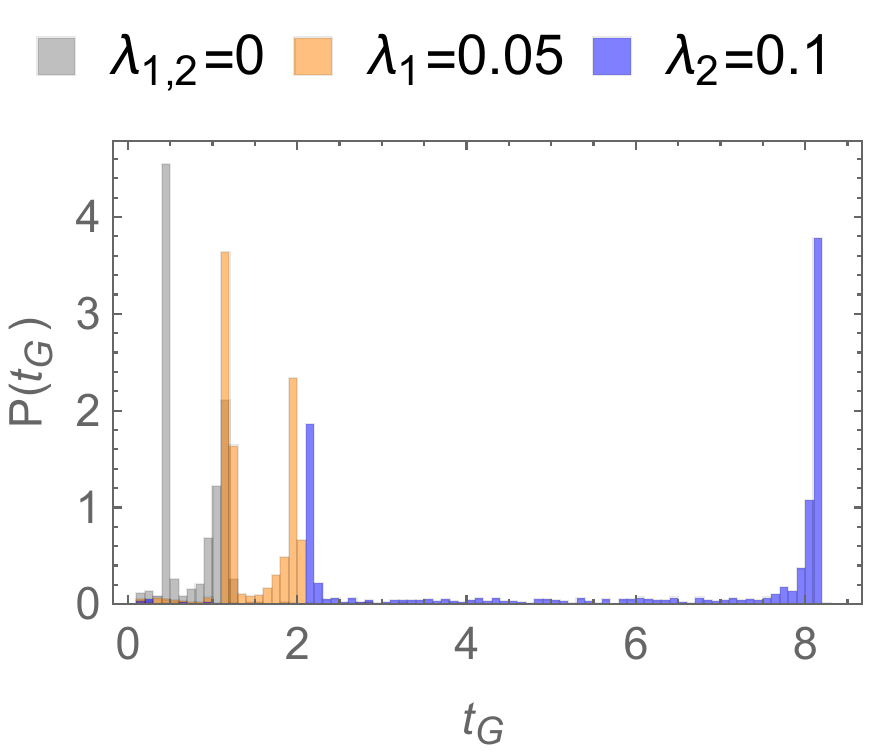}
    \includegraphics[width=0.334\textwidth]{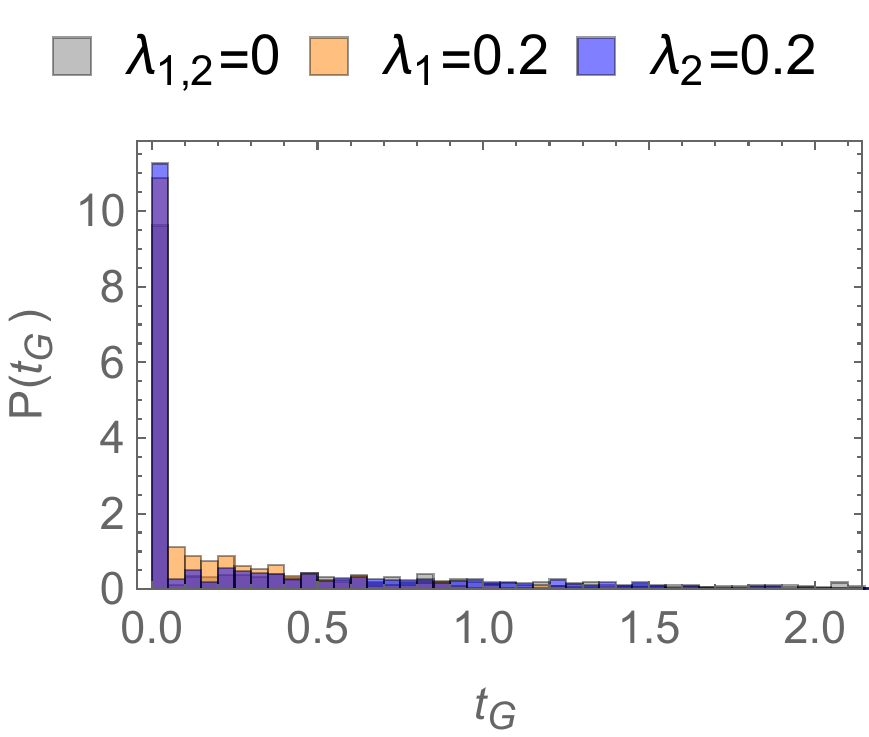}
    \caption{We show the estimated grokking-time PDFs for three fixed attention tensors: \texttt{Example 1} (left), \texttt{Example  2} (middle), and \texttt{Example  3} (right). In most cases, the grokking-time PDF is bimodal, which is in agreement with the prediction of the simple grokking model discussed in \sref{sec:grokking time n-ball}.}
    \label{fig:fixed attention grokking time}
\end{figure}

\paragraph{Grokking time} Finally, we estimate grokking-time PDF, see \fref{fig:fixed attention grokking time}. We do not expect that the prediction of \sref{sec:grokking time n-ball} will quantitatively describe the estimated grokking-time PDF. Besides the ``worst-case'' initial condition assumption, the condition $N\gg1$ is not valid. Hence, the actual value of $G$ is far from identity. However, some qualitative predictions of the $D-$dimensional ball model can still be observed. First we notice, the bimodal structure of the estimated grokking-time PDF. In the spherical model, the two peaks are a consequence of the separation between the slow and fast modes, where the dynamics of the slow modes was essentially determined by the regularisation strength $\lambda_2$. Similarly, in all three considered cases (i.e. \texttt{Example 1-3}) we can separate the eigenvalues of $G$ by size in two sets. In one set the eigenvalues are by one order of magnitude larger than in the other. However, increasing the regularisation strength $\lambda_2$ often leads to increased grokking time, which is not the case in the simple uniform ball model. The discrepancy is a consequence of the non-diagonal matrix $G$, which mixes different components of the vector $z_i(x)$. We also observe that a larger effective dimension $D_{\rm eff}$ (see Table \ref{tab:distribution data}) leads to longer grokking times and a larger class separation $\varepsilon$ to smaller grokking times. The last two observations are in agreement with the $D-$dimensional ball model discussed in \sref{sec:grokking time n-ball}.

The presented results are obtained by averaging over many initialisations of the classification tensor $B$. We sample the initial elements of $B$ from a normal distribution with zero mean and unit variance. Changing the initial distribution can impact the results. Determining the effect of the initial distribution of $B$ on the grokking-time PDF and the critical exponent is left for future research.

\subsubsection{Full model training and structure formation}
\label{sec:full model training results}
In this section, we discuss grokking and structure formation properties of the complete student model introduced in \sref{sec:student model}. We initialise the model with a random initial condition, where all the entries of the tensors $A,B$ are uncorrelated and sampled according to a normal distribution with zero mean and unit variance. We train the model with the Adam optimiser (with standard parameter setting) and learning rate 0.005.  We use the same loss as in the previous sections \eref{eq:loss general} with $L_{1,2}$ regularisation strength $\lambda_{1,2}\in[0,0.001]$. The regularisation strength is the same for the attention tensor $A$ and the classifier tensor $B$. We add a sigmoid non-linearity before the final sign non-linearity to improve the training stability and reduce the training time. We consider only open boundary conditions and set $x_{-1}=x_{M+1}=-1$. Finally, in the main text we consider only the 1-local rule. We discuss the 2-- and 3--local rules in \aref{app:3,4 local rules}.
We perform tests in three situations, namely, without regularisation ($\lambda_{1,2}=0$), with $L_2$ regularisation ($\lambda_1=0$, $\lambda_2=0.0001$), and with $L_1$ regularisation ($\lambda_1=0$, $\lambda_2=0.001$). We chose the regularisation strengths $\lambda_{1,2}$ to be the largest regularisation strengths with only few spikes in the training loss after the grokking time. To obtain zero test error it is sufficient (in almost all cases) to train only the attention parameters $A$ and fix the classification parameters $B$. This is a consequence of the gauge symmetry of the tensor attention layer~\cite{zunkovic2022Deep}. However, we will always train all model parameters. Since the full tensor-attention model is non-linear, we do not expect the theory developed in \sref{sec:toy grokking model} to be valid. On the other hand, we do observe phenomena related to neural collapse~\cite{papyan2020prevalence} and structure formation \cite{liu2022towards}.

\begin{figure}[!htb]
    \centering
    \includegraphics[width=1.\textwidth]{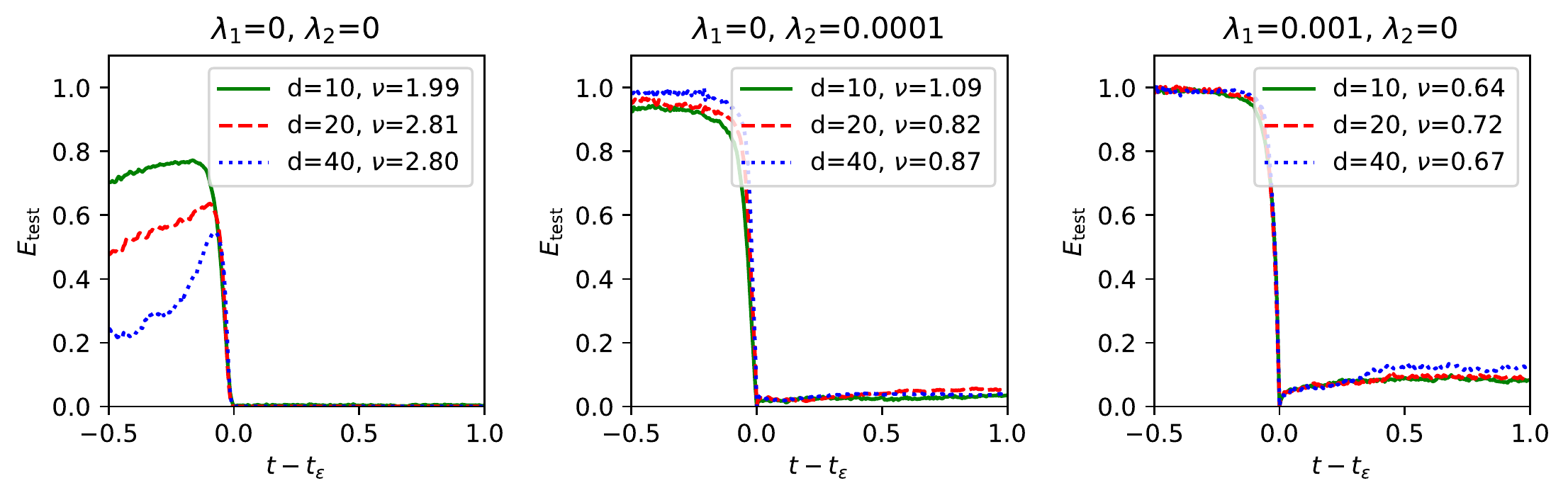}
    \caption{The average test error at the phase transition. We align the first point where the test error becomes zero (i.e. the time  $t_\epsilon$) and take the average over many ($\sim 1000$) initialisations of the model parameters. Training without regularisation results in larger critical exponent $\nu\approx 2.5$ as training with $L_1$ ($\nu\approx 0.7$) or $L_2$ ($\nu\approx0.9$) regularisation. In all experiments we used learning rate 0.005. In the legends we report the bond dimension of the trained models $d$ and the corresponding fitted critical exponent $\nu$.}
    \label{fig:error full model}
\end{figure}

\paragraph{Average test error and average effective dimension} First, we investigate the dynamics of the average test error and calculate the critical exponent $\nu$. In \fref{fig:error full model} we show that the critical exponent decreases upon increasing regularisation. Larger regularisation leads to a sharper transition to zero test error, in contrast with the linear case studied in the \sref{sec:grokkin probability ball model} and in the \sref{sec:constant attention tensors}, where the critical exponent was found to be independent of the regularisation strengths $\lambda_{1,2}$. The test error drops to zero at the grokking transition. Following the grokking transition, the test error is non-zero and experiences fluctuations. These fluctuations can be detected as sharp increases in the training loss and are more common in models with large regularisation. Therefore, the $L_{1,2}$ regularised models have larger average test error after the grokking transition.

\begin{figure}[!htb]
    \centering
    \includegraphics[width=1.\textwidth]{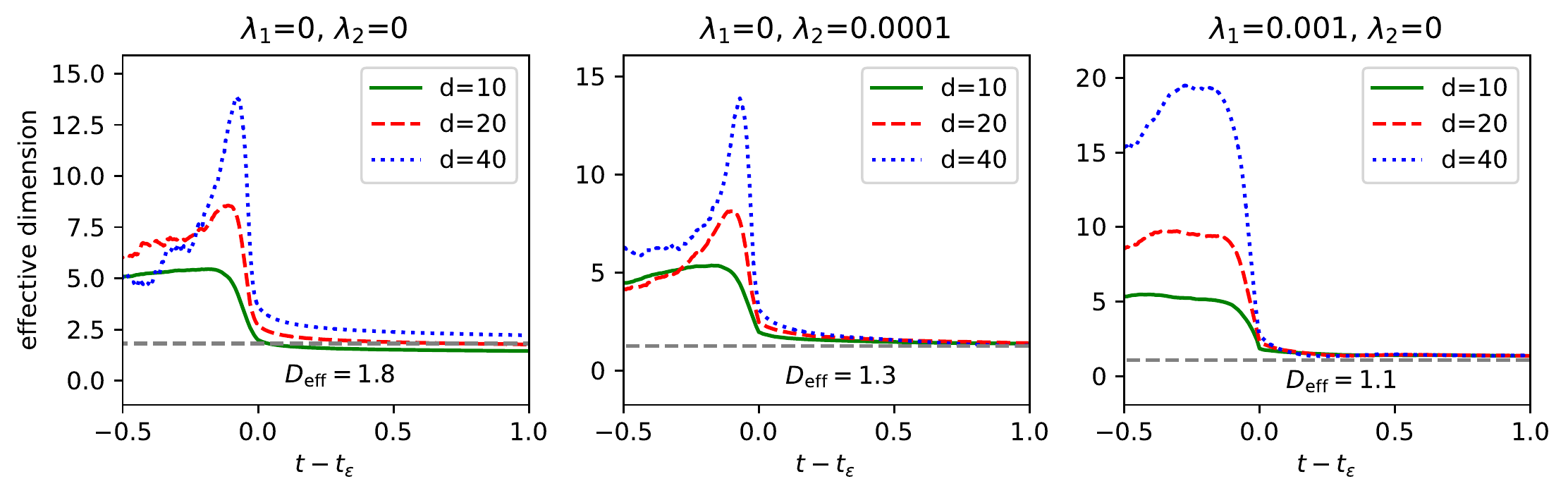}
    \caption{The average effective dimension corresponding to the error in \fref{fig:error full model}. Larger regularisation results in a smaller effective dimension $D_{\rm eff}$ after the grokking transition. The horizontal dashed line corresponds to the average of the minimal effective dimension of data with zero test error (over all samples with fixed $d$).}
    \label{fig:effective dimension full model}
\end{figure}

The shape of the average test error close to transition point $t_\epsilon$ (or the critical exponent $\nu$) depends only slightly on the model size (bond dimension $d$). This suggests that the effective dimension of the mapped data $D_{\rm eff}$ is independent of the model size. We confirm this by calculating the effective dimension of the features $z(i)$. Since we study only open boundary conditions, we consider only the effective dimension of the left context vectors $v^{\rm L}H^{\rm L}_{\rm N}(i)$. The right context vectors $H^{\rm R}_{\rm N}(i)v^{\rm R}$ have the same properties because of the model symmetry. As shown in \fref{fig:effective dimension full model}, the average effective dimension drops significantly just before the grokking transition. We observe that regularisation significantly decreases the effective dimension of the mapped vectors $v^{\rm L}H^{\rm L}_{\rm N}(i)$. The effective dimension is smallest with $L_1$ regularisation, which is expected since the $L_1$ regularisation enforces sparsity while the $L_2$ regularisation enforces smoothness. 

\begin{figure}[!htb]
    \centering
    \includegraphics[width=0.49\textwidth]{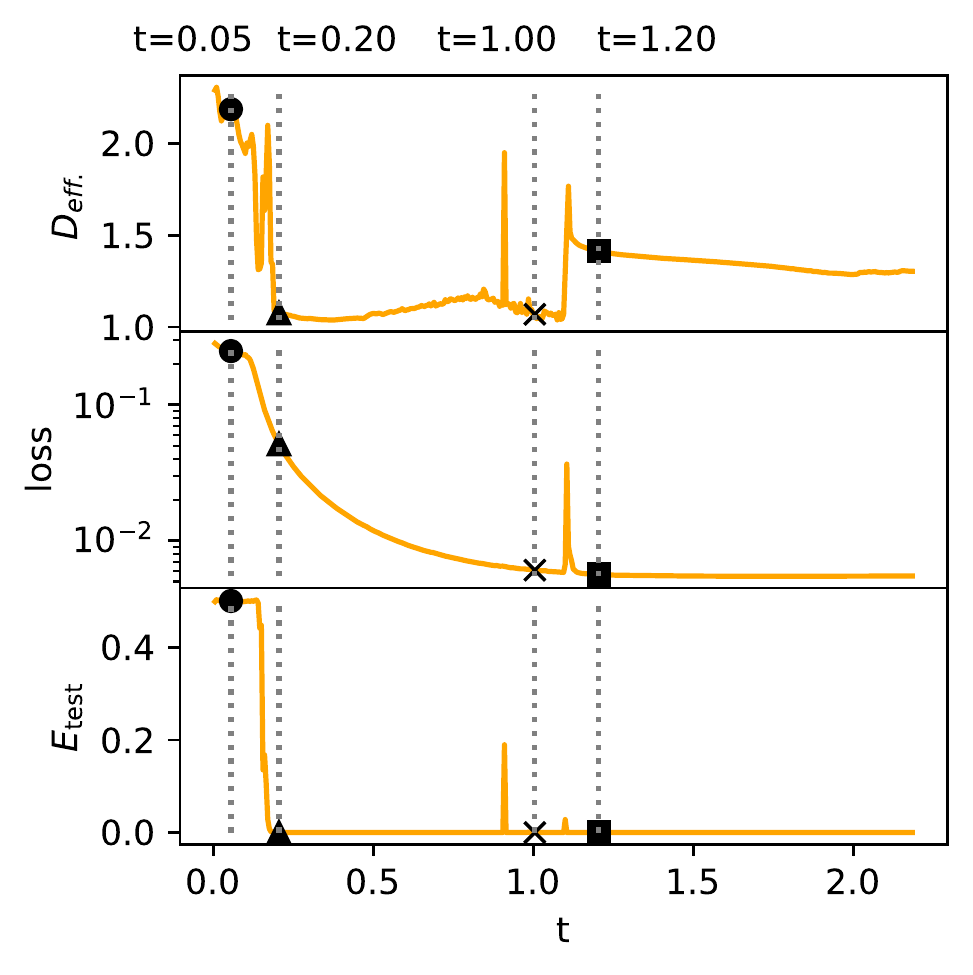}
    \includegraphics[width=0.5\textwidth]{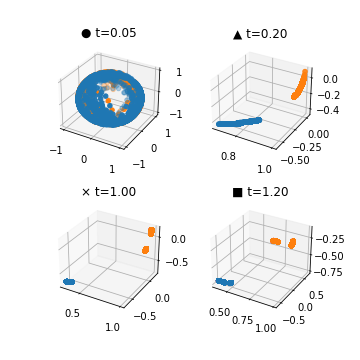}
    \caption{Several emergent structures in the feature space (Example 1). The left plots show the effective dimension $D_{\rm eff}$ (top), train loss (middle), and test error (bottom). The black markers show the value of the plotted quantities at specific times marked by vertical dotted lines and written on the top of the left panels. The right panels show the structure of the features at the marked times. We observe that an essentially one dimensional feature distribution with two distinct islands of features splits into an almost 2D feature distribution with three isolated islands. We can detect this transition as a sharp peak in the loss and a step in the effective dimension $D_{\rm eff}$.}
    \label{fig:structure formation example 1}
\end{figure}

\begin{figure}[!htb]
    \centering
    \includegraphics[width=0.49\textwidth]{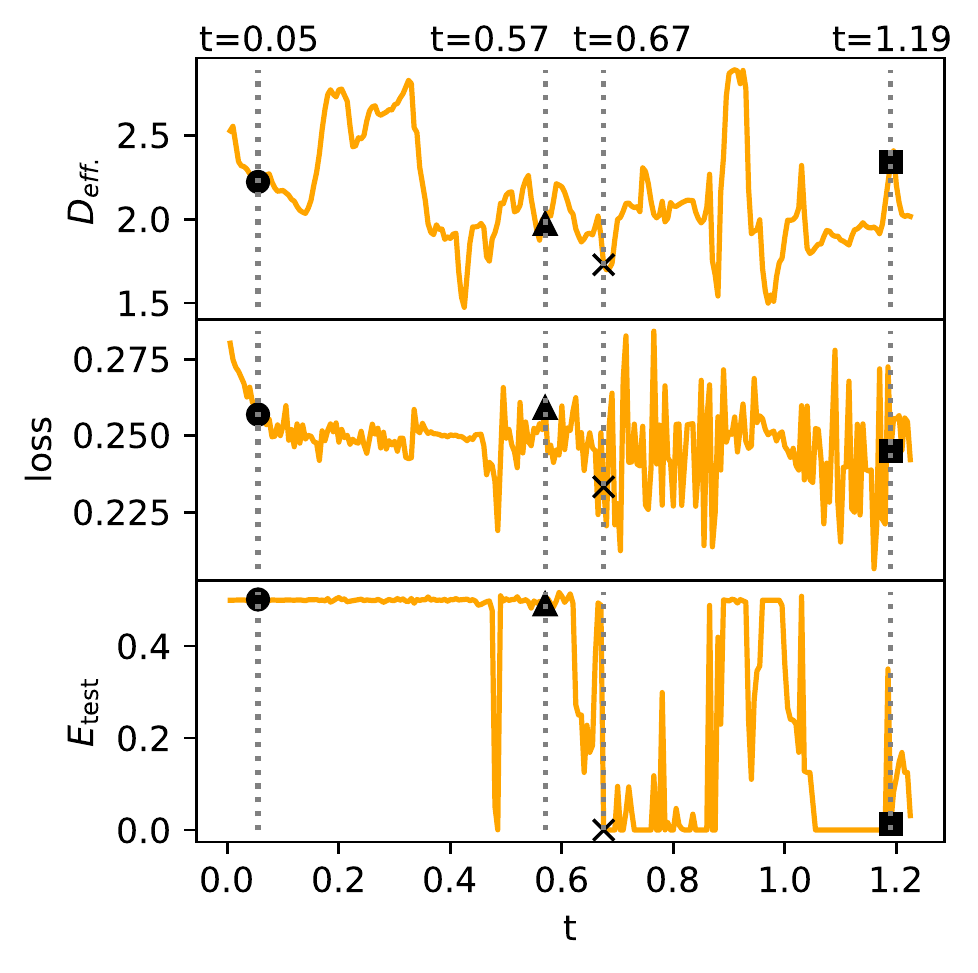}
    \includegraphics[width=0.5\textwidth]{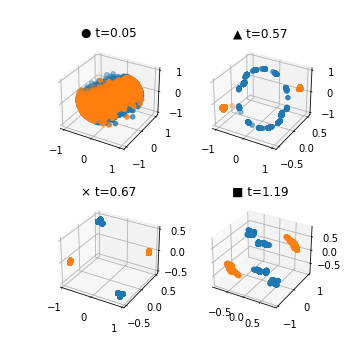}
    \caption{Several emergent structures in the feature space (Example 2). The left plots show the effective dimension $D_{\rm eff}$ (top), train loss (middle), and test error (bottom). Black markers show the value of the plotted quantities at times marked by vertical dotted lines and written on the top of the first plot. The right panels show the structure of the features at the marked times. We observe that structured data also appears in the case of high test error ($t=0.57$). The zero-test-error structure ($t=0.67$) is different compared to the example in \fref{fig:structure formation example 1}. Here, we see a 2D structure with four isolated feature islands. Finally, at time $t=1.19$ the structure starts to disappear while the test error is still considerably small (smaller than $1\%$).}
    \label{fig:structure formation example 2}
\end{figure}

\paragraph{Structure formation and grokking} We relate the decrease of the effective dimension to structure formation. As can be seen in examples shown in \fref{fig:structure formation example 1} and \fref{fig:structure formation example 2}, small effective dimension signals an emergent feature space structure which, however, can be different in each example. Similarly, in \cite{liu2022towards} the authors argue that the grokking in deep models is related to structure formation.
Our findings differ from those of \cite{liu2022towards} in that we discuss ensemble/average phenomena. The authors of \cite{liu2022towards} discuss the connection between grokking and structure formation on the single-model level. In contrast, we argue that grokking and structure formation are related on average as shown in \fref{fig:error full model} and \fref{fig:effective dimension full model}, and not for every training run individually. We call the structure formation and grokking for a single training of a model the model-wise structure formation and the model-wise grokking. We disentangle model-wise structure formation from model-wise grokking by observing specific training samples. We typically observe the appearance of a simple structure in data in the proximity of the grokking transition. This is consistent with a sharp drop of the average effective dimension at the transition (shown in \fref{fig:effective dimension full model}). Additionally, we observe that feature space structures can be different for different model parameters and initialisations.

In \fref{fig:structure formation example 1} and \fref{fig:structure formation example 2} we show the structures appearing in the features $v^{\rm L}H^{\rm L}_{\rm n}(i)$ with bond dimension $d=3$. In \fref{fig:structure formation example 1} we see that the structure of the feature space data changes also during a single run. This can be detected as a spike in the training loss or as a step-like jump in the effective dimension (see \fref{fig:structure formation example 1}). The structures can change from lower to higher dimensional and vice versa, e.f. see \fref{fig:structure formation example 1} -- the transition between $t=1$ and $t=1.2$ increases the effective dimension $D_{\rm eff.}$ of the mapped data. Appearance of geometric structures in the latent space does not necessary lead to good generalistion, i.e. small test error (see \fref{fig:structure formation example 2} at time $t=0.57$). Finally, we also show in \fref{fig:structure formation example 2} that we can have a small generalisation/test error with complex or not apparent feature space structures (see \fref{fig:structure formation example 2} at time $t=1.19$). These empirical observations suggest that grokking and structure formation are not related model-wise. That structure formation and grokking are in general two distinct phenomena is further corroborated by our simple grokking model discussed in \sref{sec:toy grokking model}, which does not require any special geometric structure (aside from the condition of linear separability).

\begin{figure}[!htb]
    \centering
    \includegraphics[width=\textwidth]{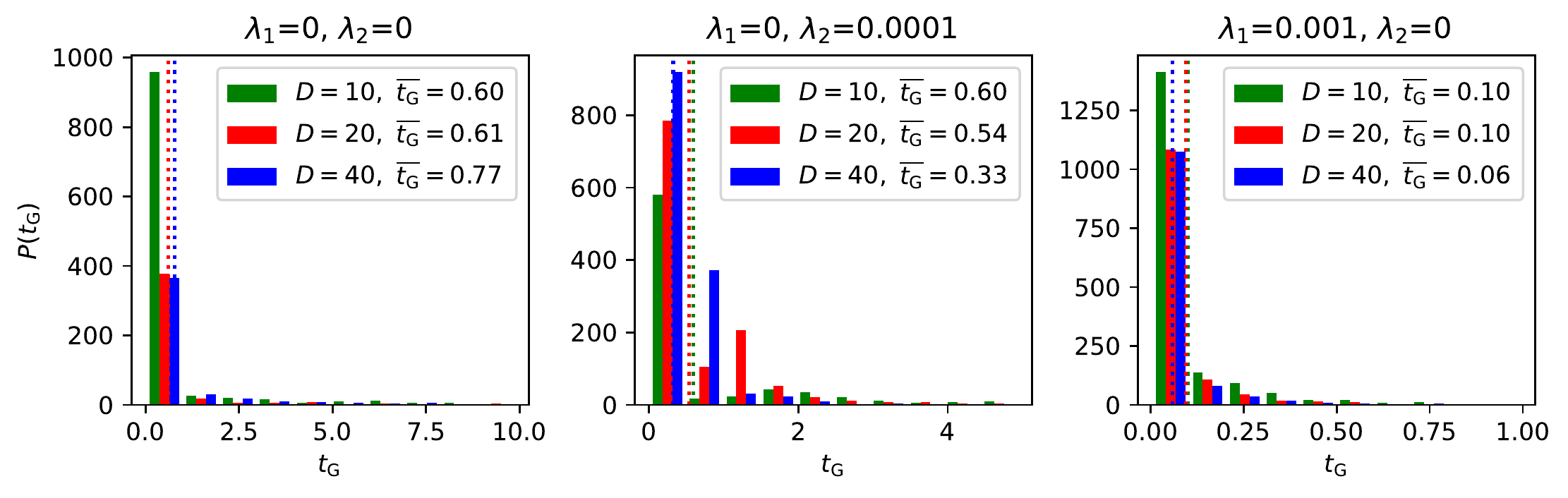}
    \includegraphics[width=\textwidth]{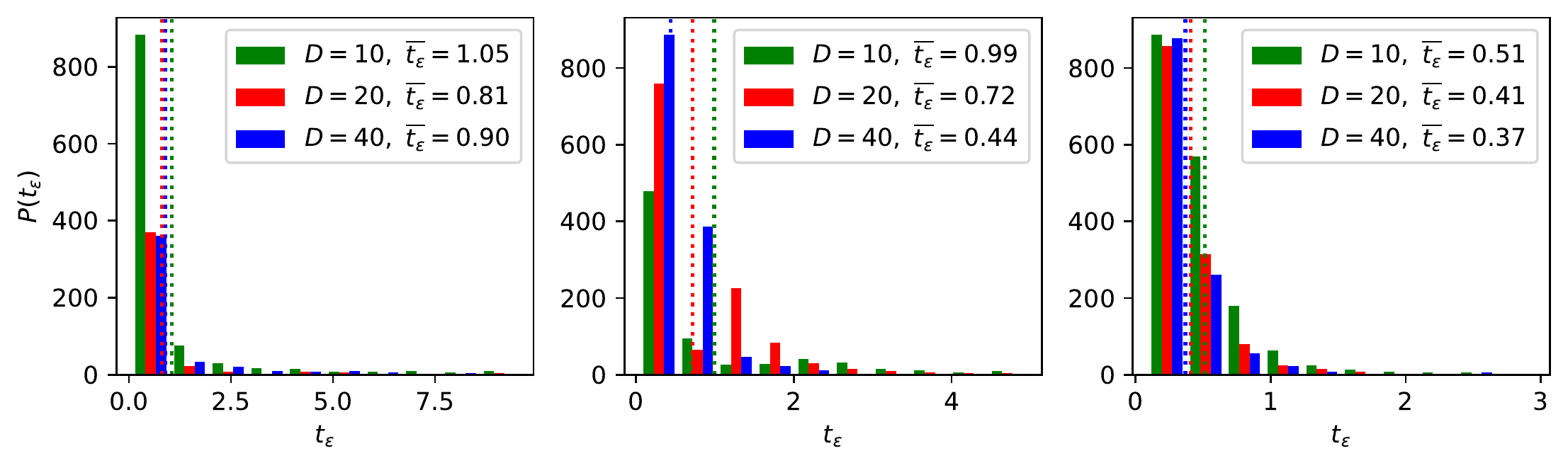}
    \caption{The estimated grokking-time PDF and $t_\epsilon$ PDF. The colors correspond to different models sizes, namely $d=10$ (blue), $d=20$ (orange), and $d=40$ (green). The vertical lines correspond to the averages reported in the legends of the panels. We find that $L_1$ regularisation reduces $t_{\rm G}$ and $t_\epsilon$. By contrast, the $L_2$ regularisation decreases only the absolute time $t_\epsilon$. In the $L_2$ case and $d=20$ we also find a clear bimodal grokking time distribution. In all cases we set the learning rate to 0.005.}
    \label{fig:grokking time full model}
\end{figure}

\paragraph{Grokking time} Finally, we estimate the PDF of the grokking times, see \fref{fig:grokking time full model} (top panels). Taking the non-regularised case as the baseline, we find that $L_1$ regularisation decreases the average grokking time $\overline{t_{\rm G}}$ significantly more than $L_2$ regularisation. Further, grokking times for $L_2$ regularised models increase with the model size. On the other hand, non-regularised and $L_1$ regularised models have roughly a model-size-independent grokking-time PDF, and hence the grokking-time average.

Since the grokking time is measured relative to the time at which the zero train error is achieved, we estimate also the PDF of times $t_\epsilon$ (zero-test-error time). We find that both $L_1$ and $L_2$ reduce the time at which the zero test error is attained. Therefore, both, the $L_1$ and the $L_2$ regularisation decrease the number of steps required for good generalisation. In addition, the $L_1$ generalisation seems to be more efficient, in the sense, that there is a shorter time interval with a large difference between training and test error.

\section{Summary and discussion}
\label{sec:discussion}
We analyse grokking from two perspectives. First, we propose a simple grokking setup (\textit{perceptron grokking}) and consider two solvable grokking models. Second, we introduce a tensor-network attention map and connect the standard statistical-mechanics teacher-student setup with the perceptron grokking setup.

\paragraph{Perceptron grokking} By studying two solvable grokking models, we show that grokking is a phase transition and calculate the critical exponent, grokking probability, and grokking-time PDF. The obtained analytic expressions allow us to determine the effect of model and training parameters on the grokking probability and the grokking-time PDF. In particular, we find a stark difference between the $L_1$- and $L_2$-regularised models. The $L_1$-regularised models have a higher grokking probability and a shorter grokking time as the $L_2$-regularised models. We also obtain a universal expression for the test-error critical exponent of spherically symmetric models, which is relevant in the transfer learning setting, where only the last layer is retrained.

\paragraph{Learning local-rules with shallow tensor networks} We use the tensor-network attention model with fixed attention tensors $A$ to test the predictions of the perceptron grokking setup on a 1D cellular-automaton rule-30 learning task. Our prediction of the critical exponent roughly agrees with the numerical estimation, thereby validating the grokking scenario on a simple problem. On the other hand, the grokking-time PDF approximation, which invokes strong assumptions, predicts the actual numerical estimate only qualitatively. 

We also perform the training of the full tensor-network student model. Similarly as in solvable perceptron grokking models, we observe a difference between the $L_1$-regularised and the $L_2$-regularised models. The former have a shorter grokking time and a lower effective dimension, which agrees with the analytic predictions of the perceptron grokking models. Therefore, we expect that $L_1$ regularisation leads to improved generalisation properties (e.g. smaller test error) compared to $L_2$ regularisation also in a more general classification setting. 

In the case of training the full tensor-network student model we discuss the connection between grokking and structure formation. We determine the grokking transition by observing the average test error. We show that the average effective dimension of the feature-space data sharply decreases at the grokking transition. By observing specific models, we also find that small effective dimensions correspond to particular feature-space structures. We accordingly relate grokking with structure formation on the ensemble level. By contrast, we find several models with zero test error without apparent feature-space structures and vice versa. This shows that in specific (though rare) cases, the test error drops to zero even if no structure is present in the data. Similarly, simple structures can appear during training also when the test error does not vanish. We thus separate the grokking and the structure formation on the level of individual training runs.

As a distinct feature of the full tensor-network training we highlight the spikes in the training loss. We observe that spikes become more frequent with larger $L_{1,2}$ regularisation. We also relate the training-loss spikes to the changes in the feature-space structures, which may become less or more complex during training. We can assess the shape of the feature-space structures by observing the effective dimension, which shows a step-like behaviour whenever we observe a training-loss spike. Typically less complex structures correspond to a smaller effective dimension. These findings can be relevant for deep-neural-network training where training-loss spikes are also observed. Frequent training-loss spikes can be avoided by using a smaller regularisation. Moreover, we can determine whether the model parameters should be reverted by monitoring the feature-space effective dimension. For example, we can revert the model only if the training-loss spikes correspond to increased effective dimension.

Finally, the proposed tensor-network map connects the grokking phenomena, which have so far been observed only in deep models, with the standard teacher-student learning setup. The considered local tensor-network rule learning setup is an extreme example of a learning rule. The standard teacher-student mean-field setup is the opposite extreme. It would be interesting to study if the proposed grokking setup and the tensor-network map can be extended to study algebraically decaying rules which interpolate between the two extremes. Extending the presented theory to deep neural networks appears to be difficult within the proposed framework.

\section*{Acknowledgement}
The authors received support from Sloveinan research agency (ARRS) project J1-2480. Computational resources were provided by SLING – Slovenian national supercomputing network. We thank Marko Robnik \v{S}ikonja for reading the first version of the draft and providing useful comments.

\bibliographystyle{unsrt}
\bibliography{main}

\appendix

\section{Grokking time in the 1D model}
\label{app:1D grokking time}
In this appendix, we provide the details of the grokking-time PDF calculation in the 1D model discussed in the main text.

First, we calculate the joint probability to find $x_{\rm min}$ and $\bar{x}$ for a given number of positive and negative training samples $N$,
\begin{align}
    \label{eq:joint xbar xmin}
    P_{N}(x_{\rm min},\bar{x})=&\prod_{k=1}^N\left[\int_0^\infty\dd \tilde{x}_k P(\tilde{x}_k)\int_0^\infty\dd x_k P(x_k)\right]\delta\left(x_{\rm min}-\min(\{x_l\})\right)\delta\left(2\bar{x}-\frac{1}{N}\sum_{l=1}^N(x_l-\tilde{x}_l)\right)\\ \nonumber
    &=2\prod_{k=1}^N\left[\int_0^\infty\dd x_k P(x_k)\right]\delta\left(x_{\rm min}-\min(\{x_l\})\right)P_N^{\rm exp}\left(\bar{x}_+-2\bar{x}\right).
\end{align}
With $\bar{x}_+$ we denote the average value of the positive samples, $\bar{x}_+=\frac{1}{N}\sum_{l=1}^Nx_l$. We calculate the PDF of the minimum given in \eref{eq:joint xbar xmin} by considering its cumulative density
\begin{align}
    \label{eq:cumulative joint xbar xmin}
    C_{N}(x_{\rm min}; \bar{x})=&2\prod_{k=1}^N\left[\int_{x_{\rm min}}^\infty\dd x_k P(x_k)\right]P^{\rm exp}_N\left(\bar{x}_+-2\bar{x}\right)\\ \nonumber
    =&2\prod_{k=1}^N\left[\int_{0}^\infty\dd x_k P(x_k+x_{\rm min})\right]P^{\rm exp}_N\left(\bar{x}_++x_{\rm min}-2\bar{x}\right) \\ \nonumber
    =&2\prod_{k=1}^N\left[\int_{0}^\infty\dd x_k P(x_k)\right]P(N x_{\rm min})P^{\rm exp}_N\left(\bar{x}_++x_{\rm min}-2\bar{x}\right) \\ \nonumber
    =&2\int_{0}^\infty\dd \bar{x}_+P_N(\bar{x}_+)P(N x_{\rm min})P^{\rm exp}_N\left(\bar{x}_++x_{\rm min}-2\bar{x}\right) \\ \nonumber
    =&\frac{2^{\frac{3}{2}-N} N^{N+\frac{1}{2}} e^{-N
   x_{\rm min}} \Theta (2 \bar{x}-x_{\rm min}) (2
   \bar{x}-x_{\rm min})^{N-\frac{1}{2}}
   K_{N-\frac{1}{2}}(2 N \bar{x}-N
   x_{\rm min})}{\sqrt{\pi } \Gamma
   (n)}\\ \nonumber
   &+\frac{\sqrt{\pi } 2^{\frac{3}{2}-N}
   N^{N+\frac{1}{2}} e^{-N x_{\rm min}} \csc (\pi  N)
   \Theta (x_{\rm min}-2 \bar{x}) (x_{\rm min}-2
   \bar{x})^{N-\frac{1}{2}} K_{N-\frac{1}{2}}(N
   (x_{\rm min}-2 \bar{x}))}{\Gamma (1-N) \Gamma
   (N)^2}.
\end{align}
For $N=2$ we find
\begin{align}
    C_{N=2}(x_{\rm min}; \bar{x})=&-e^{-4 \bar{x}} (4 \bar{x}-2 x_{\rm min}+1)
   (\Theta (2 x_{\rm min}-4 \bar{x})-1)\\ \nonumber
   &-e^{4\bar{x}-4 x_{\rm min}} (4 \bar{x}-2
   x_{\rm min}-1) \Theta (2 x_{\rm min}-4 \bar{x}) 
\end{align}
We obtain the PDF by taking the derivatives of the cumulative probabilities \eref{eq:cumulative joint xbar xmin} with respect to $x_{\rm min}$, namely $P_{N}(\bar{x},x_{\rm min})=-\frac{\partial C_{N}(x_{\rm min};\bar{x})}{\partial x_{\rm min}}$. For $N=2$ we get
\begin{align}
    P_{N=2}(\bar{x},x_{\rm min})=&2 e^{-4 \bar{x}} \Theta (2 \bar{x}-x_{\rm min})-2
   e^{4 \bar{x}-4 x_{\rm min}} (8 \bar{x}-4
   x_{\rm min}-1) \Theta (x_{\rm min}-2 \bar{x}).
\end{align}
Next, we calculate the joint probability for the grokking time $t_{\rm G}$ and the average $\bar{x}$ 
\begin{align}
    P_{N,\epsilon,\lambda}(t_{\rm G},\bar{x})=&\frac{\partial x_{\rm min}}{\partial t_{\rm G}}P_{N}(\bar{x},x_{\rm min}(t_{\rm G},\epsilon,\bar{x}_\lambda)).
\end{align}
While in principle we can derive a closed-form expression for in arbitrary $N$, they are not particularly informative and we thus write here only the expressions for $N=2$ and $\lambda_2=0$ (to shorten the notation we use $t$ instead of $t_{\rm G}$)
\begin{align}
    P_{N=2,\epsilon,\lambda_1}(t,\bar{x})=&2 e^{t-4 \bar{x}} (\epsilon_\lambda -\bar{x}) \Theta \left(\bar{x}-\epsilon_\lambda  \tanh
   \left(\frac{t}{2}\right)\right)
   \\ \nonumber
   &+2 (\bar{x}-\epsilon_\lambda ) e^{4 \left(e^{t} (\bar{x}-\epsilon_\lambda )+\epsilon_\lambda
   \right)+t} \left(4 e^{t} (\bar{x}-\epsilon_\lambda )+4 \bar{x}+4 \epsilon_\lambda -1\right)
   \Theta \left(\epsilon_\lambda  \tanh \left(\frac{t}{2}\right)-\bar{x}\right).
\end{align}
Finally, we integrate out the average of the samples $\bar{x}$ and obtain the grokking-time PDF.

\section{Grokking probability in the D-dimensional ball model}
\label{app:grokking ball N>>1}
In this section we derive the grokking probability in the $D-$dimensional ball model in the limit of many training samples, i.e. $N\gg1$. The condition for zero test error is
\begin{align}
    \frac{w^\lambda_1}{||w^\lambda||_2}>\frac{1}{\varepsilon},
\end{align}
and can be rewritten as
\begin{align}
    (\epsilon^2-1)(w^\lambda_1)^2\geq (w^\lambda_2)^2+(w^\lambda_3)^2+\ldots (w^\lambda_D)^2,\quad\mbox{~and~} \quad (w^\lambda_1)>0.
\end{align}
The stationary solution $ w^\lambda=G^{-1}a$ is
\begin{align}
    G=&\frac{1}{2N}\sum_{i=1}^{2N} \tilde{x}^i\otimes \tilde{x}^i+\lambda_2\mathds{1}_D =\frac{1}{2N}\sum_{i=1}^Nx^i\otimes x^i+\epsilon\otimes\epsilon + \epsilon\otimes\bar{x}+\bar{x}\otimes\epsilon+\lambda_2\mathds{I}_D , \\ \nonumber
     a=&\frac{1}{2N}\sum_{i=1}^{2N} y^i \tilde{x}^i -\lambda_1\sgn(w) = \bar{x}-\lambda_1\sgn(w),
\end{align}
where $\bar{x}=\frac{1}{2N}\sum_{i=1}^{2N}y^ix^i$. In the limit $N\gg1$ we can separate two contributions to the matrix $G=A+B$ where
\begin{align}
    A=&\lambda_D\mathds{I}_D+\epsilon\otimes\epsilon, \\ \nonumber
    B=&\frac{1}{2N}\sum_{i=1}^{2N}x^i\otimes x^i-\frac{1}{D+2}\mathds{I}_D+\epsilon\otimes\bar{x}+\bar{x}\otimes\epsilon,
\end{align}
where $\lambda_D=\frac{1}{D+2}+\lambda_2$. In the limit $N\gg1$ we have $||B||_F=\mathcal{O}(1/\sqrt{N})$, hence we can approximate the inverse of the matrix $G$ as,
\begin{align}
    G^{-1}\approx A^{-1}-A^{-1}BA^{-1}.
\end{align}
The stationary solution can thus be approximated by
\begin{align}
    \label{eq:w lambda full}
    w^\lambda \approx& A^{-1}a - A^{-1}BA^{-1}a \\ \nonumber
    \approx & A^{-1}(\bar{x}+\epsilon-\sgn(w)\lambda_1) - A^{-1}BA^{-1}(\epsilon+\sgn(w)\lambda_1),
\end{align}
where we have kept only the first nontrivial order in $1/\sqrt{N}$. We will separately consider the case $\lambda_1=0$ and the case $\lambda_1>0$.

\subsection{\textbf{Case } $\mathbf{\lambda_1=0}$} By explicitly evaluating the above expression, \eref{eq:w lambda full}, and assuming $\lambda_1=0$, we find
\begin{align}
    \label{eq:general w}
    w_1^\lambda=\beta+\alpha_1\bar{x}_1+\alpha_2\overline{x^2_1},\quad w_{j>1}^\lambda=\alpha_3\bar{x}_j+\alpha_4\overline{x_1x_j},
\end{align}
where
\begin{align}
    \label{eq:alpha beta}
    \beta=&\frac{\varepsilon}{\lambda_D+\varepsilon^2}+\frac{\varepsilon}{(\lambda_D+\varepsilon^2)^2(D+2)},  \\ \nonumber \alpha_1=&\frac{1}{\lambda_D+\varepsilon^2}-\frac{2\varepsilon^2}{(\lambda_D+\varepsilon^2)^2},\\ \nonumber \alpha_2=&-\frac{\varepsilon}{(\lambda_D+\varepsilon^2)^2},\\ \nonumber
    \alpha_3=&\frac{1}{\lambda_D}-\frac{\varepsilon}{\lambda_D(\lambda_D+\varepsilon^2)} ,\\ \nonumber
    \alpha_4=&-\frac{\varepsilon}{\lambda_D(\lambda_D+\varepsilon^2)}.
\end{align}

The first few nontrivial moments of the uniform distribution in a $D-$dimensional ball are reported in Table~\ref{tab:ball statistics}. 
\begin{table}[!htb]
    \centering
    \begin{tabular}{c|c|c}
        Statistics & Mean & Second moment \\
        \hline 
        $x_j$ & 0 & $\frac{1}{D+2}$ \\ 
        $x_j^2$ & $\frac{1}{D+2}$ & $\frac{3}{8+6D+D^2}$ \\
        $x_ix_j$, $i\neq j$ & 0 & $\frac{1}{8+6D+D^2}$
    \end{tabular}
    \caption{First nontrivial moments of the uniform ball distribution. All odd moments vanish.}
    \label{tab:ball statistics}
\end{table}
Considering the variances and the means in Table \ref{tab:ball statistics}, we find that (in the limit $N\gg1$) all random variables appearing in \eref{eq:general w} to be normally distributed,
\begin{align}
    \bar{x}_1\sim~&\mathcal{N}\left(0,\frac{1}{2N(D+2)}\right), \\ \nonumber
    \overline{x^2_1}\sim~& \mathcal{N}\left(\frac{1}{D+2},\frac{D+1}{N(D+2)^2(D+4)}\right),\\  \nonumber
    \overline{x_1x_{j>1}}\sim~& \mathcal{N}\left(0,\frac{1}{2N(8+6D+D^2)}\right).
\end{align}
The distributions are independent since all the necessary covariances vanish. 

The sum of independent normal distributions is again a normal distribution, leading to 
\begin{align}
    \label{eq:w distribution w1}
    w_1\sim~&\mathcal{N}_1(w_1)=\mathcal{N}\left(\beta +\frac{\alpha_2}{D+2},\frac{\alpha_1^2}{2N(D+2)}+\frac{\alpha_2^2(D+1)}{N(D+2)^2(D+4)}\right), \\ 
    \label{eq:distribution wj}
    w_{j>1}\sim~& \mathcal{N}\left(0,\frac{\alpha_3^2}{2N(D+2)}+\frac{\alpha_4^2}{2N(8+6D+D^2)}\right).
\end{align}
The grokking probability is then given by
\begin{align}
    \label{eq:full grokking probability}
    P_{E(\infty)=0}=\int_0^\infty\dd w_1 \mathcal{N}_1(w_1) \int_{0}^{(\epsilon^2-1)w_1^2\left(\frac{\alpha_3^2}{2N(D+2)}+\frac{\alpha_4^2}{2N(8+6D+D^2)}\right)}\dd r\,\chi^2_{D-1}(r). 
\end{align}
In the limit $N\gg\lambda_D\gg1$ we make following approximations
\begin{align}
    \label{eq:simplified w}
    \beta\approx\frac{\varepsilon}{\lambda_D+\epsilon^2},\quad \alpha_1\approx\frac{1}{\lambda_D+\epsilon^2},\quad\alpha_2=0,\quad \alpha_3\approx\frac{1}{\lambda_D},\quad \alpha_4=0.
\end{align}
With these simplifications, \eref{eq:full grokking probability} reduces to the grokking probability obtained in the main text, see \eref{eq:grokking probability D ball}.

\subsection{\textbf{Case } $\mathbf{1\gg\lambda_1>0}$} By explicitly evaluating the expression in \eref{eq:w lambda full} we find
\begin{align}
    \label{eq:general w 2}
    w_1^\lambda=&\begin{cases}\beta-\frac{\lambda_1\sgn(w_1^\lambda)}{\lambda_D+\varepsilon^2}+\alpha_1\bar{x}_1+\alpha_2\overline{x^2_1} &,~ (\lambda_D+\varepsilon^2)|\beta+\alpha_1\bar{x}_1+\alpha_2\overline{x^2_1}|>\lambda_1 \\ 0 & \mbox{,~else}\end{cases},\\ \nonumber
    w_{j>1}^\lambda=&\begin{cases}-\frac{\lambda_1\sgn(w_j^\lambda)}{\lambda_D}+\alpha_3\bar{x}_j+\alpha_4\overline{x_1x_j} &,~ \lambda_D|\alpha_3\bar{x}_j+\alpha_4\overline{x_1x_j}|>\lambda_1 \\ 0 &\mbox{,~else}\end{cases},
\end{align}
with $\alpha_j$ and $\beta$ given in \eref{eq:alpha beta}. The number of non-vanishing components of the stationary solution $w^\lambda$ depends on the value of $\lambda_1$. Therefore, we get (in the $N\gg1$ limit) an additional sum over the number of non-zero elements in the $w^\lambda$,
\begin{align}
    P_{E(\infty)=0}=&\int^\infty_{\frac{\lambda_1}{\lambda_{2,D}+\varepsilon^2}}\dd w_1\mathcal{N}_1(w_1)\Bigg[ (1-p_\lambda)^{D-1} \\ \nonumber
    &+\sum_{k=1}^{D-1}\binom{D-1}{k}p_\lambda^k(1-p_\lambda)^{D-1-k}\int_{0}^{(\epsilon^2-1)\left(w_1- \frac{\lambda_1}{\lambda_{2,D}+\varepsilon^2}\right)^2\left(\frac{\alpha_3^2}{2N(D+2)}+\frac{\alpha_4^2}{2N(8+6D+D^2)}\right)}\dd r R_k(r) \Bigg],
\end{align}
where $p_\lambda=1-\text{erf}\left(\frac{\lambda_1}{\lambda_D}/\sqrt{\frac{\alpha_3^2}{2N(D+2)}+\frac{\alpha_4^2}{2N(8+6D+D^2)}}\right)$ is the probability that $|w_{j>1}|$ (sampled from \eref{eq:distribution wj})is larger than $\lambda_1/\lambda_D$. With $R_k(r)$ we denote the PDF of the sum of squares of $k$ random variables sampled from the truncated normal distribution. As in the main text (see \sref{sec:grokkin probability ball model}), we can calculate a lower bound on the grokking probability by discarding the sum over $k>1$,
\begin{align}
    \label{eq:grokking probability D ball bound full}
    P_{E(\infty)=0}\geq&(1-p_\lambda)^{D-1}\int^\infty_{\frac{\lambda_1}{\lambda_{2,D}+\varepsilon^2}}\dd w_1\mathcal{N}_1(w_1)\\ \nonumber
    \approx&\frac{(1-p_\lambda)^{D-1}}{2}\left(1+\text{erf}\left(\left( \beta +\frac{\alpha_2}{D+2}-\frac{\lambda_1}{\lambda_{2,D}+\epsilon^2}\right)/\sqrt{\frac{\alpha_1^2}{2N(D+2)}+\frac{\alpha_2^2(D+1)}{N(D+2)^2(D+4)}} \right)\right).
\end{align}
Also in the more general case, we find the same difference between the $L_1$ and $L_2$ regularisations as discussed in the main text.

\section{Fixed attention}
\label{app:fixed attention}
Below we specify the attention tensors for the discussed examples, see \sref{sec:constant attention tensors} in the main text:
\begin{itemize}
    \item \texttt{Example  1}: 
    \begin{align*}
        A_0=\left(
\begin{array}{cc}
 0.782735 & 0.225481 \\
 -0.21562 & 0.290028 \\
\end{array}
\right),\quad A_1=\left(
\begin{array}{cc}
 1.17554 & -0.275503 \\
 1.18283 & -0.157563 \\
\end{array}
\right),
    \end{align*}
    \item \texttt{Example 2}: 
    \begin{align*}
        A_0=\left(
\begin{array}{cc}
 1.6749 & -1.29059 \\
 0.285324 & -0.708621 \\
\end{array}
\right),\quad A_1=\left(
\begin{array}{cc}
 0.0462428 & -0.0797724 \\
 -0.509457 & 0.922777 \\
\end{array}
\right),
    \end{align*}
    \item \texttt{Example 3}: 
    \begin{align*}
        A_0=\left(
\begin{array}{cc}
 1.37336 & -0.465853 \\
 -1.10382 & 0.720113 \\
\end{array}
\right),\quad A_1=\left(
\begin{array}{cc}
 0.128517 & 0.166033 \\
 0.634426 & 1.13816 \\
\end{array}
\right).
    \end{align*}
\end{itemize}

\section{Additional results for the 2--local and the 3--local rules}
\label{app:3,4 local rules}
In this appendix we present a similar analysis as the one in \sref{sec:full model training results}, but for the 2--local rule and the 3--local rule. In general we observe a similar behavior as in the 1--local case discussed in the main text.

First, we discuss the test error shown in \fref{fig:error full model rule 2,3}. Again, we find that the critical exponent $\nu$ is roughly independent on the model-size and is smaller for larger rule range $K$.
\begin{figure}[!htb]
    \centering
    \includegraphics[width=1.\textwidth]{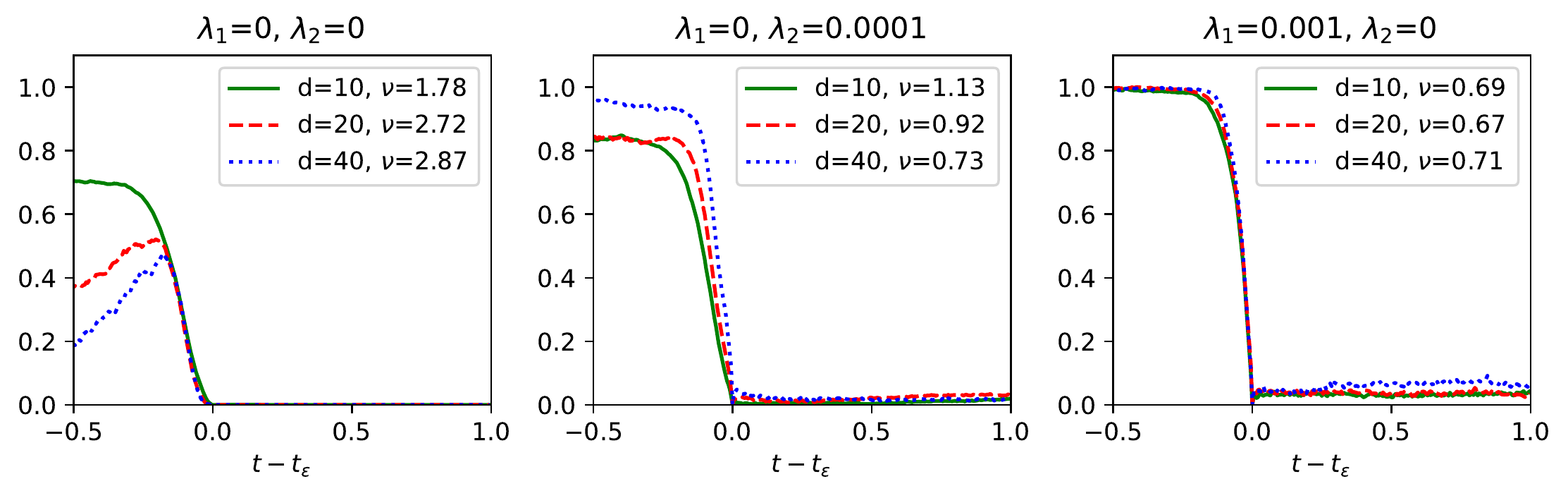}
    \includegraphics[width=1.\textwidth]{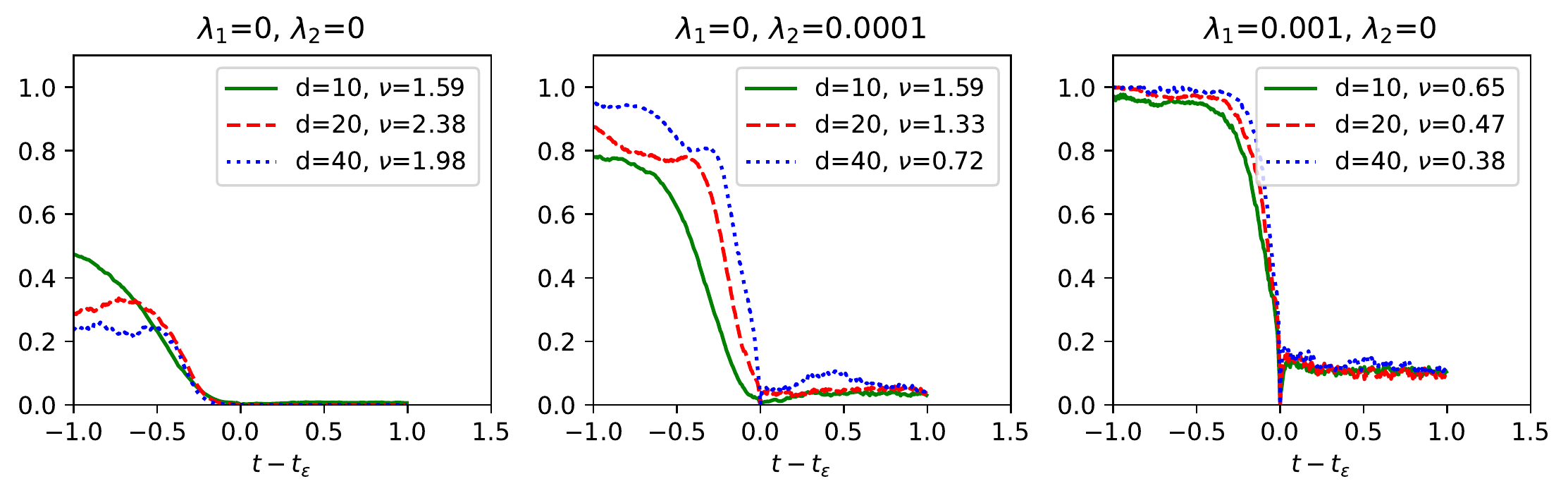}
    \caption{The average error at the phase transition for the 2--local rule (first row) and the 3--local rule (second row). We align the first point where the test error becomes zero (i.e. the time  $t_\epsilon$) and take the average over many ($\sim 1000$) initialisations of the model parameters. As in the 1--local rule case the training without regularisation results in larger critical exponent $\nu\approx 2$ as training with $L_1$ or $L_2$ regularisation. In all experiments we used learning rate 0.005.}
    \label{fig:error full model rule 2,3}
\end{figure}

As in the 1--local case, the grokking transition corresponds to a sharp decrease in the effective dimension of the $H^{\rm L}(i)$ attention vectors, shown in \fref{fig:effective dimension full model rule 2,3}. In the considered rule-learning scenario, the smallest effective dimension is determined by the locality of the rule and is expected to increase exponentially with $K$~\cite{zunkovic2022Deep}. In contrast to the 1--local case, we find that larger regularisation does not necessary correspond to a smaller effective dimension. However, this is only the case for smaller instances where the bond dimension is very close to the effective dimension (or the smallest possible bond dimension with zero test error). For the larger instances we again find that larger regularisation corresponds to smaller effective dimension $D_{\rm eff}$. In this case we also find that models with $L_1$ regularisation have a slightly smaller average effective dimension compared to models with $L_2$ regularisation.
 \begin{figure}[!htb]
    \centering
    \includegraphics[width=1.\textwidth]{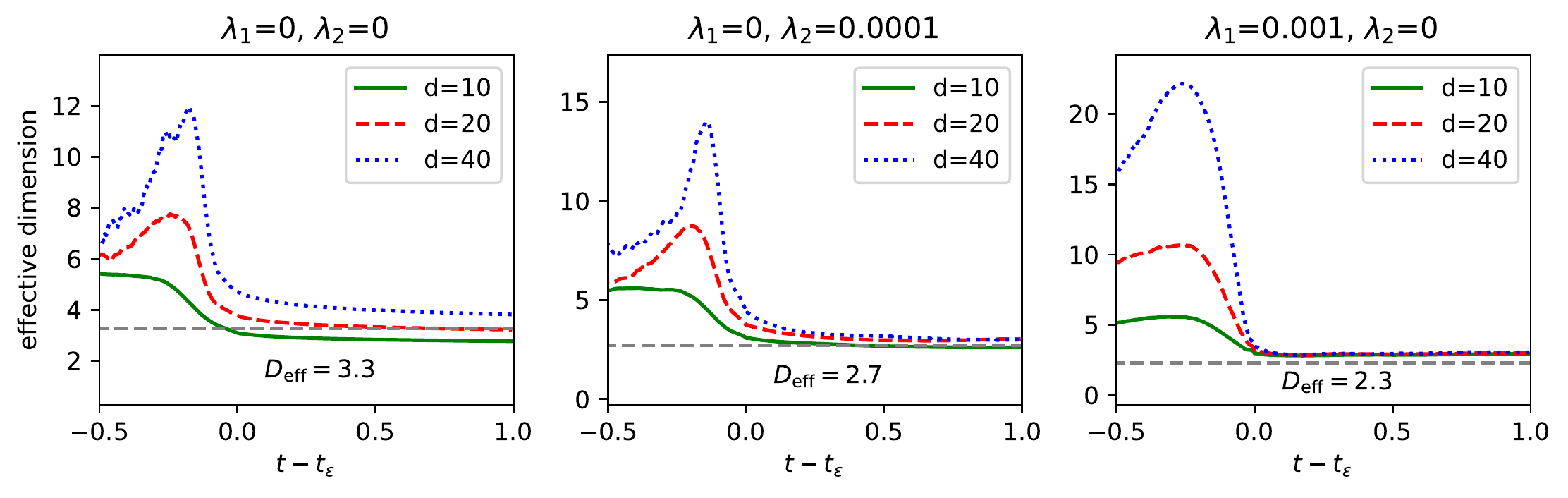}
    \includegraphics[width=1.\textwidth]{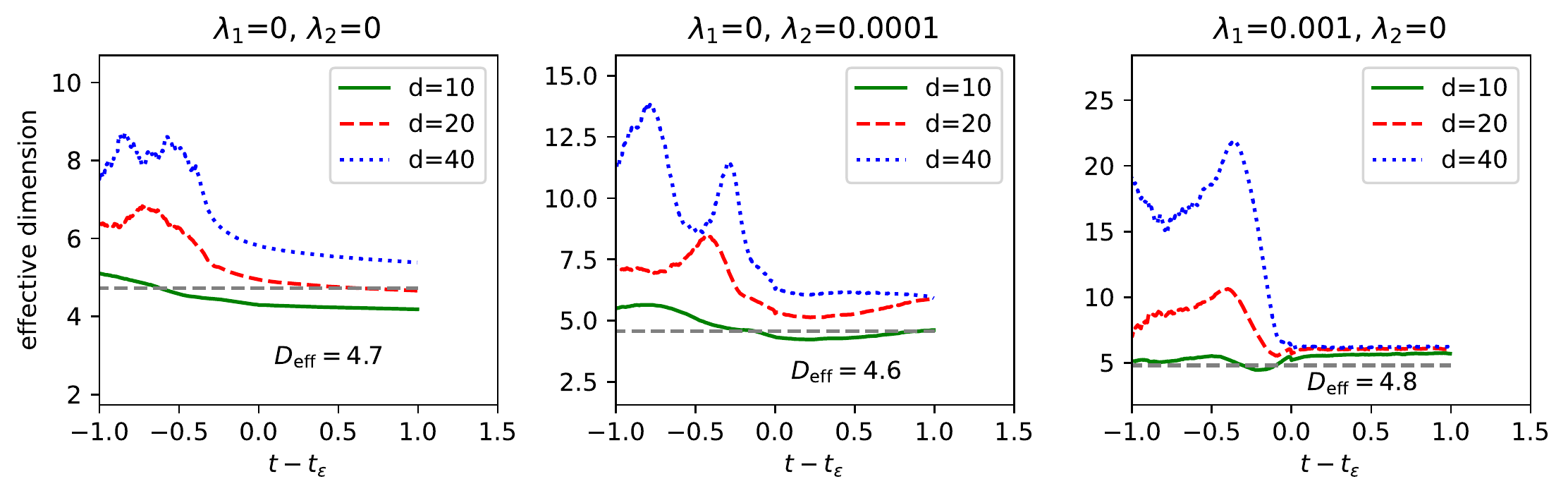}
    \caption{The average effective dimension of the 2--local rule (first row) and the 3--local rule (second row) corresponding to the error in \fref{fig:error full model rule 2,3}. The horizontal dashed line corresponds to the average (over the bond dimension $d$) of the minimal effective dimension of data with zero test error (over all samples with fixed $d$).}
    \label{fig:effective dimension full model rule 2,3}
\end{figure}

Finally, we estimate the grokking-time PDF and the generalisation-time ($t_\epsilon$) PDF for the 2--local and the 3--local rule, shown in \fref{fig:grokking time full model rule 2,3} and \fref{fig:t_epsilon density full model rule 2,3}.
\begin{figure}[!htb]
    \centering
    \includegraphics[width=\textwidth]{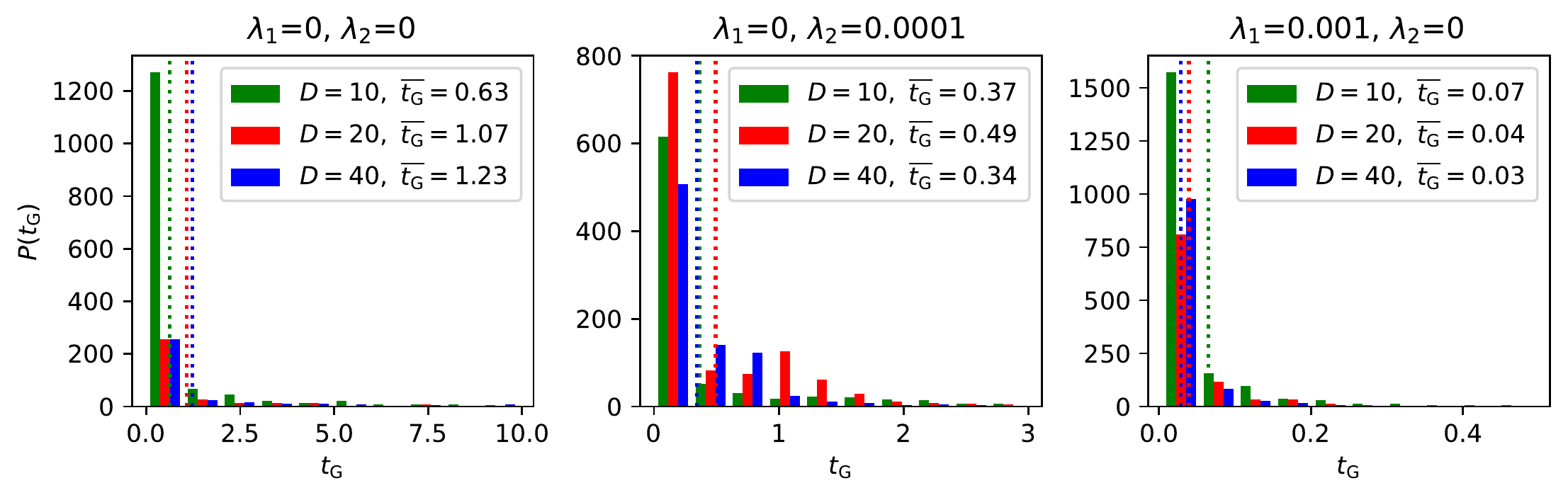}
    \includegraphics[width=\textwidth]{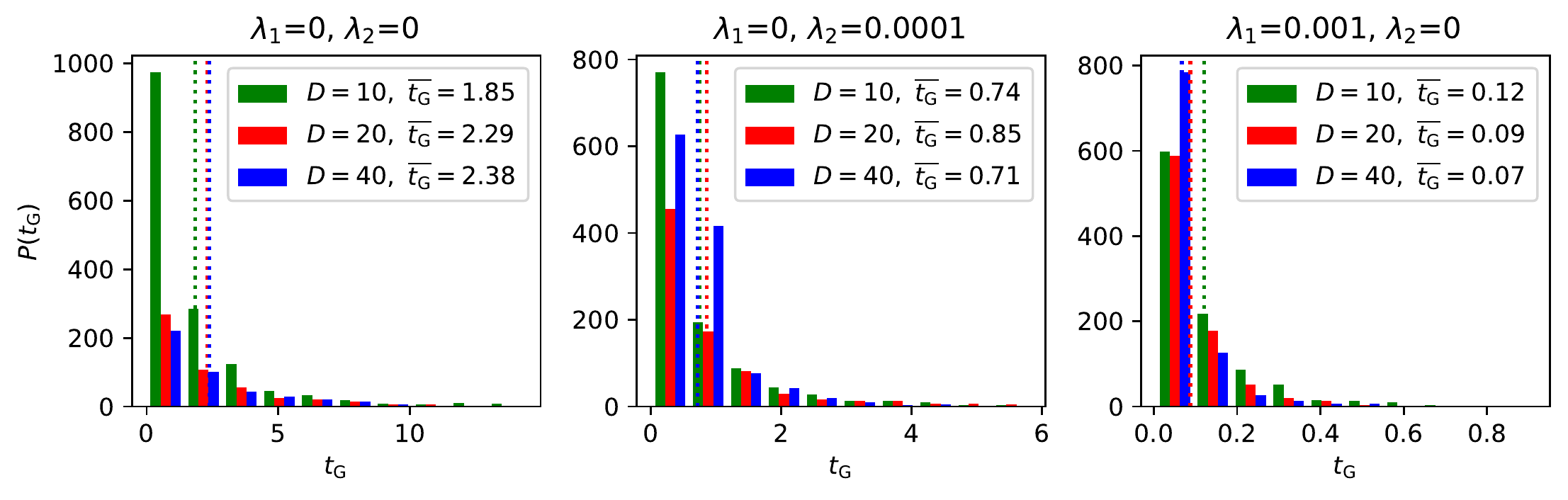}
    \caption{The estimated grokking-time PDF for the 2--local rule (first row) and the 3--local rule (second row). The colors correspond to different models sizes, namely $d=10$ (blue), $d=20$ (orange), and $d=40$ (green). The vertical lines correspond to the averages reported in the legends of the panels. We find that grokking time increases with increased rule range $K$. As in the 1--local case, we find that $L_1$ regularisation reduces $t_{\rm G}$ in all cases. In contrast, the use of $L_2$ regularisation can in some cases increase the average grokking time. In the $L_2$ case we also find a clear bimodal grokking time distribution. In all cases we set the learning rate to 0.005.}
    \label{fig:grokking time full model rule 2,3}
\end{figure}

\begin{figure}[!htb]
    \centering
    \includegraphics[width=\textwidth]{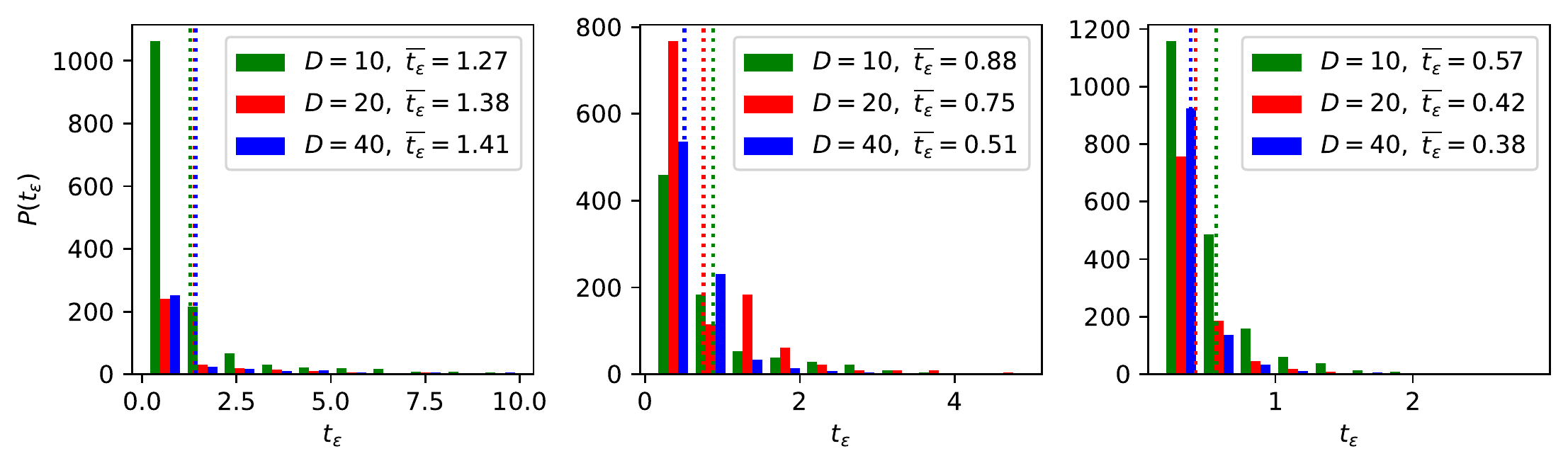}
    \includegraphics[width=\textwidth]{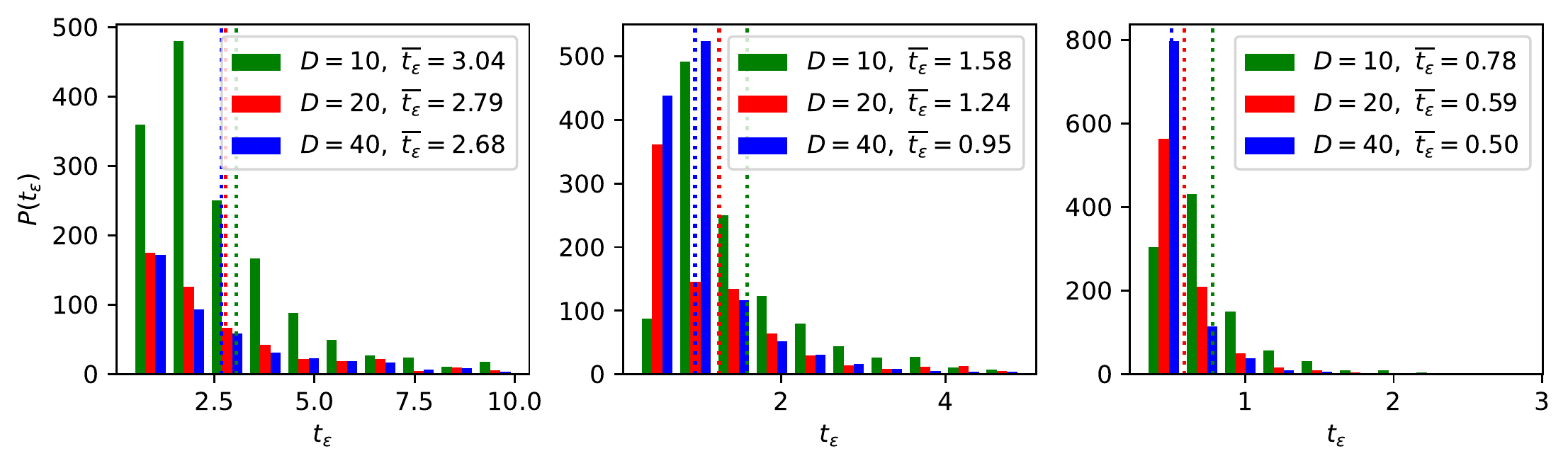}
    \caption{The estimated $t_\epsilon$ PDF for the 2--local rule (first row) and the 3--local rule (second row). The colors correspond to different models sizes, namely $d=10$ (blue), $d=20$ (orange), and $d=40$ (green). The vertical lines correspond to the averages reported in the legends of the panels. In all cases we set the learning rate to 0.005.}
    \label{fig:t_epsilon density full model rule 2,3}
\end{figure}

\end{document}